\documentclass[ALICE,manyauthors]{cernphprep}
\usepackage[comma,square,numbers,sort&compress]{natbib}
\usepackage{hyperref}
\usepackage{lineno}
\usepackage{xspace}
\usepackage{comment}
\usepackage{multirow}
\usepackage{array,booktabs}
\usepackage{amsmath}
\usepackage{orcidlink}

\usepackage{bm}

\usepackage{xcolor}
\definecolor{myOrange}{rgb}{1,0.5,0.}
\definecolor{myGreen}{rgb}{0.0,0.6,0.1}

\newcommand{\removetext}[1]{} 

\newcommand{\pp}           {\ensuremath{\mathrm{pp}}\xspace}

\newcommand{\PbPb}         {\mbox{Pb--Pb}\xspace}

\newcommand{\AuAu}         {\mbox{Au--Au}\xspace}

\newcommand{\aaa}{\mbox{A--A}\xspace}

\newcommand{\s}            {\ensuremath{\sqrt{s}}\xspace}
\newcommand{\snn}          {\ensuremath{\sqrt{s_{\mathrm{NN}}}}\xspace}
\newcommand{\pt}           {\ensuremath{p_{\rm T}}\xspace}

\newcommand{\nineH}        {$\sqrt{s}~=~0.9$~Te\kern-.1emV\xspace}
\newcommand{\seven}        {$\sqrt{s}~=~7$~Te\kern-.1emV\xspace}
\newcommand{\twoH}         {$\sqrt{s}~=~0.2$~Te\kern-.1emV\xspace}
\newcommand{\twosevensix}  {$\sqrt{s}~=~2.76$~Te\kern-.1emV\xspace}
\newcommand{\five}         {$\sqrt{s}~=~5.02$~Te\kern-.1emV\xspace}
\newcommand{\twosevensixnn}{$\sqrt{s_{\mathrm{NN}}}~=~2.76$~Te\kern-.1emV\xspace}
\newcommand{\fivenn}       {$\sqrt{s_{\mathrm{NN}}}~=~5.02$~Te\kern-.1emV\xspace}

\newcommand{\MeVc}         {Me\kern-.1emV/$c$\xspace}
\newcommand{\TeV}          {Te\kern-.1emV\xspace}
\newcommand{\GeV}          {Ge\kern-.1emV\xspace}
\newcommand{\MeV}          {Me\kern-.1emV\xspace}
\newcommand{\GeVmass}      {Ge\kern-.2emV/$c^2$\xspace}
\newcommand{\MeVmass}      {Me\kern-.2emV/$c^2$\xspace}

\newcommand {\etajet} {\ensuremath{\eta_{\mathrm{jet}}}}
\newcommand {\Ntrig} {\ensuremath{N_{\mathrm{trig}}}}

\newcommand {\dphi}        {\ensuremath{\Delta\varphi}\xspace}

\newcommand{\sqrtsNN}{\ensuremath{\sqrt{s_\mathrm{NN}}}}
\newcommand{\sqrts}{\ensuremath{\sqrt{s}}}

\newcommand{\pizero}{\ensuremath{\pi^0}}

\newcommand{\pT}{\ensuremath{p_\mathrm{T}}}

\newcommand{\pTjet}{\ensuremath{p_\mathrm{T, jet}}}
\newcommand{\pTjetch}{\ensuremath{p_\mathrm{T, ch~jet}}}

\newcommand{\pTraw}{\ensuremath{p_\mathrm{T, ch~jet}^\mathrm{raw}}}
\newcommand{\pTrawi}{\ensuremath{p_\mathrm{T, ch~jet}^{\mathrm{raw,}i}}}

\newcommand{\pTreco}{\ensuremath{p_\mathrm{T, ch~jet}^\mathrm{reco}}}
\newcommand{\pTrecoi}{\ensuremath{p_\mathrm{T, ch~jet}^{\mathrm{reco,}i}}}

\newcommand{\pTjetpart}{\ensuremath{p_\mathrm{T, ch~jet}^\mathrm{part}}}
\newcommand{\pTjetdet}{\ensuremath{p_\mathrm{T, ch~jet}^\mathrm{det}}}

\newcommand{\Ajet}{\ensuremath{A_\mathrm{jet}}}
\newcommand{\Ajeti}{\ensuremath{A_\mathrm{jet}^{i}}}
\newcommand{\rhoAi}{\ensuremath{\rho\Ajeti}}
\newcommand{\rhoA}{\ensuremath{\rho\Ajet}}

\newcommand{\cRef}{\ensuremath{c_\mathrm{Ref}}}

\newcommand{\drho}{\ensuremath{\Delta\rho}}

\newcommand{\pTtrig}{\ensuremath{p_\mathrm{T}^\mathrm{trig}}}

\newcommand{\gev}{\ensuremath{\mathrm{GeV/}c}}

\newcommand{\kT}{\ensuremath{k_\mathrm{T}}}
\newcommand{\antikT}{anti-\ensuremath{k_\mathrm{T}}}

\newcommand{\rr}{\ensuremath{R}}

\newcommand{\pTlow}{\ensuremath{p_\mathrm{T,low}}}
\newcommand{\pThigh}{\ensuremath{p_\mathrm{T,high}}}

\newcommand{\dNjetdDphidpT}{\ensuremath{\frac{{\rm d}^{2}N_{\mathrm{jet}}}{ {\mathrm{d}\pTjet} {\mathrm{d}\dphi}}}}

\newcommand{\dNjetcorr}{\ensuremath{\mathrm {d}}{^2}N_\mathrm{jet}^\mathrm{corr}}

\newcommand{\IAA}{\ensuremath{{I}_{\rm AA}}}

\newcommand{\IAApT}{\ensuremath{{I}_{\rm AA}(\pTjetch)}}
\newcommand{\IAAphi}{\ensuremath{{I}_{\rm AA}(\dphi)}}

\newcommand{\Qsq}{\ensuremath{{Q}^{2}}}

\newcommand{\ztilde}{\ensuremath{\tilde{z}}}

\newcommand{\qhat}{\ensuremath{\hat{q}}}

\newcommand{\zvtxtrack}{\ensuremath{z_\mathrm{vtx}^\mathrm{track}}}
\newcommand{\zvtxspd}{\ensuremath{z_\mathrm{vtx}^\mathrm{SPD}}}

\newcommand{\Drecoil}{\ensuremath{\Delta_\mathrm{recoil}}}

\newcommand{\DrecoilpTphi}{\ensuremath{\Delta_\mathrm{recoil}(\pTjet,\dphi)}}
\newcommand{\DrecoilpT}{\ensuremath{\Delta_\mathrm{recoil}(\pTjet)}}
\newcommand{\Drecoilphi}{\ensuremath{\Delta_\mathrm{recoil}(\dphi)}}

\newcommand{\DrecoilpTreco}{\ensuremath{\Delta_\mathrm{recoil}(\pTreco)}}

\newcommand{\DrecoilpTchphi}{\ensuremath{\Delta_\mathrm{recoil}(\pTjetch,\dphi)}}
\newcommand{\DrecoilpTch}{\ensuremath{\Delta_\mathrm{recoil}(\pTjetch)}}

\newcommand{\sigmadphi}{\ensuremath{\sigma_{\dphi}}} 

\newcommand{\AAtoh}{\ensuremath{\mathrm{AA}\rightarrow{\mathrm{h}}}}
\newcommand{\AAtohjet}{\ensuremath{\mathrm{AA}\rightarrow{\mathrm{h+jet}}}}

\newcommand{\pTh}{\ensuremath{p_\mathrm{T,h}}}
\newcommand{\TTSig}{\ensuremath{\mathrm{TT}_\mathrm{sig}}}
\newcommand{\TTRef}{\ensuremath{\mathrm{TT}_\mathrm{ref}}}

\newcommand{\TT}[2]{\ensuremath{\mathrm{TT}\{#1,#2\}}}

\begin{document}

\begin{titlepage}
\PHyear{2023}       
\PHnumber{188}      
\PHdate{28 August}  

\title{Measurements of jet quenching using semi-inclusive hadron+jet distributions in pp and central \PbPb\ collisions at $\mathbf{\sqrt{\textit{s}_{\mathrm{\mathbf{NN}}}}} \mathbf{\;= 5.02}$ TeV}
\ShortTitle{Semi-inclusive h+jet production in \pp\ and \PbPb\ collisions}  
\Collaboration{ALICE Collaboration\thanks{See Appendix~\ref{app:collab} for the list of collaboration members}}
\ShortAuthor{ALICE Collaboration} 

\begin{abstract}
The ALICE Collaboration reports measurements of the semi-inclusive distribution of charged-particle jets recoiling from a high transverse momentum (high \pT) charged hadron, in \pp\ and central \PbPb\ collisions at center-of-mass energy per nucleon--nucleon collision $\sqrtsNN=5.02$ TeV. The large uncorrelated background in central \PbPb\ collisions is corrected using a data-driven statistical approach, which enables precise measurement of recoil jet distributions over a broad range in \pTjetch\ and jet resolution parameter \rr. Recoil jet yields are reported for $\rr=0.2$, 0.4, and 0.5 in the range $7<\pTjetch<140$ \gev\ and $\pi/2<\dphi<\pi$, where \dphi\ is the azimuthal angular separation between hadron trigger and recoil jet. The low \pTjetch\ reach of the measurement explores unique phase space for studying jet quenching, the interaction of jets with the quark--gluon plasma generated in high-energy nuclear collisions. Comparison of \pTjetch\ distributions from \pp\ and central \PbPb\ collisions probes medium-induced jet energy loss and intra-jet broadening, while comparison of their acoplanarity distributions explores in-medium jet scattering and medium response. The measurements are compared to theoretical calculations incorporating jet quenching.
\end{abstract}

\end{titlepage}

\setcounter{page}{2} 



\section{Introduction} 
\label{sect:Intro}

Nuclear matter under conditions of extreme temperature and pressure forms a quark--gluon plasma (QGP), the deconfined state of matter whose dynamics are governed by interactions between sub-hadronic constituents~\cite{Busza:2018rrf,Harris:2023tti}. The QGP filled the early universe a few microseconds after the Big Bang, and is generated and studied today using collisions of atomic nuclei at the CERN Large Hadron Collider (LHC) and the Brookhaven Relativistic Heavy Ion Collider (RHIC). Experimental measurements at these facilities, together with their comparison to theoretical calculations, have shown that the QGP is a fluid with very low specific viscosity~\cite{Heinz:2013th,JETSCAPE:2020shq,JETSCAPE:2020mzn,Nijs:2020roc} that is opaque to the passage of energetic color charges~\cite{Majumder:2010qh,Cunqueiro:2021wls}.

In hadronic collisions, jets arise in hard (high-momentum-transfer \Qsq) interactions of quarks and gluons (partons) from the projectiles. The scattered partons are initially virtual and come on-shell through gluon radiation, which generates a parton shower. The shower subsequently hadronizes, and the jet can be observed in a detector as a collimated spray of hadrons. Jet reconstruction algorithms have been developed which apply both to experimental data and to theoretical calculations based on perturbative quantum chromodynamics (pQCD), providing well-controlled theory--data comparisons~\cite{Cacciari:2011ma}. Jet production and substructure have been measured extensively in proton--proton (\pp) collisions, and pQCD calculations are found to be in excellent agreement with such measurements over a wide kinematic range~\cite{Abelev:2006uq,Adamczyk:2016okk,Abelev:2013fn,Aad:2014vwa,Khachatryan:2016mlc,CMS:2016jip,ATLAS:2017ble,Acharya:2019jyg,ATLAS:2011myc,ALICE:2018ype,ATLAS:2019rqw,ALICE:2020pga}. Jets in \pp\ collisions therefore provide incisive probes of QCD.

In heavy-ion collisions at collider energies, jets are generated concurrently with the QGP. Following a high-\Qsq\ partonic interaction, the evolving parton shower interacts with the expanding and cooling QGP. These secondary interactions proceed via elastic (collisional) and inelastic (radiative) processes, which modify jet production and structure relative to jets generated in vacuum (``jet quenching''). Experimentally observable consequences of jet quenching include energy transport out of the jet cone (energy loss); modification of intra-jet structure; and jet deflection. Extensive jet quenching measurements have been carried out with nuclear collisions at RHIC and the LHC (see Refs.~\cite{Majumder:2010qh,Cunqueiro:2021wls,Apolinario:2022vzg,ALICE:2022wpn} and references therein). 

The measurement of reconstructed jets in heavy-ion collisions is challenging, however, due to the large background in the complex environment of such collisions. Initial studies of jet quenching therefore utilized high-\pT\ hadron production and correlations ~\cite{PHENIX:2001hpc,PHENIX:2003qdj,PHENIX:2003qdw,PHENIX:2003djd,PHENIX:2008saf,STAR:2002ggv,STAR:2003fka,STAR:2002svs,STAR:2003pjh,STAR:2006vcp}, which are more readily measurable with high precision in such an environment. High-\pT\ hadrons are leading fragments of jets, and inclusive high-\pT\ hadron yield suppression is the hallmark of partonic energy loss due to jet quenching. Comparison of inclusive hadron suppression data with theoretical models has been used to constrain the in-medium jet transport parameter \qhat, which characterizes the magnitude of energy loss in jet quenching models~\cite{Burke:2013yra,Casalderrey-Solana:2014bpa,Chien:2015vja,Andres:2016iys,Noronha-Hostler:2016eow,Zigic:2018ovr,Ru:2019qvz,Xie:2019oxg,JETSCAPE:2021ehl}. However, observed high-\pT\ hadrons are expected to arise predominantly from jets which experience relatively little medium-induced energy loss, due to the interplay of the steeply falling inclusive jet energy spectrum shape, the hadron fragmentation function, and energy loss~\cite{Baier:2002tc,Zhang:2007ja,Renk:2012ve,Bass:2008rv}. Deeper insight into the mechanisms underlying jet quenching and the response of the QGP to the passage of energetic partons requires measurements incorporating reconstructed jets. 

Significant progress has been made over the past decade in the measurement of reconstructed jets in heavy-ion collisions in terms of inclusive jet production, di-jet correlations, and trigger--jet coincidence observables~\cite{Abelev:2013kqa,Adam:2015ewa,Adam:2015hoa,Adam:2015doa,Adam:2016jfp,Acharya:2017goa,Acharya:2017okq,Acharya:2018uvf,Acharya:2019djg,Acharya:2019jyg,Acharya:2019tku,ATLAS:2010isq,ATLAS:2012tjt,ATLAS:2014cpa,ATLAS:2014ipv,ATLAS:2017xfa,ATLAS:2018dgb, ATLAS:2018gwx,ATLAS:2022zbu,CMS:2011iwn,CMS:2012ytf,CMS:2012ulu,CMS:2015hkr,CMS:2016uxf,Adamczyk:2017yhe,STAR:2020xiv,ALICE:2023waz,STAR:2023pal,STAR:2023ksv}. Model studies incorporate both jet and hadronic observables, for a more comprehensive study of jet quenching (e.g.~\cite{Casalderrey-Solana:2018wrw,Cunqueiro:2021wls,Ehlers:2022ulm}). 

In this paper and its companion Letter~\cite{ShorthJetpaper}, the ALICE Collaboration reports new measurements of the semi-inclusive distribution of charged-particle jets recoiling from a high-\pT\ hadron (``h+jet'') ~\cite{Adam:2015doa,Adamczyk:2017yhe,Acharya:2017okq} in \pp\ and central \PbPb\ collisions at center-of-mass energy per nucleon--nucleon collision $\sqrtsNN=5.02$ TeV. The analysis utilizes the approach developed in Ref.~\cite{Adam:2015doa}, which provides data-driven correction of the complex uncorrelated background for jet measurements in central collisions of nuclei (\aaa). This approach enables systematically well-controlled jet measurements over a broad range, including low jet transverse momentum (\pTjet) and large resolution parameter \rr. These measurements extend significantly the \pTjet\ reach at both high and low \pTjet\ relative to that in Ref.~\cite{Adam:2015doa}, down to the lowest \pTjet\ values that are interpretable in terms of perturbatively generated jets ($\pTjet\approx10$ \gev; see also Ref.~\cite{Adamczyk:2017yhe}).

Corrected semi-inclusive recoil jet yields measured in \pp\ and central \PbPb\ collisions are compared, in order to explore jet quenching. The \rr\ and \pTjet-dependence of recoil yields is reported, which is sensitive to jet energy loss and medium-induced intra-jet broadening~\cite{Adam:2015doa,Adamczyk:2017yhe,STAR:2020xiv,Acharya:2019jyg}. Distributions in \dphi, the trigger--recoil jet azimuthal separation (acoplanarity), are also reported, which probe in-medium multiple scattering~\cite{Chen:2016vem,Mueller:2016gko,Vaidya:2020cyi}, scattering from QGP quasi-particles~\cite{DEramo:2012uzl,DEramo:2018eoy}, or the response of the QGP medium to energy loss~\cite{Casalderrey-Solana:2014bpa,KunnawalkamElayavalli:2017hxo,JETSCAPE:2022jer}. The measurements are also compared with theoretical models incorporating jet quenching.

The low \pTjet\ reach of this measurement, which to date is unique for reconstructed jet measurements in heavy-ion collisions at the LHC, is notable. 
Jet measurements in heavy-ion collisions, which impose a lower threshold in jet \pT\ of a few 10s of \gev, are subject to a selection bias in the reported jet population (for instance, in the relative fraction of quark and gluon jets), which complicates their interpretation in terms of jet quenching~\cite{Brewer:2018dfs,Du:2021pqa}. The measurements reported here are much less affected by this bias, however, because of their much lower \pTjet\ threshold. For acoplanarity measurements, low \pTjet\ is advantageous because medium-induced effects are expected to be largest in relative terms in that range, and in-vacuum broadening due to Sudakov radiation is smallest at low \pTjet~\cite{Chen:2016vem}. 

These measurements at low \pTjet\ and large \rr\ in \PbPb\ collisions require the determination of a trigger-correlated signal in a large and complex background. This paper details the data-driven procedures used to carry out such measurements and determine their uncertainties. However, it is also valuable to carry out a qualitative cross-check of the entire framework, to ascertain the degree to which the correlated signal reported at low \pTjet\ and large \rr\ is already present in the raw data, prior to application of the correction procedures, and is not generated solely by the corrections. This cross-check is also presented.

The manuscript is organized as follows: 
Sec.~\ref{sect:DetectorDataset} presents the detector and datasets; 
Sec.~\ref{sect:Analysis} presents the analysis algorithms and observables; 
Secs.~\ref{sect:dRecoilPbPb} and~\ref{sect:dRecoilpp} present the raw distributions in \PbPb and \pp collisions, respectively; 
Sec.~\ref{Sect:simulations} presents the theoretical models and simulations used for correction of the data and for physics studies;
Sec.~\ref{sect:Corrections} presents the correction procedures; 
Sec.~\ref{sect:SysUncert} presents the systematic uncertainties; 
Sec.~\ref{sect:Closure} presents the closure test for the \PbPb analysis; 
Sec.~\ref{sect:Results} presents the results; and Sec.~\ref{sect:Summary} presents a summary and outlook. 

\section{Detector and datasets}
\label{sect:DetectorDataset}

The ALICE experiment and its performance are described in Refs.~\cite{ALICE:2008ngc,ALICE:2014sbx}. 
The ALICE central barrel consists of detectors for charged-particle tracking, particle identification, and electromagnetic calorimetry, inside a large solenoidal magnet a with field strength of 0.5 T. The tracking in this analysis is carried out by the Inner Tracking System (ITS)~\cite{ALICE:2010tia}, a six-layer silicon detector with radial distance 3.9--43 cm from the beamline, and the Time Projection Chamber (TPC)~\cite{ALICE:TPC}, a gaseous detector with radial distance 85--247 cm from the beamline. Both detectors provide precise charged-particle tracking for track $\pT>0.2$ \gev\ within a pseudorapidity ($\eta$) coverage of $\eta<0.9$.

\subsection{Datasets}
\label{sect:datasets}

Online triggering for the datasets used in this analysis was based on signals in the V0A and V0C forward scintillation detectors~\cite{ALICE:2013axi}, collectively referred to as V0. The V0A acceptance is $2.8<\eta<5.1$ and that of V0C is $-3.7<\eta<-1.7$, over the full azimuth.

\paragraph{pp data:} The data  used in this analysis for \pp\ collisions at $\sqrts=5.02$ TeV were recorded during the 2015 and 2017 LHC running periods, with a minimum bias (MB) trigger that required a coincidence signal in V0A and V0C. 

Offline event selection requires the presence of a primary vertex constructed from at least two tracklets, which are track segments formed by pairing hits in the Silicon Pixel Detector (SPD), the two layers of the ITS which are closest to the beam line; a primary vertex formed by tracks from the full tracking system with position in the beam direction $|\zvtxtrack|<10$ cm relative to the nominal center of ALICE; and consistency in the location of the two vertices, $|\zvtxtrack-\zvtxspd|<0.5$ cm. In-bunch event pileup is suppressed by rejecting events where multiple vertices are reconstructed, while out-of-bunch pileup is rejected based on V0 timing.

After event selection cuts, 100M \pp\ collision events from the 2015 data taking period and 940M events from the 2017 data taking period are accepted. Detailed study of the features of the two datasets finds excellent consistency. They are combined and analyzed together, leading to a total of 1.04B \pp\ events which corresponds to an integrated luminosity of $20$ $\mathrm{nb^{-1}}$. 

\paragraph{\PbPb\  data:} The data used in this analysis for \PbPb\ collisions at $\sqrtsNN=5.02$ TeV were recorded during the 2018 LHC heavy-ion run. Minimum-bias events were triggered online based on the coincidence of signals in the V0A and V0C detectors. A separate trigger class based on the V0 signal amplitude was used to collect a larger sample of central \PbPb\ collisions. Offline event selection requires $|\zvtxtrack|<10$ cm relative to the nominal center of ALICE.
Same-bunch collision pileup is negligible in the Pb--Pb sample, while out-of-bunch pileup within the SPD readout time was removed using V0 timing information. Additional selection based on the correlation between the number of SPD tracklets and the number of TPC clusters was applied, to suppress pileup of collisions from different bunch crossings that occur within the TPC readout time~\cite{ALICE:2014sbx}.

Events are characterized offline by ``centrality,'' which is defined in terms of the percentile of the \PbPb\ hadronic cross section using the summed V0A and V0C (V0) signal amplitudes~\cite{ALICE-PUBLIC-2018-011}. The \PbPb\ analysis focuses on ``central'' collisions, corresponding to 10\% of the \PbPb\ hadronic cross section with the largest V0 signal amplitude. After event selection, the central \PbPb\ dataset comprises 89M events, corresponding to an integrated luminosity of $0.12$ $\mathrm{nb^{-1}}$.

\subsection{Track reconstruction}
\label{sect:tracking}

Charged-particle tracking is performed offline using hits in the ITS and TPC, both covering $|\eta|<0.9$ over the full azimuth. The SPD had spatially non-uniform and time-varying coverage during the recording of these data. In order to ensure uniform and stable tracking efficiency in the analysis, ``hybrid'' tracks are therefore employed for both the \pp\ and \PbPb\ analyses. The hybrid track population consists of two exclusive sets of tracks: ``global tracks'', which are tracks with at least one SPD hit and good track-fit residuals in the ITS, but without a primary vertex constraint; and ``complementary'' tracks, which do not have any SPD hits, are constrained by the primary vertex, and likewise have good track-fit residuals in the ITS. Both sets of tracks are required to have at least 70 crossed pad rows and at least
80\% of the geometrically findable space-points in the TPC. Tracks accepted for the analysis have $|\eta|<0.9$ over the full azimuth, and transverse momentum of $\pT>0.15$ \gev. 

The tracking efficiency is estimated from a full detector simulation. For \pp\ collisions, the tracking efficiency is 60\% for $\pT=0.15$~\gev, increasing to 80\% for $\pT>0.4$ \gev~\cite{Acharya:2019tku}. For central \PbPb\ collisions, the tracking efficiency is lower than that in \pp\ collisions by up to 2\%. The momentum resolution in \pp\ collisions is better than 3\% for hybrid tracks for $\pT<1$ \gev, increasing linearly to 10\% at $\pT=100$~\gev~\cite{ALICE:2014sbx}.
In central \PbPb\ collisions, the momentum resolution worsens by $10-15$\% at high \pt relative to the momentum resolution in \pp\ due to the high-multiplicity environment.

\section{Analysis}
\label{sect:Analysis}

The analysis strategy and procedures are based on those developed in Ref.~\cite{Adam:2015doa}. Their main features are discussed in this section.

\subsection{Event selection}
\label{Sect:EvtSelection}

Event selection requires the presence of a high-\pT\ charged-hadron trigger particle in a defined \pT\ interval, $\pTlow<\pTtrig<\pThigh$ \gev, denoted \TT{\pTlow}{\pThigh} (``Trigger Track''). If an event contains more than one hadron in the trigger interval, one is chosen at random. With this definition, the \pT\ dependence of the TT population is the same as that of the inclusive charged-hadron yield distribution. 

The principal analysis is carried out using charged-hadron triggers in \TT{20}{50} (signal distribution, denoted \TTSig). Uncorrelated background yield is corrected using a lower-\pTtrig\ interval, corresponding to \TT{5}{7} (reference distribution, denoted \TTRef; see Sec.~\ref{sect:Drecoil}). The \PbPb\ and \pp\ datasets are each divided randomly into two distinct subsets of unequal numbers of events, with one for selecting the \TTSig\ population and the other for \TTRef. The relative fraction of the population in each subset is chosen to maximize the statistical precision of the corrected distributions. For the \PbPb\ dataset, 95\% of events are assigned to the \TTSig\ population and 5\% are assigned to the \TTRef\ population, while for the \pp\ dataset the corresponding fractions are 90\% and 10\%.

\subsection{Jet reconstruction}
\label{Sect:JetReco}
Several types of jets are used in the analysis, which are distinguished by labeling their assigned \pT\ as follows~\cite{Adam:2015doa}:

\begin{itemize}

\item For real data, \pTraw\ refers to the raw output of the jet-reconstruction algorithm; \pTreco\ denotes \pTraw\ after event-wise subtraction for the uncorrelated background energy (Eq.~\ref{eq:pTreco}); and \pTjetch\ denotes \pT\ for fully corrected jet distributions.

\item For simulations, \pTjetpart\ refers to jets reconstructed from generated charged particles (particle level), and \pTjetdet\ refers to jets built from reconstructed charged tracks from the simulated data (detector level);

\item Generic reference to a jet without specification of its level of correction or simulation is denoted \pTjet.

\end{itemize}

The measured distributions are two-dimensional functions of \pTjetch\ and \dphi, so the same labeling likewise applies to \dphi\ distributions. However, for simplicity the jet type label (reco, part, or det) of the \dphi\ distributions is suppressed and can be deduced from context.

Jet reconstruction is carried out on TT-selected events. Jet reconstruction for both \pp\ and \PbPb\ collisions is performed using charged-particle tracks with $|\eta|<0.9$ and $\pT>0.15$ \gev\ over the full azimuth, using the Fastjet implementation of the \kT\ and \antikT\ algorithms with boost-invariant \pT-recombination scheme~\cite{FastJet,Cacciari:2008gp,Cacciari:2011ma}.
The momentum resolution is limited for high-\pT\ tracks, and 
jet candidates containing a track constituent with $\pT>100$~\gev\ are rejected. Less than 1\% of jet candidates in the range $100<\pTreco<140$ \gev\ are rejected by this cut, with negligible effect on the corrected physics distributions.

For both collision systems, jet reconstruction is carried out twice on each event. The first reconstruction pass utilizes the \kT\ algorithm and accepts jets with $|\etajet|<0.9-\rr$, where \etajet\ is the jet centroid calculated by the \pT-weighted vector sum of its constituent momenta. The first-pass jet population is used to determine $\rho$, the event-wise estimate of the background energy density~\cite{Cacciari:2007fd}. For \PbPb\ collisions, $\rho$ is defined as

\begin{equation}
\rho = \mathrm{median} \left\lbrace \frac{\pTrawi}{\Ajeti} \right\rbrace,
\label{eq:rho}
\end{equation}

\noindent
where \pTrawi\ and \Ajeti\ are the raw (uncorrected) jet \pT\ and area~\cite{Cacciari:2008gn} of the $i^\mathrm{th}$ jet in the event, respectively. Jet area is calculated using ghosts with area 0.005~\cite{Cacciari:2008gn}. For \pp\ collisions a modified definition appropriate for sparse events~\cite{CMS:2012rmf} is utilized.

\begin{equation}
\rho = C \times \mathrm{median} \left\lbrace \frac{\pTrawi}{\Ajeti} \right\rbrace;\ \ C = \frac{\sum_i \Ajeti}{A_\mathrm{total}},
\label{eq:rhosparse}
\end{equation}

in which $i$ enumerates reconstructed jet candidates, and $A_\mathrm{total}=1.8\times2\pi$ corresponds to the total detector acceptance. The two hardest jets in the event are excluded from the median calculation~\cite{Cacciari:2007fd} in both Eq.~\ref{eq:rho} and~\ref{eq:rhosparse}.

The second jet reconstruction pass utilizes the \antikT\ algorithm ~\cite{Cacciari:2011ma} with $\rr =0.2, 0.4$, and $0.5$. The jet centroid is likewise calculated as the \pT-weighted vector sum of constituents, with acceptance $|\etajet|<0.9-\rr$ over the full azimuth. An additional selection on the jet area is applied to suppress unphysical jet candidates~\cite{Adam:2015doa}, requiring that $\Ajet>0.07$ for $\rr=0.2$, $\Ajet>0.4$ for $\rr=0.4$, and $\Ajet>0.6$ for $\rr=0.5$.

The raw jet \pT\ is then corrected event-wise for the estimated background density $\rho$ according to~\cite{Cacciari:2007fd}

\begin{equation}
\pTrecoi = \pTrawi - \rhoAi,
\label{eq:pTreco}
\end{equation}

\noindent
where $\rho$ for the event is calculated using either Eq.~\ref{eq:rho} or Eq.~\ref{eq:rhosparse}. This adjustment accounts largely for event-wise variation in the overall level of background, which can be sizable for central \PbPb\ collisions due to the broad distribution of charged-particle multiplicity within the $0-10$\% centrality class. However, it does not account for local background fluctuations, which likewise are sizable. Such residual fluctuations are corrected by unfolding ensemble-level distributions, as discussed in Sec.~\ref{sect:Corrections}.

\subsection{Semi-inclusive distributions}
\label{Sect:SemiInclDistr}

For each TT-selected event set, recoil jet candidates are tabulated in bins of \pTreco\ and \dphi, and the distribution is normalized to the number of triggers, \Ntrig. This normalized distribution is semi-inclusive, since event selection is based solely upon the presence of an inclusively-distributed high-\pT\ trigger track, without requiring the presence of jets with specific properties in the recoil region. It is therefore equivalent to the ratio of hard cross sections~\cite{Adam:2015doa},

\begin{equation}
\frac{1}{\Ntrig}
\frac{\dNjetcorr}{\mathrm{d}\pTjet\mathrm{d}\dphi}\Bigg\vert_{\pTtrig\in{\mathrm{TT}}} = \left(
\frac{1}{\sigma^{\AAtoh}}
\frac{\rm{d}^2\sigma^{\AAtohjet}}{\mathrm{d}\pTjet\mathrm{d}\dphi}\right) 
\Bigg\vert_{\pTh\in{\mathrm{TT}}},
\label{eq:hJetDefinition}
\end{equation}

\noindent
where $\dNjetcorr / \mathrm{d}\pTjet\mathrm{d}\dphi$ represents the differential yield of recoil jets, AA denotes \pp\ or \PbPb\ collisions, $\sigma^{\AAtoh}$ is the cross section to generate a hadron within the \pT\ interval of the selected TT class, and $\mathrm{d}^2\sigma^{\AAtohjet}/\mathrm{d}\pTjet\mathrm{d}\dphi$ is the differential cross section for coincidence production of a hadron in the TT interval and a recoil jet. Both cross sections in the ratio are perturbatively calculable in \pp\ collisions~\cite{deFlorian:2009fw,Adam:2015doa}.

In central \PbPb\ collisions, the measured recoil jet population contains correlated jet candidates that originate from the same high-\Qsq\ interaction as the trigger track. In addition, the population contains uncorrelated combinatorial jet candidates arising from the random overlap of hadrons originating from multiple soft (low-\Qsq) interactions, and uncorrelated but physical jet candidates arising from a different high-\Qsq\ interaction than the trigger (multi-partonic interactions)~\cite{Adam:2015doa,Adamczyk:2017yhe}. The uncorrelated yield can be sizable, especially for jets with large \rr\ at low \pTjet\ in central \PbPb\ collisions. 

\begin{figure}[tb]
    \begin{center}
    \includegraphics[width = 0.95\textwidth] {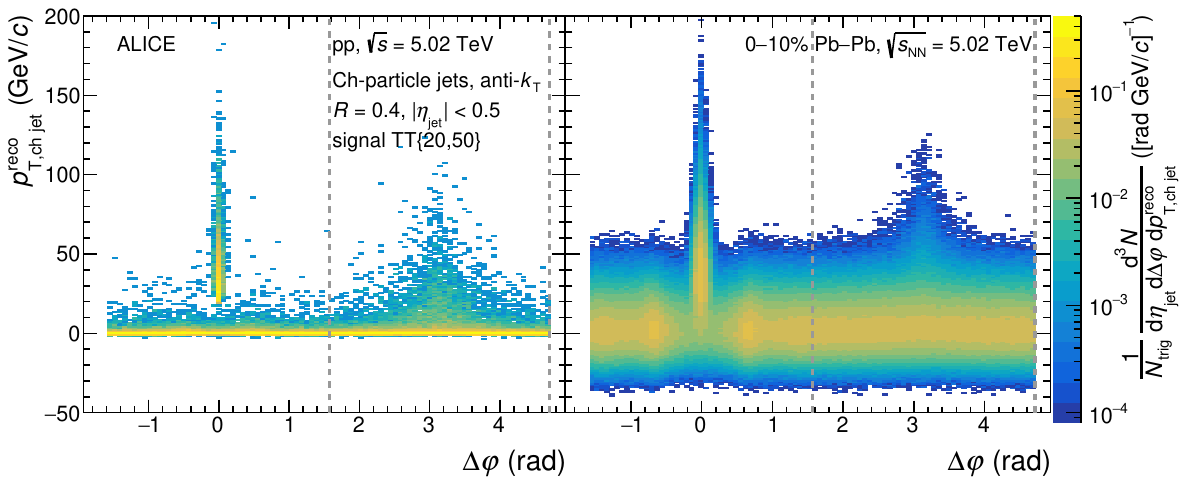}
  \end{center}
    \caption{Trigger-normalized recoil jet distributions ($\rr=0.4$) as a function of \dphi and \pTreco, in \pp collisions (left) and in \PbPb collisions (right) at $\snn=5.02$ \TeV, for \TT{20}{50} \gev. The azimuthal region of the analysis is indicated by the vertical dashed lines.} 
    \label{fig:raw-pt-deltaphi}
\end{figure}

Figure~\ref{fig:raw-pt-deltaphi} shows the trigger-normalized recoil jet distributions for \TT{20}{50} in \pp\ (left panel) and central \PbPb\ (right panel) collisions for $\rr=0.4$. While the distributions are displayed over the full range in \dphi, this analysis focuses on the recoil region $\pi/2 < \dphi < 3\pi/2$, as indicated by the vertical dashed lines. 
In \PbPb collisions there is significant yield in the region $\pTreco<0$. This yield arises because $\rho$ is the median jet \pT-density in the event, i.e.~approximately half of the acceptance has local \pT-density less than $\rho$. As discussed in Refs.~\cite{Adam:2015doa,Adamczyk:2017yhe}, this yield originates predominantly from background fluctuations and enables data-driven normalization of background yield. This population is therefore not rejected; all jet candidates are accepted in the analysis.

\subsection{Definition of the $\pmb{\Drecoil}$ observable}
\label{sect:Drecoil}

The goal of this analysis is to measure the trigger-normalized recoil jet distribution over a broad phase space, including low \pTjet\ and large \rr. However, in practice the semi-inclusive yield contains both  trigger-correlated and uncorrelated contributions to the recoil jet yield. Uncorrelated  background yield is especially large relative to correlated signal for low \pTjet\ and large \rr\ in central \PbPb\ collisions. The uncorrelated background distribution cannot be modeled accurately, and well-controlled background correction requires a fully data-driven approach. 

The choice of observable for this analysis is motivated by the observation that, by definition, the trigger-normalized uncorrelated jet yield is independent of \pTtrig, and can therefore be removed by subtracting trigger-normalized recoil jet yields obtained with two different TT ranges. The observable \Drecoil~\cite{Adam:2015doa}, which is designed for this purpose, is the difference between two semi-inclusive distributions with widely differing \pTtrig\ ranges: the signal distribution, denoted \TTSig, and the reference distribution, denoted \TTRef,

\begin{equation}
\DrecoilpTphi =
\frac{1}{\Ntrig}\dNjetdDphidpT\Bigg\vert_{\pTtrig\in{\TTSig}} -
\cRef\times \frac{1}{\Ntrig}\dNjetdDphidpT\Bigg\vert_{\pTtrig\in{\TTRef}},
\label{eq:DRecoil}
\end{equation}

\noindent
where \cRef\ is a normalization factor whose value is determined from the data. Scaling the \TTRef\ distribution by \cRef\ is needed to account for the effect of the correlated recoil jet yield at large positive \pTreco, which is smaller in the \TTRef\ population, on the magnitude of the normalized distribution at small and negative \pTreco~\cite{Adam:2015doa} (see also Sec.~\ref{sect:PbPbcRef}). The \DrecoilpTphi\ is normalized to unit $\etajet$ (notation not shown).

While the subtraction in \Drecoil\ removes the large uncorrelated jet yield, the \TTRef\ population contains an admixture of trigger-correlated yield which is also removed from the measurement by the subtraction. As noted in Sec.~\ref{Sect:EvtSelection}, the \TTRef\ \pT-range is \TT{5}{7}. This \pT-interval is chosen to minimize the \TTRef\ correlated component, while still having high enough trigger \pT\ that its inclusive production cross section is perturbatively calculable in \pp\ collisions. The \Drecoil\ distribution is therefore not that of a single semi-inclusive recoil distribution, but rather the difference of two such distributions, both of which are perturbatively calculable; the \Drecoil\ distribution is likewise perturbatively calculable. 

In order to assess the effect of the subtraction of the \TTRef\ correlated yield for the choice $\TTRef=\TT{5}{7}$, the analysis was also carried out with $\TTRef=\TT{8}{9}$. While small differences are observed in the central values of the corrected results, all such differences are smaller than the systematic uncertainties of the measurement. This variation is however not an uncertainty; the choice of \TTRef\ defines the observable. This cross-check shows rather that the physics conclusions from the analysis are not significantly dependent upon the specific choice of \TTRef.

This paper reports the following projections of \DrecoilpTphi:
\begin{itemize}
\item \DrecoilpT: projection onto \pTjet\ for $|\dphi-\pi|<0.6$; 
\item \Drecoilphi: projection onto \dphi\ for various intervals in \pTjet.
\end{itemize}

\section{Measurement of \pmb{\Drecoil}: central \PbPb\ collisions}
\label{sect:dRecoilPbPb}

For jet measurements in central \PbPb\ collisions at low \pTjet\ and large \rr, where the uncorrelated background is much larger than the correlated signal yield, the two terms in Eq.~\ref{eq:DRecoil} are similar in magnitude. Measurement of the correlated signal in this region therefore requires accurate determination of a small difference between two large numbers. This in turn requires precise {\it relative} calibration of the two terms, for both \pT-scale and yield. 

This calibration is based on the observation, discussed above, that the jet yield in the region $\pTreco<0$ is strongly dominated by background fluctuations, whose contribution is common to the two terms in Eq.~\ref{eq:DRecoil}. 

Specifically, the calibrated distributions of the two terms in Eq.~\ref{eq:DRecoil} are required to be consistent within statistical uncertainties over a significant range in the left-most part of the  \pTreco\ distribution~\cite{Adam:2015doa,Adamczyk:2017yhe}. As shown in Sec.~\ref{sect:PbPbcRef}, the distributions vary significantly in this region; this requirement is therefore highly restrictive, providing strong and purely data-driven constraints on the calibration.

\subsection{Relative $\mathbf{p_\mathrm{T}}$-scale calibration: $\Delta\mathbf{\rho}$}
\label{sect:rhoaligngment}

The two terms in Eq.~\ref{eq:DRecoil} correspond to different TT-selected populations, which have significantly different recoil jet \pT\ spectra, as detailed in Sec.~\ref{sect:PbPbcRef}. Since the presence of correlated hard jets reduces the acceptance for uncorrelated, soft combinatorial jets~\cite{Adam:2015doa}, this difference can generate different $\rho$ distributions (Eq.~\ref{eq:rho})~\cite{Adamczyk:2017yhe}. This in turn will affect the \pT-scale calibration, since Eq.~\ref{eq:DRecoil} is a function of \pTreco, which includes the area-based adjustment \rhoA~(Eq.~\ref{eq:pTreco}).

In Ref.~\cite{Adamczyk:2017yhe}, comparison of the $\rho$ distributions in the same event (SE) and mixed event (ME) populations shows that their shapes are similar but are displaced by a shift $\drho=60$ MeV/$c$. Shifting the ME $\rho$ distribution by $\drho=60$ \MeV/$c$ significantly improves the agreement of the shapes of the SE and ME \pTreco\ distributions in the region $\pTreco<0$, thereby validating this procedure to calibrate the relative \pT-scale.

\begin{figure}[tb]
\begin{center}
\includegraphics[width = 0.60\textwidth]{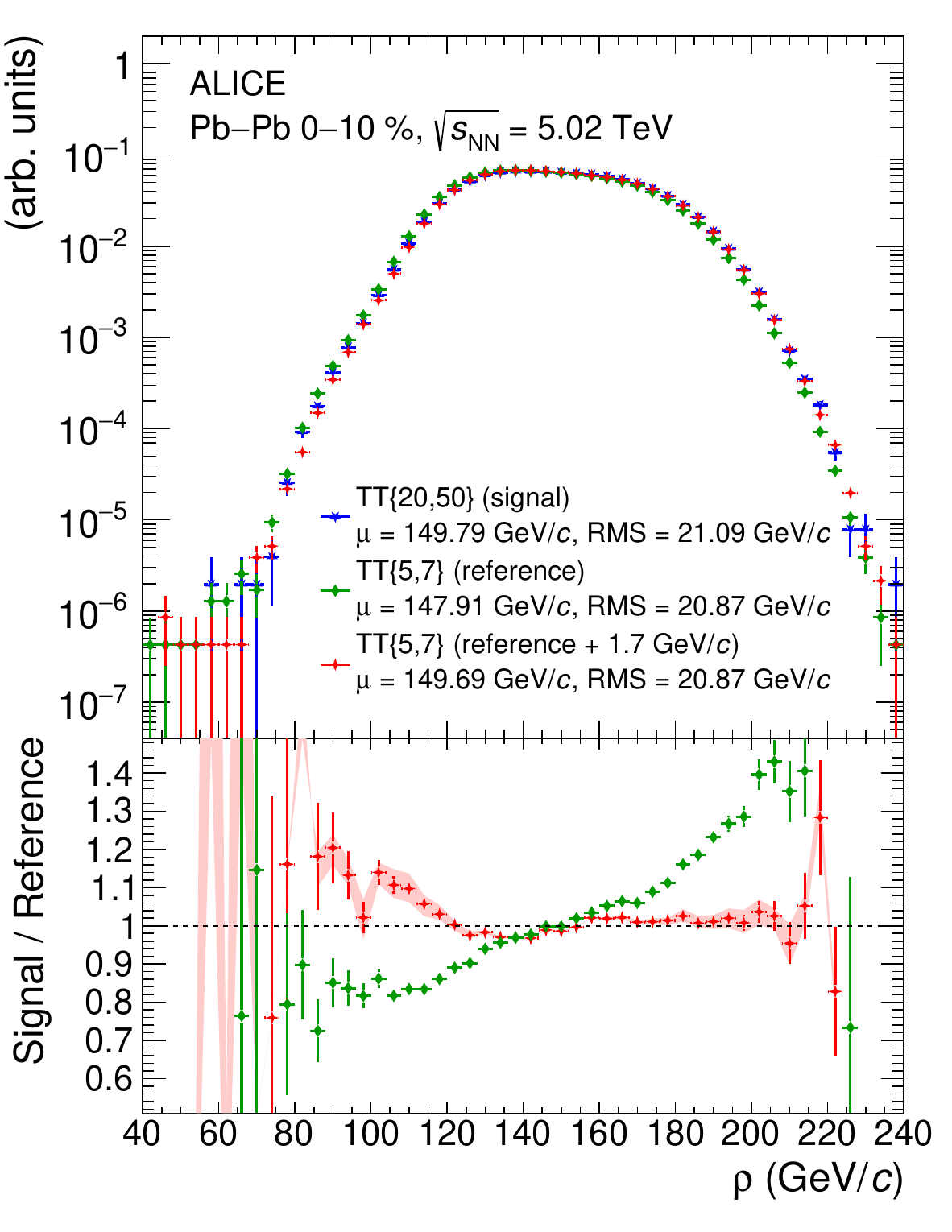}
\end{center}
\caption{$\rho$ distributions for central \PbPb\ collisions at $\sqrtsNN=5.02$ TeV. Upper panel: \TTSig\ and \TTRef\ distributions, and the \TTRef\ distribution shifted by $\drho=1.7$ \gev\ (Sec.~\ref{sect:rhoaligngment}). Lower panel: ratio of \TTRef\ and shifted \TTRef\ to \TTSig\ distribution. The mean ($\mu$) and RMS of the distributions are given in the legend. The vertical lines on the data points are the statistical uncertainties, and the shaded band on the ratio of the \TTSig\ over the shifted \TTRef\ distributions represents the systematic uncertainty of the procedure.}
\label{fig:rho-shift}
\end{figure}

Figure~\ref{fig:rho-shift} shows the $\rho$ distributions for the \TTSig\ and \TTRef\ event selections in this analysis, for central \PbPb\ collisions. While the shapes are similar, the mean of the \TTSig-selected distribution is larger. Shifting the \TTRef-selected distribution to larger values uniformly by $\drho=1.7$ \gev\ flattens the ratio of the \TTSig\ and \TTRef\ distributions  with high precision over much of the measured range (Fig.~\ref{fig:rho-shift}, lower panel), and this flatness persists when applying additional shifts in the \TTRef\ distribution up to $\pm 0.1$ \gev. A shift of $\drho=1.7\pm{0.1}$ \gev\ is therefore the optimal $\rho$ calibration for this analysis.

Note that this \drho\ calibration procedure was not applied in the ALICE Run 1 analysis of this observable~\cite{Adam:2015doa}, thereby limiting its \pTjet\ range and precision. 

\subsection{Yield calibration: $\mathbf{c_\mathrm{Ref}}$}
\label{sect:PbPbcRef}

\begin{figure}[tbhp]
\begin{center}
\includegraphics[width = 0.97\textwidth] {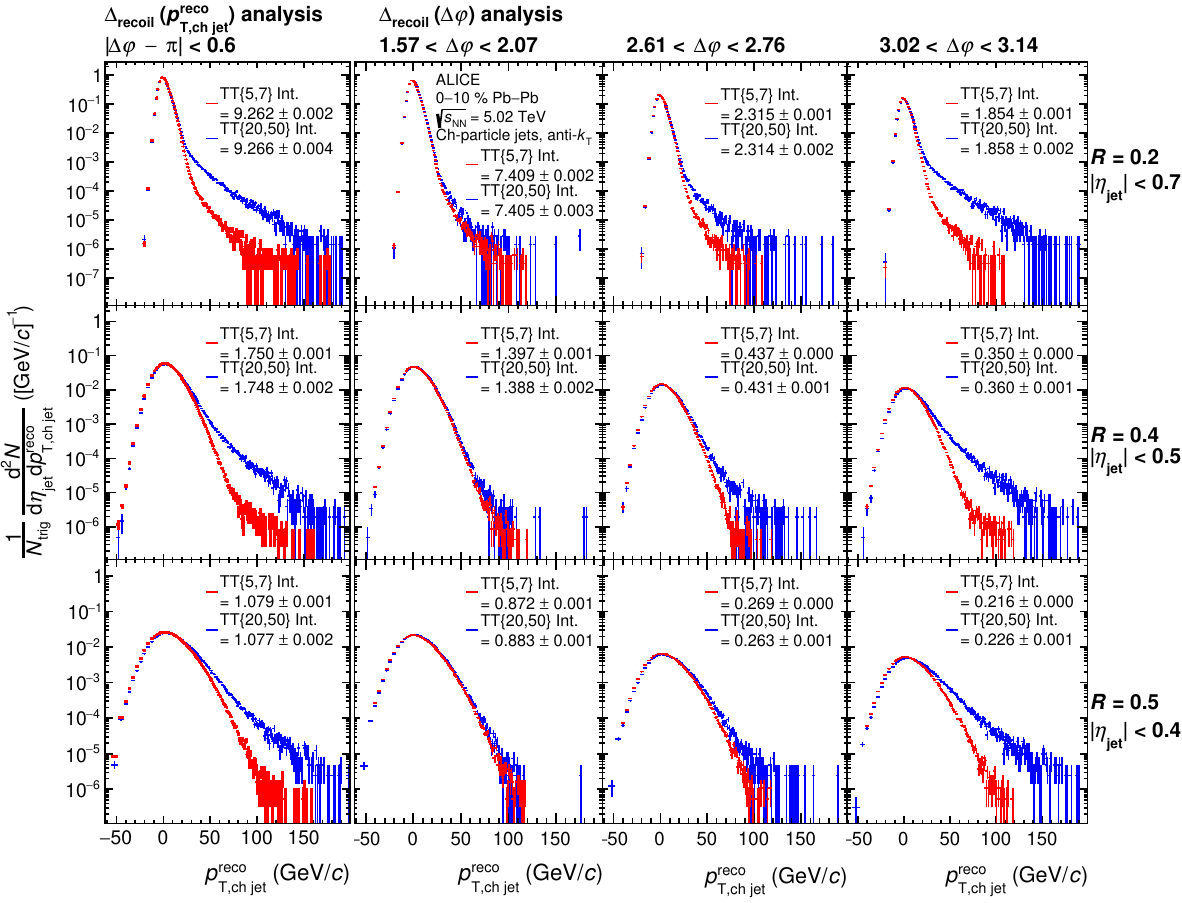}
\end{center}
\caption{Trigger-normalized semi-inclusive recoil jet distributions for \TTSig\ and \TTRef-selected populations in central \PbPb\ collisions at $\sqrtsNN=5.02$ \TeV, for $\rr=0.2$ (upper panels), 0.4 (middle panels), and 0.5 (lower panels). The \drho\ correction (Sec.~\ref{sect:rhoaligngment}) has been applied to the \TTRef\ distributions. The left column shows distributions in the \dphi\ acceptance of the \DrecoilpT\ analysis. The remaining columns show distributions in selected \dphi\ bins for the \Drecoilphi\ analysis, with the second column having the largest deviation from $\dphi=\pi$ and the rightmost column at $\dphi\approx\pi$.
}
\label{fig:cref-PbPb-pt}
\end{figure}

Figure~\ref{fig:cref-PbPb-pt} shows the trigger-normalized semi-inclusive recoil jet distributions for \TTSig\ and \TTRef-selected events in central \PbPb\ collisions for $\rr=0.2$, 0.4, and 0.5, for a \dphi\ interval corresponding to that of the \DrecoilpT\ analysis (left panels), and for selected \dphi\ bins of the \Drecoilphi\ analysis (middle and right panels). The \TTSig\ and \TTRef-selected distributions are similar in the region $\pTreco<0$, even though a systematic difference is visible in all panels of Fig.~\ref{fig:cref-PbPb-pt}. This is consistent with what has been found previously for central \aaa\ collisions in Refs.~\cite{Adam:2015doa,Adamczyk:2017yhe}. 

\begin{figure}[tbhp]
\begin{center}
\includegraphics[width = 0.97 \textwidth]
{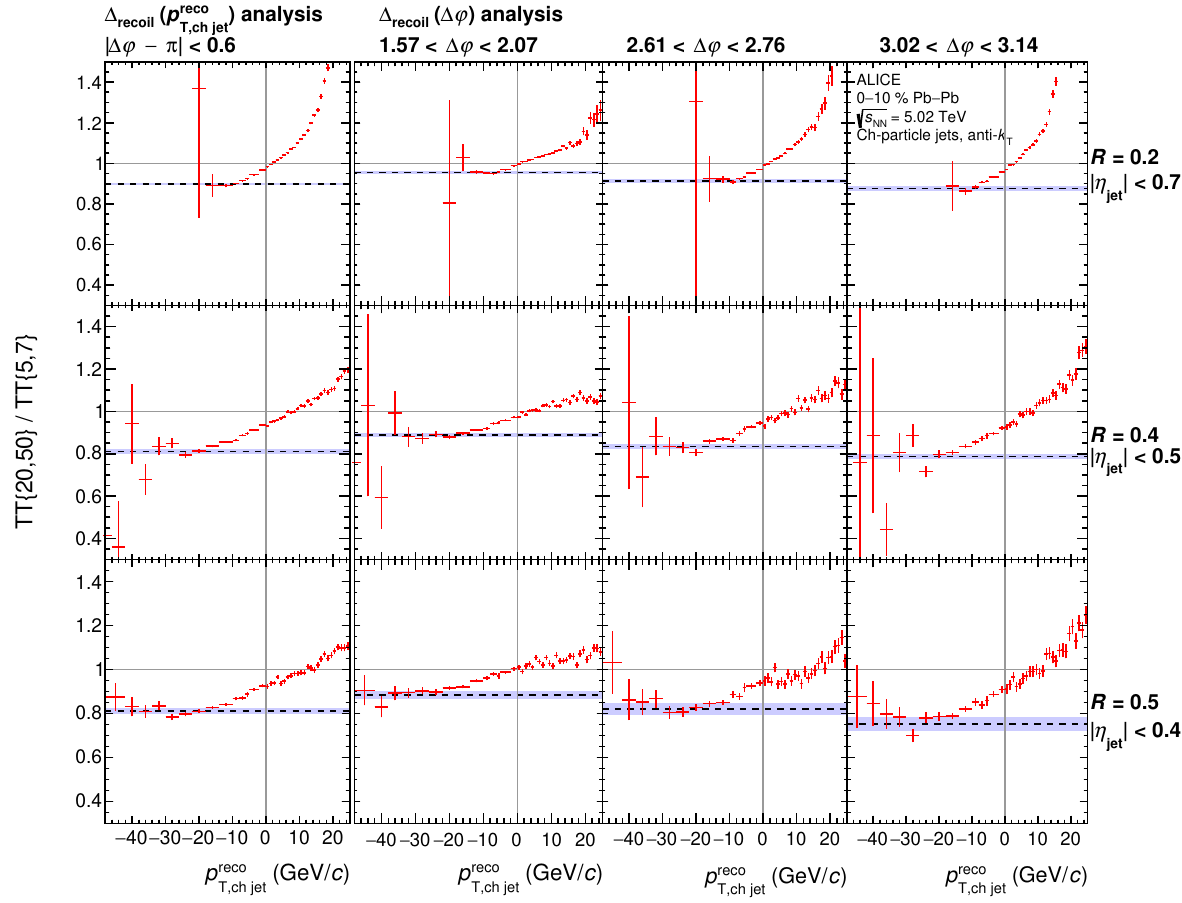}
\end{center}
\caption{Ratio of \TTSig\ and \TTRef-selected distributions from Fig.~\ref{fig:cref-PbPb-pt}, with the same panel layout. 
The horizontal dashed line and band show the value of \cRef\ and its uncertainty. The vertical solid grey line is at $\pTreco=0$, and horizontal solid grey line is at unity. See text for details.}
\label{fig:cref-PbPb-dphi}
\end{figure}

The integrals of the \TTSig\ and \TTRef-selected distributions reported in each panel are consistent to within a few percent, with many of them consistent at the per-mil level (see also Refs.~\cite{Adam:2015doa,Adamczyk:2017yhe}). This invariance, combined with the harder tail of the \TTSig-selected recoil jet distribution at large positive \pTreco, implies that the \TTSig\ recoil distribution is lower than the \TTRef\ recoil distribution in the region $\pTreco<0$. Figure~\ref{fig:cref-PbPb-dphi} shows the ratio of the \TTSig\ and \TTRef-selected distributions for each panel of Fig.~\ref{fig:cref-PbPb-pt}. The ratio in the negative \pTreco\ region is indeed less than unity in all cases, as expected. 

Notably, for $\rr=0.4$ and 0.5, the ratio is independent of \pTreco\ within statistical uncertainties for a significant range in \pTreco, i.e.~the \TTSig\ and \TTRef-selected distributions have the same shape within uncertainties in this region, over which they each vary by several orders of magnitude. For $\rr=0.2$, this invariance may also be present but cannot be observed clearly in the negative \pTreco\ region because of the significantly smaller background fluctuations than for larger-\rr\ distributions. 

Consistency in the shape of the \TTSig\ and \TTRef-selected distributions over a significant range in the negative \pTreco\ region confirms that the yield in this region is dominated by uncorrelated background. To account for the difference in their overall magnitude, the \TTRef\ distribution is renormalized by a factor \cRef, as indicated in Eq.~\ref{eq:DRecoil}~\cite{Adam:2015doa}. The value of \cRef\ is determined for each $\dphi$ bin using a two-parameter linear fit to the left-most points of the ratio, up to a cutoff value of \pTreco. The cutoff value is varied and the slope and its error from the fit are determined. The procedure is terminated at the largest cutoff in \pTreco\ for which the slope parameter is still consistent with zero within $2\sigma$, indicating the range in \pTreco\ over which the uncorrelated jet yield dominates both the \TTSig\ and \TTRef-selected distributions.
The value of \cRef\ is then determined by evaluating the fit function in the middle of its range.

\begin{figure}[tbh]
\begin{center}
\includegraphics[width = 0.65 \textwidth]{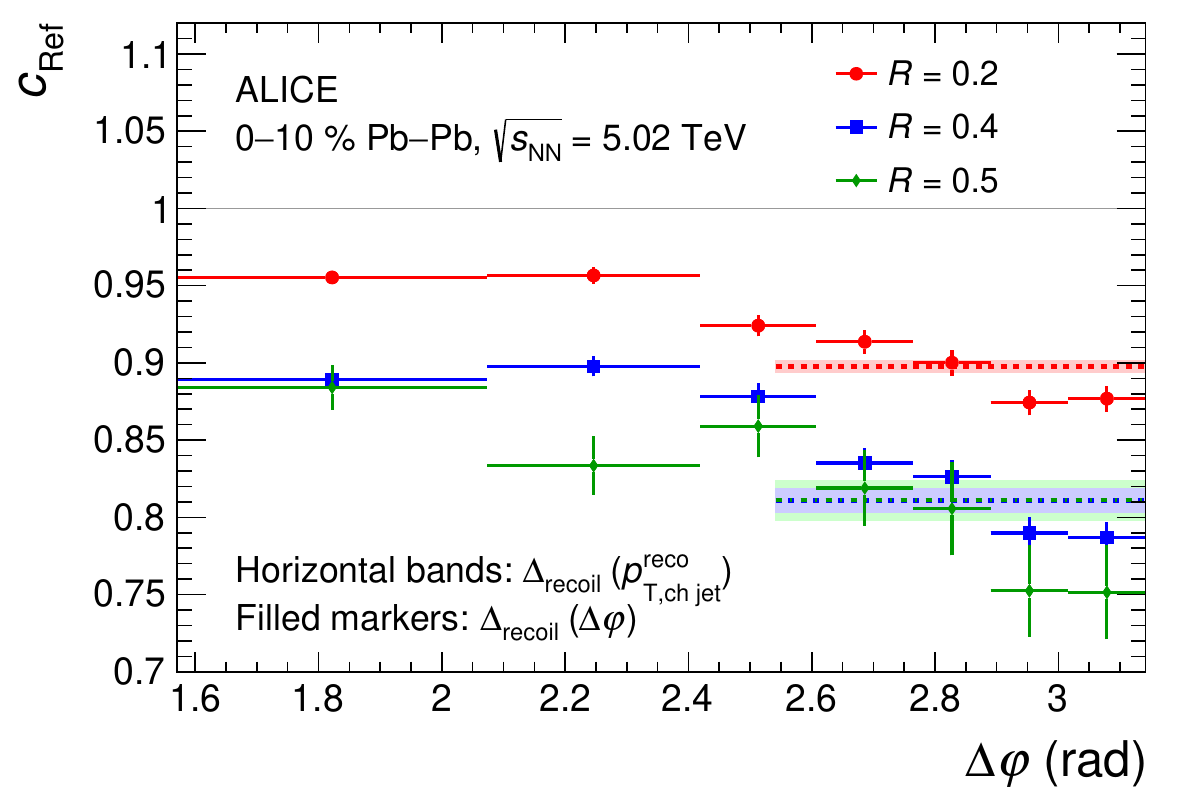}
\end{center}
\caption{Scaling factor \cRef\ of \Drecoilphi\ for $\rr=0.2$, 0.4, and 0.5. The vertical bars indicate the statistical uncertainties. The value of \cRef\ for the \DrecoilpT\ analysis, which integrates over the region $|\dphi - \pi| < 0.6$, is displayed as a horizontal bar for each value of \rr, with shaded bands indicating statistical uncertainty.}
\label{fig:cref-vs-dphi}
\end{figure}

Figure~\ref{fig:cref-vs-dphi} shows the trend of \cRef\ as a function of \dphi, which is used in the \Drecoilphi\ analysis, for $\rr\ = 0.2$, 0.4, and 0.5. The value of \cRef\ decreases as \dphi\ approaches $\pi$. This is because the largest difference in the \TTSig\ and \TTRef\ distributions at high \pTreco\ occurs at $\dphi = \pi$, where the largest correlated yield is present, resulting in a larger vertical offset between \TTSig\ and \TTRef\ at $\pTreco<0$. Conversely, the smallest difference occurs at $\dphi = \pi/2$, where a smaller vertical offset between \TTSig\ and \TTRef\ at $\pTreco<0$ occurs. The \cRef\ values for the \DrecoilpT\ analysis are indicated in the figure as horizontal bands. 

\subsection{Qualitative estimate of low-$\mathbf{p_\mathrm{T,ch\,jet}}$ kinematic reach}
\label{sect:EstimatedKine}

The measurement in central \PbPb\ collisions of recoil jet distributions at low \pTjet, large \rr, and large azimuthal deviation from the back-to-back configuration $\dphi\approx\pi$ is especially challenging, because of the especially small correlated signal relative to large uncorrelated background. The following sections detail the correction procedures and systematic uncertainties for such measurements. However, it is useful to first look at the raw data for a qualitative assessment of this analysis challenge. Specifically, to what extent can a non-zero correlated signal already be seen in the raw data, in the kinematic region expected to contribute to the low \pTjetch\ region of the fully corrected recoil jet distribution?

Figure~\ref{fig:cref-PbPb-dphi} can be used to address this question. In each panel, a statistically significant difference between the \TTSig\ and \TTRef-selected distributions is evident at negative values of \pTreco.  However, \pTreco\ still includes the effect of residual background fluctuations, which are subsequently corrected by unfolding (Sec.~\ref{sect:Corrections}). The lowest value of \pTreco\ at which this difference is significant therefore does not indicate directly the low-\pTjet\ range achievable in fully-corrected distributions. Nevertheless, that range can be estimated parametrically.

Section~\ref{sect:unfolding} describes the embedding procedure to quantify the magnitude of \pTjet\ smearing from residual background fluctuations. The \pTjet-smearing calculated by this procedure is characterized by an RMS of $\approx6$ \gev\ for $\rr=0.2$ and $\approx17$ \gev\ for $\rr=0.5$. In Fig.~\ref{fig:cref-PbPb-dphi}, statistically significant differences between \TTSig\ and \TTRef-selected distributions occur in the \pTreco\ ranges $[-10,-5]$ \gev\ for $\rr=0.2$, and $[-20,-15]$ \gev\ for $\rr=0.5$. These values of \pTreco\ are below zero by amounts similar to the RMS values of \pTjet-smearing, and therefore have contributions predominantly from yield in the true recoil jet distribution at very low (positive) values of \pTjet. Based on these considerations of uncorrected data, significant measurements of the corrected recoil jet yield at low \pTjetch\ are expected. Section~\ref{sect:Results} shows that this is indeed achieved by the analysis employing the full correction procedure.

\subsection{\pmb{\Drecoil} distributions}
\label{sect:DrecoilPbPb}

\begin{figure}[tb!]
\begin{center}
\includegraphics[width = 0.54\textwidth] {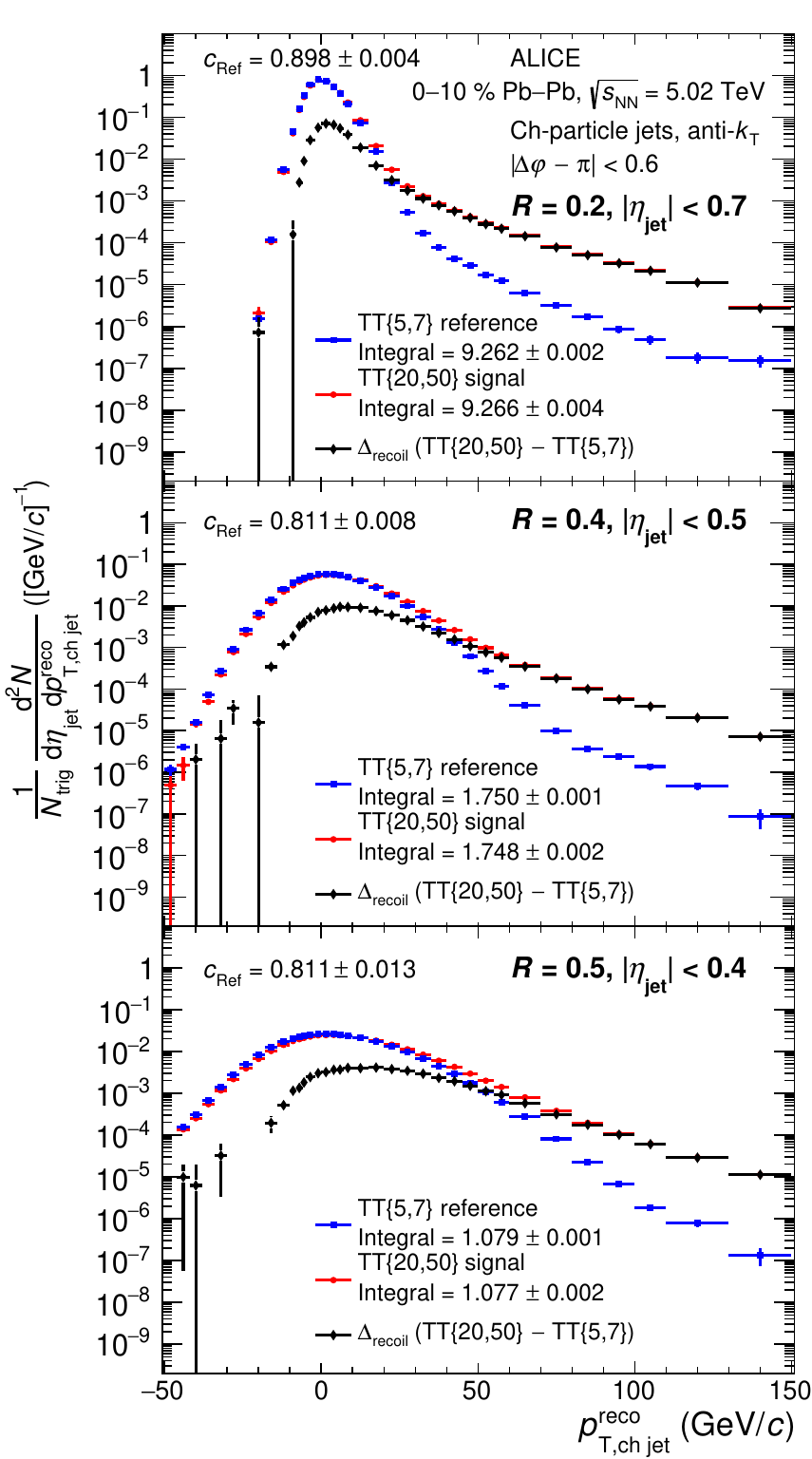}
\includegraphics[width = 0.45\textwidth] {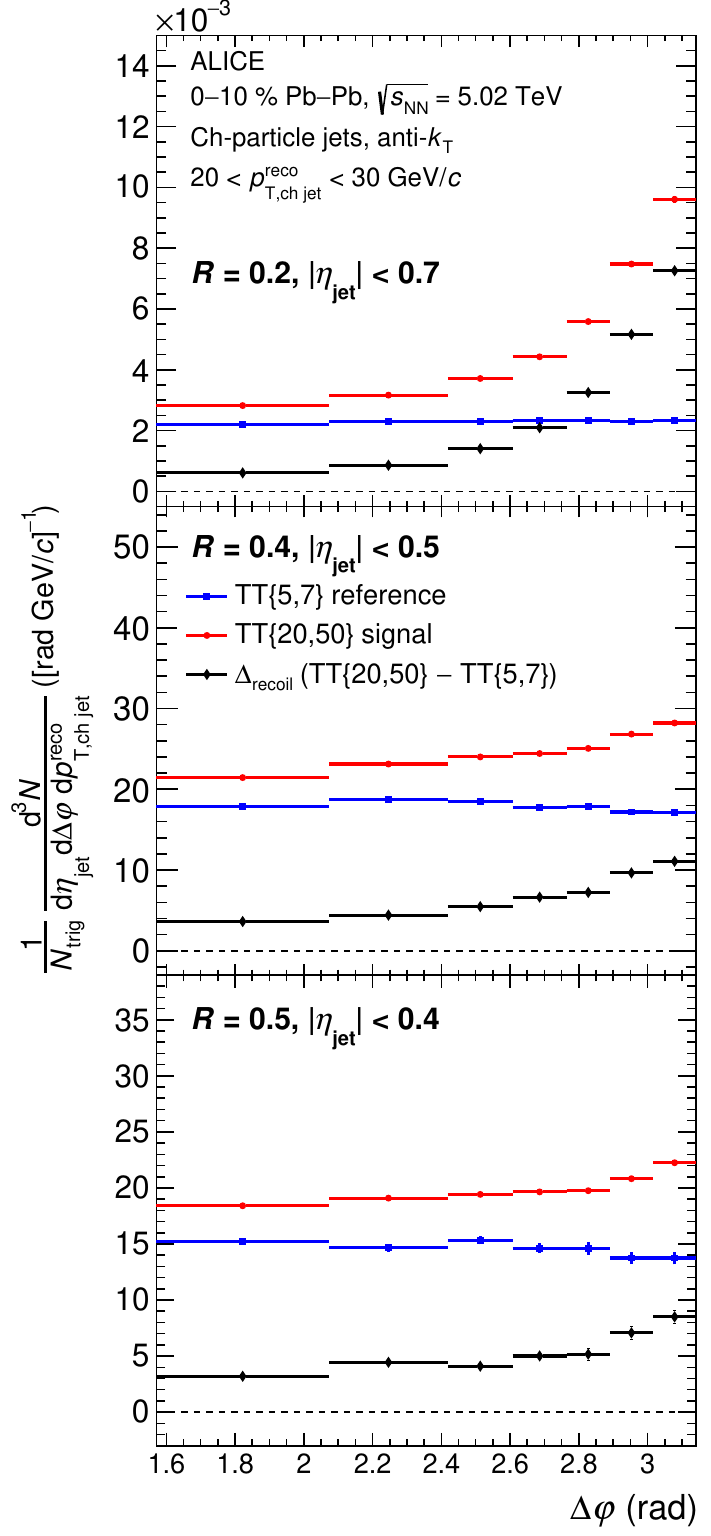}
\end{center}
\caption{Trigger-normalized semi-inclusive recoil jet distributions for \TTSig\ and \TTRef-selected populations in central \PbPb\ collisions at $\sqrtsNN=5.02$ \TeV, for $\rr=0.2$ (top), 0.4 (middle), and 0.5 (bottom). The \TTRef\ distribution has the \drho\ calibration applied and is scaled by \cRef. The resulting \Drecoil\ distribution is also shown. Left panels: Distributions as a function of \pTreco\ in the \dphi\ acceptance of the \DrecoilpT\ analysis. Right panels: Distributions as a function of \dphi, for $\pTreco\in[20,30$] \gev. Data points with a negative value for \Drecoil\ are not shown, but all such points are consistent with zero within statistical error.}
\label{fig:raw-PbPb-Drecoil}
\end{figure}

Figure~\ref{fig:raw-PbPb-Drecoil} shows the trigger-normalized recoil distributions for \TTSig\ and \TTRef-selected populations in central \PbPb\ collisions for $\rr = 0.2$, 0.4, and 0.5. The \TTRef\ distribution is fully calibrated, i.e. the ~\drho\ correction is applied and its amplitude is scaled by \cRef. The resulting \DrecoilpTreco\ (left panels) and \Drecoilphi\ (right panels) distributions are shown. The \Drecoilphi\ distributions shown correspond to the $20 < \pTreco < 30$~\gev\ region. Since \Drecoil\ is the difference between two terms, it can take negative values; however, the vertical axes of the left panels of Fig.~\ref{fig:raw-PbPb-Drecoil} use a logarithmic scale and cannot display negative values. Points with negative values for \Drecoil\ are consequently not shown, but in every case are consistent with zero within statistical uncertainties.

Because the \TTSig\ and \TTRef-selected distributions at low \pTreco\ are closely similar in magnitude and vary rapidly, a mismatch in the \pTreco\ scale of the two distributions would in turn generate a rapidly varying structure in their difference. However, the \Drecoil\ distributions are seen to vary smoothly as a function of \pTreco, providing additional validation of the calibration techniques presented above.

\section{Measurement of \pmb{\Drecoil}: pp collisions}
\label{sect:dRecoilpp}

In this analysis, medium-induced effects are determined by comparing measurements of \PbPb\ collisions to those of \pp\ collisions. The observable \Drecoil\ was developed for precise, data-driven correction of the large background accompanying jet measurements in \PbPb\ collisions, which is not present in \pp\ collisions, and a more conventional approach could suffice for the analysis of \pp\ data. Nevertheless, for accurate comparison of the two systems, measurements of \Drecoil\ and its projections are likewise reported for \pp\ collisions. However, since the uncorrelated background is much smaller in \pp\ than in central \PbPb\ collisions, the calibrations discussed in Sec.~\ref{sect:rhoaligngment} and~\ref{sect:PbPbcRef} are simpler. Specifically, the magnitude of $\rho$  is much smaller in \pp\ collisions, and the \drho\ calibration is not required.

\begin{figure}[tb!]
\begin{center}
\includegraphics[width = 0.48\textwidth]{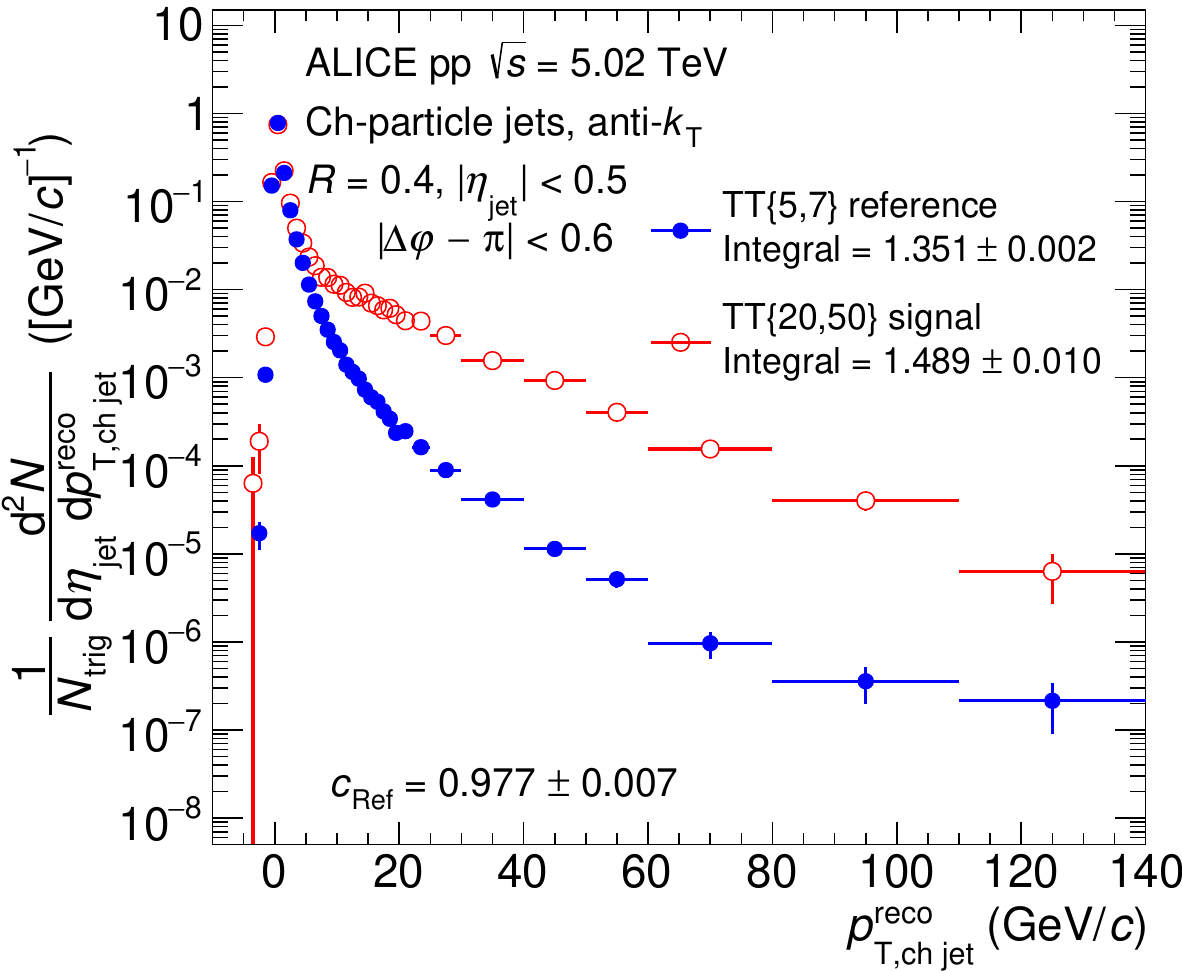}
\includegraphics[width = 0.48\textwidth]{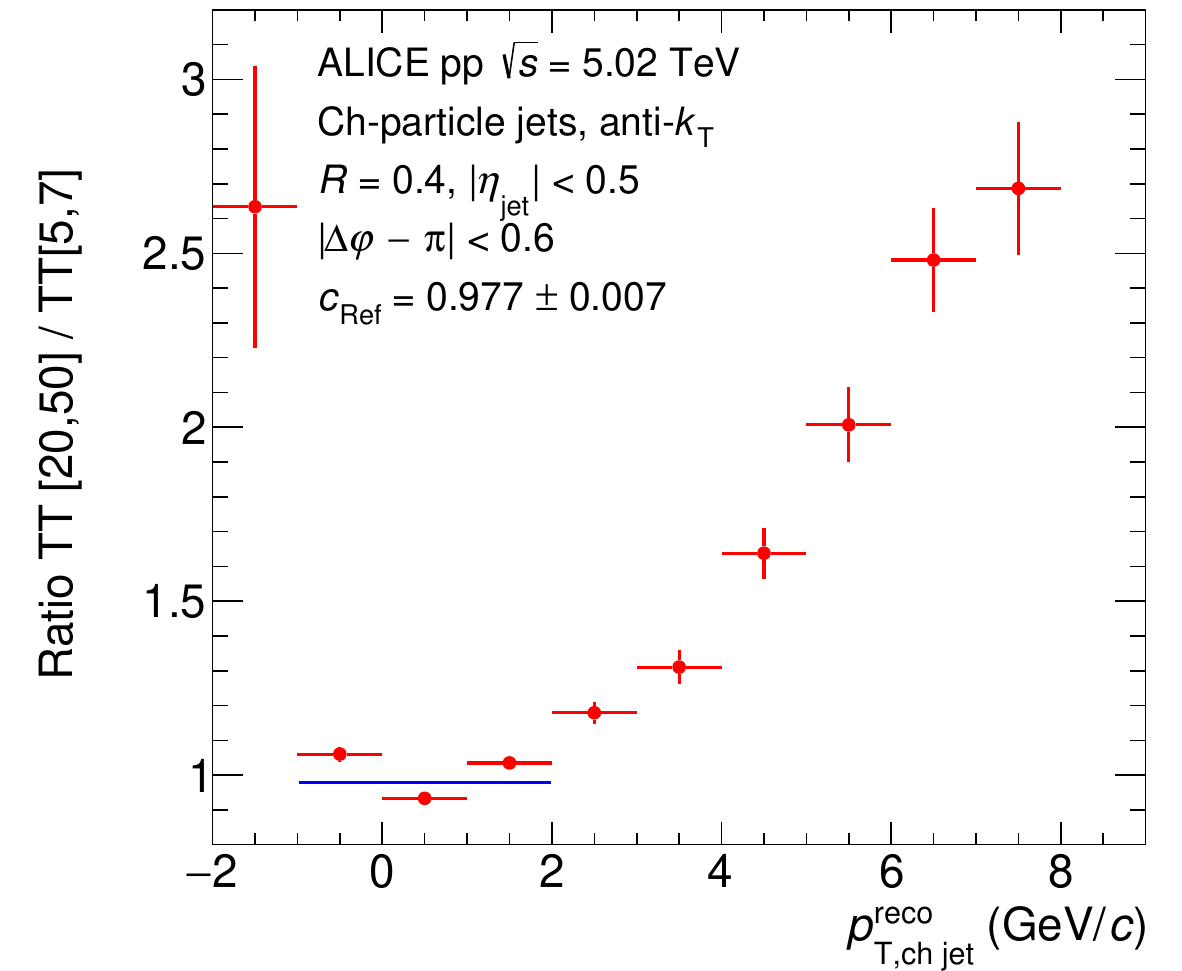}
\end{center}
\caption{Left: trigger-normalized semi-inclusive recoil jet distributions with $\rr=0.4$ for \TTSig\ and \TTRef-selected populations in \pp\ collisions at $\s=5.02$ \TeV; Right: ratio of the two distributions. The horizontal blue line indicates the fit to the ratio close to $\pTreco=0$ for the determination of \cRef, the value of which is also given in the figures.}
\label{fig:ppcref}
\end{figure}

Figure~\ref{fig:ppcref} shows the uncorrected distribution of semi-inclusive recoil jet distributions for \TTSig\ and \TTRef-selected populations in \pp\ collisions at $\sqrts=5.02$~TeV for $\rr=0.4$, together with their ratio to determine the value of \cRef. The principal value of \cRef\ is determined using a two-parameter linear fit in a narrow range around $\pTreco=0$. The fit range is varied to estimate the systematic uncertainty. The value of \cRef\ obtained with this procedure varies between 0.92 and 1.0, depending on \rr. The dependence of \cRef\ on \dphi\ for the \Drecoilphi\ analysis is negligible, and the same value of \cRef\ is used for all \dphi\ bins for \pp\ collisions.

\begin{figure}[tb!]
    \begin{center}
    \includegraphics[width = 0.54\textwidth]{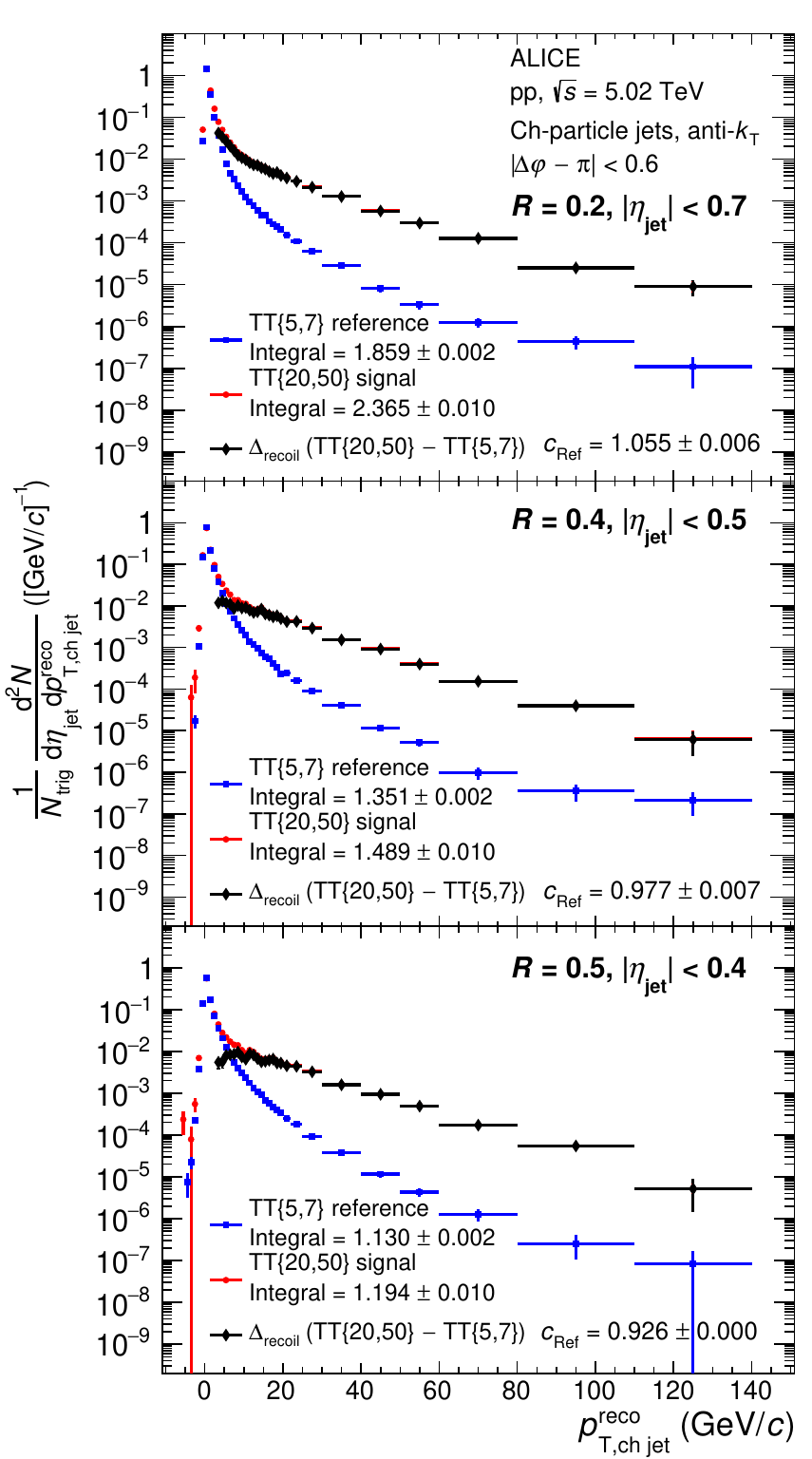}
    \includegraphics[width = 0.45\textwidth]{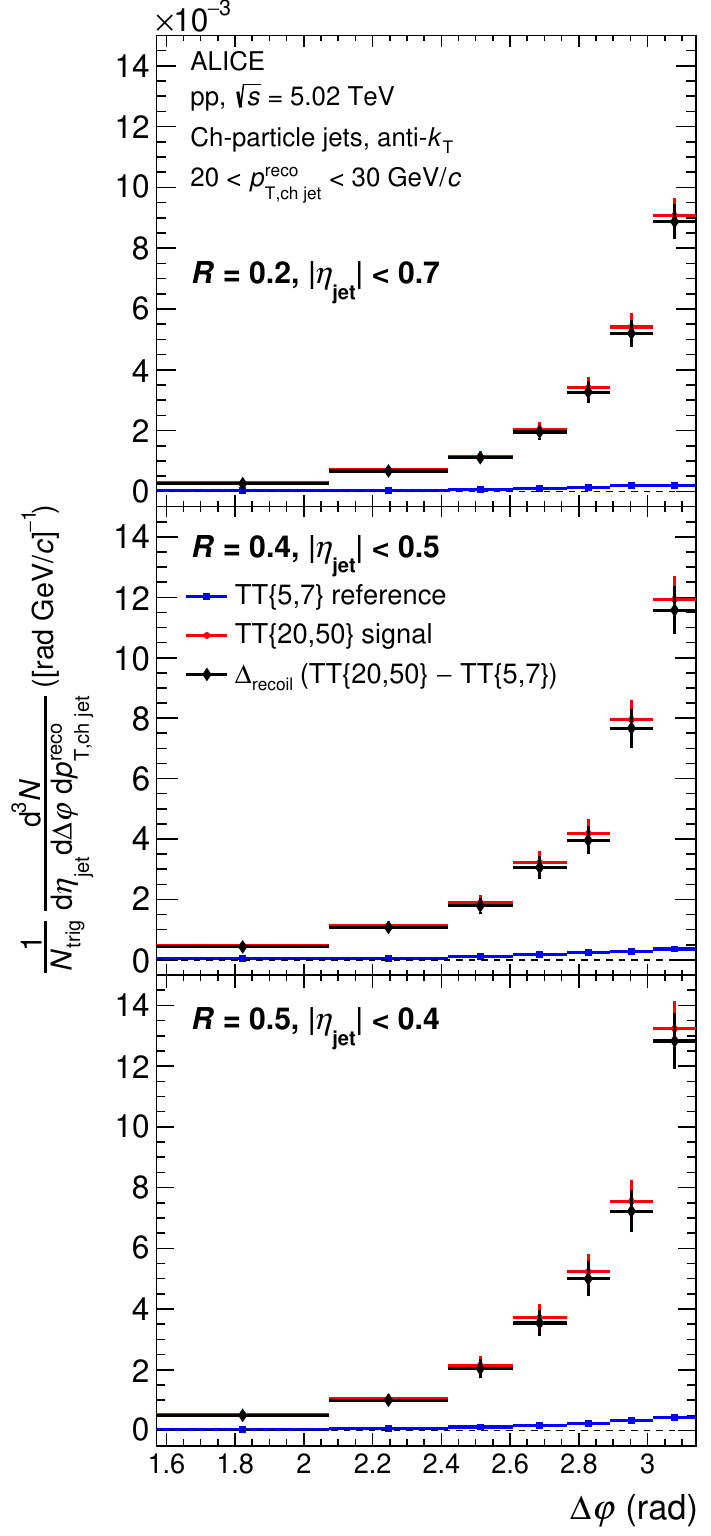}
\end{center}
\caption{Trigger-normalized semi-inclusive recoil jet distributions for \TTSig\ and \TTRef-selected populations in \pp collisions at $\sqrts=5.02$ \TeV, for $\rr=0.2$ (top), 0.4 (middle), and 0.5 (bottom). The \TTRef\ distribution is scaled by \cRef. The resulting \Drecoil\ distribution is also shown. Left panels: Distributions as a function of \pTreco\ in the \dphi\ acceptance of the \DrecoilpT\ analysis. Right panels: Distributions as a function of \dphi, for $\pTreco\in[20,30$] \gev. \TTSig\ and \TTRef\ distributions in left middle panel are the same as in Fig.~\ref{fig:ppcref}, left panel.
}
\label{fig:RawPP}
\end{figure}

Figure~\ref{fig:RawPP} shows the trigger-normalized recoil distributions for \TTSig\ and \TTRef-selected populations, and the corresponding \DrecoilpTreco\ and \Drecoilphi\ distributions, for \pp\ collisions at $\sqrts=5.02$ TeV. The distributions correspond to the same \pTreco\ and \dphi regions as the \PbPb\ distributions shown in Fig.~\ref{fig:raw-PbPb-Drecoil}. The \TTRef\ component is negligible except in a narrow region around $\pTreco=0$, so that the \Drecoil\ distributions closely match those of the \TTSig-selected population.

The integrals over \pTreco\ of the \TTSig\ and \TTRef-selected distributions  do not agree as well for \pp\ collisions in Fig.~\ref{fig:ppcref} and ~\ref{fig:RawPP} as those for central \PbPb\ collisions in Fig.~\ref{fig:cref-PbPb-pt}. As discussed in Refs.~\cite{Adam:2015doa,Adamczyk:2017yhe}, the close correspondence of these integrals in central \aaa\ collisions arises from the combined effects of high particle density, sufficient for the reconstructed jet population to fill the entire experimental acceptance, and the resilience of jet area for jets reconstructed by the \antikT\ algorithm to the event environment~\cite{Cacciari:2008gp}. The average number of reconstructed jets in an event is therefore due largely to geometry, and not the kinematics of the specific jet population, resulting in the observed invariance of the integral for the \TTSig\ and \TTRef-selected distributions. 
As discussed in Sect.~\ref{sect:PbPbcRef}, this invariance is used for precise determination of the value of \cRef. In contrast, \pp\ collision events are relatively sparse, and their reconstructed jet population does not fill the experimental acceptance (e.g. the integral for  the \TTSig\ distribution with $\rr=0.4$ is 1.75 for central \PbPb\ and 1.50 for \pp\ collisions). The agreement of the integrals for the \TTSig\ and \TTRef-selected distributions is consequently poorer, because there is no constraint on the total jet number due to events filling the acceptance. However, as also shown in Fig.~\ref{fig:ppcref} and ~\ref{fig:RawPP}, the uncorrelated background contribution in \pp\ collisions is much smaller than in \PbPb\ collisions, and the required precision for the value of \cRef\ is correspondingly lower. To summarize, the data-driven procedure to determine \cRef\ is most precise where the precision is most needed, i.e. for central \PbPb\ collisions. 

\section{Theoretical calculations}
\label{Sect:simulations}

Monte Carlo (MC) event generators are used both for physics studies and for the simulation of the detector response to correct the measured distributions for instrumental effects. 

Simulations used to correct for instrumental effects are carried out with the PYTHIA8 MC generator (Monash 2013 tune)~\cite{Sjostrand:2014zea,Skands:2014pea}). The detector response is simulated using GEANT~3.21~\cite{Brun:1082634}. In \PbPb\ collisions, additional corrections for the residual background fluctuations are carried out by embedding PYTHIA8 \pp events into \PbPb data events, as explained in detail in Sect.~\ref{sect:Corrections}.

For the simulation of \pp\ collisions for physics studies, the following MC generators are used: 

\begin{itemize}
    \item PYTHIA8: leading order (LO) pQCD with logarithmic corrections;
    \item POWHEG~\cite{Nason:2004rx,Frixione:2007vw,Alioli:2010xd,Buckley:2016bhy} with  CT14NLO PDFs~\cite{Dulat:2015mca}: calculation of scattering processes via jet pair production~\cite{Alioli:2010xa} up to next-to-leading order (NLO), matched to parton shower and hadronization from PYTHIA8;
    \item JETSCAPE PP19 tune~\cite{JETSCAPE:2019udz}: based on PYTHIA8, with modified parton shower.
\end{itemize}

For model calculations in \PbPb\ collisions, the following predictions are used. These models are based on PYTHIA8 (except JEWEL which is based on PYTHIA6~\cite{Sjostrand:2006za}) to generate hard processes, but differ in their treatment of jet--medium interactions and response of the QGP medium to the traversing jet:

\begin{itemize}
    \item JETSCAPE~\cite{Putschke:2019yrg} with \PbPb\ tune~\cite{JETSCAPE:2022jer}: multi-stage, modular MC generator for full simulation of heavy-ion collisions. Partonic evolution in the QGP is modelled using MATTER at high virtuality~\cite{Majumder:2013re,Majumder:2009zu} and LBT at low virtuality~\cite{He:2015pra,Wang:2013cia};
    \item JEWEL~\cite{Zapp:2008gi,Zapp:2013vla}: MC generator which includes both collisional and radiative parton energy loss mechanisms in a pQCD approach. Calculations are carried out in two different ways: (i) including recoiling partons from the medium in the jet finding and subtracting the recoil partons 4-momentum from \pTjet~\cite{KunnawalkamElayavalli:2017hxo}) (``recoils on, 4MomSub''); and (ii) not including recoiling partons in the jet response (``recoils off'');
    \item Hybrid Model~\cite{Casalderrey-Solana:2014bpa}: MC generator which incorporates both weakly- and strongly-coupled elements of jet quenching by describing the pQCD jet dynamics using DGLAP evolution, and the soft jet-medium interaction using a holographic description based on the AdS/CFT correspondence. Model predictions optionally include the effects of  Moli\`ere elastic scattering~\cite{DEramo:2018eoy} and wake effects.
\end{itemize}   

The acoplanarity distributions are also compared to an analytical calculation:

\begin{itemize}
    \item pQCD@LO: LO pQCD calculation with Sudakov resummation~\cite{Chen:2016vem} based on the framework in Refs.~\cite{Mueller:2016gko,Mueller:2016xoc}. Azimuthal broadening due to gluon radiation in vacuum is treated separately from medium-induced broadening, with the latter controlled by the in-medium jet transport coefficient $\hat{q}$. 
\end{itemize} 
\section{Corrections}
\label{sect:Corrections}

This section presents the corrections for instrumental and background effects. The semi-inclusive distributions reported here have two components: the high-\pT\ trigger hadron, which is used for event selection, and the reconstructed jets in the selected events. The corrections for each component are discussed in the next subsections.

\subsection{Trigger hadrons}
\label{sect:CorrTrigHadron}

As discussed in Ref.~\cite{Adam:2015doa}, high-\pT\ charged hadrons rather than jets are chosen as the trigger for this analysis because they are measured in central \PbPb\ collisions with high precision event-by-event, without the need for corrections to the complex accompanying background. Tracking efficiency at high-\pT\ is independent of \pT~\cite{ALICE:2018vuu}, so that the loss of tracks due to inefficiency is equivalent in this analysis simply to a reduction in integrated luminosity without imposing a bias on the hadron selection. Correction for the trigger hadron tracking efficiency is therefore not required. The effect of track momentum resolution on the selection of trigger hadrons near threshold was found to be negligible in Ref.~\cite{Adam:2015doa}, so is also not considered here. 

A potential additional bias is the correlation of the high-\pT\ hadron distribution with that of the event plane (EP). However, for central \PbPb\ collisions at $\sqrtsNN=2.76$ TeV this bias was found to be negligible~\cite{Adam:2015doa}. Its magnitude was rechecked for this analysis and again found to be negligible, and this effect is not considered further here.

The azimuthal angle ($\varphi$) resolution of charged tracks is better than $0.5$ $\mathrm{m\,rad}$ for tracks with $\pT>5$~\gev, and as such no correction for angular smearing of the trigger track is warranted.

\subsection{Reconstructed jet distributions}
\label{sect:CorrJets}

The measured charged-particle jet distributions are corrected for the effects of both detector response and residual background fluctuations. The detector response corresponds to the effects of tracking inefficiencies, track \pT-resolution, and weak-decay background, all of which modify the jet momentum and axis when the jet is reconstructed in both \pp and \PbPb collisions. Local fluctuations in the background also smear the reconstructed jet \pT\ and \dphi\ in \PbPb collisions.

Corrections for detector response and residual background fluctuations are carried out using an unfolding procedure. For the \DrecoilpT\ analysis unfolding is done in one dimension, \pTjet, while for the \Drecoilphi\ analysis unfolding is done in two dimensions,  \pTjet\ and \dphi. \IAApT\ for the two cases (see Secs.~\ref{sect:ppPbPb_oneD} and~\ref{sect:Acoplanarity}) are consistent within experimental uncertainties for all \rr, when projected in a common \dphi\ acceptance.

\subsubsection{Parameterized detector performance}
\label{sect:paramperf}

While the unfolding procedure is carried out using the full response matrices described below, key parameters that characterize the detector performance for jet reconstruction are summarized here for reference. These quantities are not used for correction of the data.

\paragraph{$\mathbf{p_\mathrm{T,jet}}$ resolution and median $\mathbf{p_\mathrm{T,jet}}$ shift:}

The detector effects which smear the \pTjet\ distribution are characterized in \pp\ simulations with the relative difference between the \pTjet\ at detector level and particle level, $(\pTjetdet - \pTjetpart)/\pTjetpart$ ~\cite{Acharya:2019tku}. Table ~\ref{tab:resolution-pt} shows the width and median shift of this distribution for $\rr=0.2$ and 0.5 in selected \pTjetpart\ intervals. 
Since a value of \pTjetdet\ larger than \pTjetpart\ can only arise from \pTjet\ resolution effects, and the distribution is not symmetric about zero, the \pTjet\ resolution is determined by fitting a Gaussian function to the distribution for $(\pTjetdet - \pTjetpart)/\pTjetpart > 0$ while fixing the mean of the fit to zero, since resolution effects are symmetric. The median relative \pTjetpart\ shift, which is non-zero due to detector inefficiencies, is also reported. These values are representative of the instrumental effects in both \pp\ and \PbPb\ collisions.

\begin{table}[!htb]
\small
	\caption{The resolution and median of the relative smearing of \pTjet\ due to detector effects, $S = (\pTjetdet - \pTjetpart)/\pTjetpart$, for $\rr=0.2$ and 0.5 in selected \pTjetpart\ intervals. Values are expressed in percentages.}
	\begin{center}
		\def\arraystretch{1.2}\tabcolsep=2pt    
		\begin{tabular}{l|ccc|cc} 
			\toprule
            &  \multicolumn{2}{c}{$R=0.2$} & & \multicolumn{2}{c}{$R=0.5$}  \\
		\pTjetpart [\gev]	&  $\sigma^{\mathrm{RHS}}(S$) & $\mathrm{median}(S)$ & & $\sigma^{\mathrm{RHS}}(S$)  & $\mathrm{median}(S)$  \\
			\midrule
			$[5,10]$  & 0.9\% & $-$0.3\%  & & 0.7\% & $-$3\%  \\
			$[50,100]$ & 1.8\% & $-$9\% & & 1.7\% & $-$10\%   \\

			\bottomrule
		\end{tabular}
	\end{center}
	\label{tab:resolution-pt}
\end{table}

\paragraph{{\dphi} resolution:}

The resolution in \dphi, denoted \sigmadphi, is the standard deviation of the difference between the detector-level and truth-level jet values of \dphi. Table~\ref{tab:resolution-dphi} shows \sigmadphi\ for $R=0.2$ and $R=0.5$ jets for selected \pTjet\ intervals in simulations of \PbPb and \pp collisions. The resolution is due to detector effects in \pp\ collisions, and to both detector effects and background fluctuations in \PbPb\ collisions. The resolution is finer for high \pTjetpart\ and small \rr. Note that \sigmadphi\ is smaller than the width of the \dphi\ bins in the analysis, and therefore corresponds to only a small correction in \dphi.

\begin{table}[!htb]
\small
	\caption{Azimuthal difference resolution \sigmadphi\ for $\rr=0.2$ and 0.5 in \pp and \PbPb collisions in selected \pTjetpart\ intervals. Values are expressed in radians.}
	\label{tab:resolution-dphi}
	\begin{center}
		\def\arraystretch{1.2}\tabcolsep=2pt    
		\begin{tabular}{c|ccc|cc} 
			\toprule
            &  \multicolumn{2}{c}{$R=0.2$} & & \multicolumn{2}{c}{$R=0.5$}  \\
		\pTjetpart [\gev] &  $\sigma_{\dphi}$ ($\PbPb$) & $\sigma_{\dphi}$ (\pp) & & $\sigma_{\dphi}$ ($\PbPb$) & $\sigma_{\dphi}$ (\pp)  \\
			\midrule
			$[10,20]$  & 0.05 & 0.020  & & 0.13 & 0.05  \\
			$[50,100]$ & 0.013 & 0.015 & & 0.09 & 0.03   \\

			\bottomrule
		\end{tabular}
	\end{center}
\end{table}

\subsubsection{Unfolding}
\label{sect:unfolding}

In the \pp\ analysis, the response matrix for unfolding is constructed using \pp\ events simulated using the PYTHIA8~\cite{Sjostrand:2014zea} event generator and the GEANT3~\cite{Brun:1082634} transport code. Detector-level jets are matched to particle-level jets 
based on their relative separation in rapidity and azimuth, $\Delta\rr = \sqrt{\Delta\eta^2 + \Delta\varphi^2}$, and requiring $\Delta\rr$ to be less than 0.15 for $\rr=0.2$ jets, 0.25 for $\rr=0.4$ jets, and 0.35 for $\rr=0.5$. 

For \PbPb\ collisions, the response matrix accounts for both background fluctuations and detector response. It is constructed by embedding detector-level \pp\ events simulated with PYTHIA8 into real \PbPb\ data, and matching the PYTHIA8 detector-level jets with jets reconstructed from the combined event.
The matching procedure requires the PYTHIA8 detector-level jet to share constituents carrying at least $50\%$ of its total \pT\ with the PYTHIA8+\PbPb detector-level jet. 

Unfolding is carried out using the iterative Bayesian algorithm implemented in the RooUnfold package~\cite{Adye:2011gm}. For the \DrecoilpT\ analysis, unfolding is carried out in one dimension to correct \pTreco, taking advantage of the fact that the \dphi\ resolution correction is small. For the \Drecoilphi\ analysis, two-dimensional unfolding is used to correct both \pTreco\ and \dphi. Crucial to this analysis is the ability to include the full \pTreco\ range in the unfolding, which is enabled by the subtraction of the entire combinatorial background yield; the full \Drecoil\ range shown in Figs.~\ref{fig:raw-PbPb-Drecoil} and~\ref{fig:RawPP} is therefore used in the unfolding. 

The regularization parameter, which for iterative Bayesian unfolding is the number of iterations, is optimized using both a consistency test between the raw and back-folded distributions, and the requirement that unfolded distributions from successive iterations have minimal variation. The optimal regularization parameters lie between 4 and 8, depending on collision system and jet \rr.

A key element of the unfolding procedure is the choice of prior. For the \PbPb\ analysis the prior is based on the \Drecoil\ distribution calculated with JEWEL (recoils off), and for the \pp analysis the prior is based on the \Drecoil\ distribution calculated with PYTHIA8. The distributions are fitted with a smooth function to remove the effect of finite statistical precision in the MC generation. 

For the one-dimensional \DrecoilpT\ analysis, the prior \DrecoilpT\ distribution is fitted with an exponential function $\Delta(\pTjet) = p_1 \exp(-p_2 \, \pTjet) + p_3  \, (\pTjet)^{p_4}$, where $p_{1,2,3,4}$ are fit parameters. In the \pp\ analysis, this function is used as the prior, $P(\pTjet)$. In the Pb–Pb analysis, at low \pTjet\ the prior is less constrained in models and significant discrepancies between the prior and unfolded solution can lead to unstable unfolding. For this reason, additional regulation to $P(\pTjet)$ is required. The prior in the \PbPb\ analysis is therefore defined as $P(\pTjet) = K \times (1 - f(\pTjet)) + \Delta(\pTjet) \times f(\pTjet)$, where $K$ is a constant defining the magnitude of the prior at low $\pTjet$ and $f(\pTjet)$ is a cutoff function,  $f(\pTjet) = 0.5 \times ( \tanh{ [ (\pTjet  - a ) / b ] } + 1  )$, where $a$ and $b$ are constants. 
This function allows the prior to tend towards $K$ for $\pTjet<a$ and  towards \Drecoil (JEWEL) for $\pTjet>a$. This definition takes into account that jets with radius \rr\ subtend finite area and their total number in the acceptance is limited in practice. The three constants are defined separately for each jet $R$ analysis, based on an iterative procedure in which each constant is varied independently, and the values that provide good convergence of unfolding are chosen.

For the \Drecoilphi\ \pp\ and \PbPb\ analyses, the \dphi\ projections of the prior are parameterized with a function defined as $g(\dphi) = q_1 \times \exp(\frac{\dphi - \pi}{\sigma}) + q_2$, where $q_{1,2}$ and $\sigma$ are fit parameters. The function $g(\dphi)$ is used to fit the $\dphi$ distribution for each \pTjet\ bin separately. The function $g(\dphi)$ in each $\pTjet$ interval is then scaled such that the integral of $g(\dphi)$ in the region $|\Delta\varphi - \pi| < 0.6$ is equal to $P(\pTjet)$ in the same $\pTjet$ region.

The unfolding procedure is validated through a full closure test in simulation, detailed in Sec.~\ref{sect:Closure}.

\subsubsection{Jet-finding efficiency}
\label{sect:JetEff}

An efficiency correction, which is a function of \pTjetch\ and \dphi, is applied to the unfolded spectrum to account for the loss of jets outside the measured range due to smearing effects. 
In \PbPb\ collisions, this efficiency is calculated using PYTHIA8-generated \pp\ events embedded into \PbPb\ data. For the \DrecoilpT\ analysis, for $\rr=0.2$ the efficiency is greater than 99\% over the full \pTjetch\ range, while for $\rr=0.5$ the efficiency is around 92\% at $\pTjetch=10$~\gev, rising to unity at $\pTjetch=100$~\gev\ and dropping to 96\% at $\pTjetch=130$~\gev. For the \Drecoilphi\ analysis, this efficiency is around 93\% for $\rr=0.2$ and above 70\% for $\rr=0.5$ in the region $\dphi\approx\pi/2$, increasing as $\dphi$ approaches $\pi$. 
In \pp\ collisions, the efficiency is calculated using PYTHIA8 simulations. For the \DrecoilpT\ analysis, the efficiency for all \rr\ is greater than 97\% for $\pTjetch=10$~\gev, and consistent with unity at $\pTjetch=140$~\gev. For the \Drecoilphi\ analysis, the efficiency is about 97\% for $\rr=0.2$ (95\% for $\rr=0.5$) at $\dphi\approx\pi/2$, increasing to 99\% for $\rr=0.2$ (96\% for $\rr=0.5$) at $\dphi\approx\pi$.

An additional efficiency correction is applied after unfolding to account for the probability of reconstructing a particle-level jet, i.e. for the matching efficiency when constructing the response matrix. This factor is calculated using PYTHIA8 simulations of \pp\ events, as the ratio of detector-level and particle-level jet yield. It is found to be independent of \dphi. In \PbPb\ collisions, this efficiency is around 90\% at $\pTjetch = 10$~\gev\ and around 98\% at $\pTjetch = 100$~\gev. In pp collisions, it is approximately 92\% at $\pTjetch = 10$~\gev\ and 100\% at $\pTjetch = 100$~\gev. 

\section{Systematic Uncertainties}
\label{sect:SysUncert}

For the \pp\ and \PbPb\ analyses, the systematic uncertainties are due to the tracking-efficiency uncertainty, the uncertainty on the scaling factor \cRef, and the unfolding uncertainties, which include uncertainties due to the choice of prior, the choice of regularization parameter, the \pTjet\ binning choice, and the unfolding method. For the \PbPb analysis, additional sources include the uncertainty due to the \drho\ correction, the jet matching, and the unfolding non-closure. Tables~\ref{tab:Syspp} and~\ref{tab:SysPbPb} present a summary of the systematic uncertainties in \pp and \PbPb collisions, respectively.

\begin{table}[htbp]
\small
\caption{
Relative systematic uncertainties (in \%) for the analysis of \pp\ collision data for \DrecoilpT\ and \Drecoilphi\ at $\dphi=\pi$ and $\pi/2$. Uncertainties are given for the lowest and highest measured values of \pTjetch. Where the uncertainty is less than 0.1\%, it is specified as `negl.' in the table. Where the uncertainty is not applicable, it is specified as `X'.}
	\begin{center}
		\def\arraystretch{1.2}\tabcolsep=2pt    
		\begin{tabular}{lcccccccc} 
			\toprule
			& \multicolumn{2}{c}{\DrecoilpT} & & \multicolumn{2}{c}{\Drecoilphi; $\dphi\approx\pi/2$} & & \multicolumn{2}{c}{\Drecoilphi; $\dphi\approx\pi$} \\
			\cmidrule{2-3} \cmidrule{5-6} \cmidrule{8-9} 
			\multicolumn{1}{c}{$R=0.2$}  & [7, 10]~\gev \phantom{xx}  & [110, 140] \phantom{xx}  & & [10, 20] \phantom{xx} & [50, 100] \phantom{xx} & & [10, 20] \phantom{xx} & [50, 100] \phantom{xx} \\
			\midrule
			Tracking efficiency         & 0.2\% & 8.6  &  & 14.5 & 7.9 &  & 4.1 & 17.2  \\
			$\cRef$                     & 1.0 & negl.  &  & 0.8  & 0.5  &  & 0.4 & negl.  \\
			Prior                       & 8.1 & 7.5  &  & negl.  & 0.5  &  & negl. & 0.3  \\
			Regularization parameter    & 0.3 & 1.0  &  & 0.3  & 4.4  &  & 0.1 & 0.3  \\
			Binning                     & 1.0 & 0.2  &  & negl.  & 3.3  &  & 1.9 & 0.9  \\
   	      Unfolding method         & 0.6 & 6.0  &  & X & X &  & X & X \\
			Total uncertainty           & 8.2 & 12.3 &  & 14.6 & 9.7 &  & 4.6 & 17.3 \\
			\midrule
			\multicolumn{1}{c}{$R=0.4$}  & [7, 10]~\gev \phantom{xx}  & [110, 140] \phantom{xx}  & & [10, 20] \phantom{xx} & [50, 100] \phantom{xx} & & [10, 20] \phantom{xx} & [50, 100] \phantom{xx} \\
			\midrule
			Tracking efficiency         & 3.5\% & 10.0 &  & 11.9 & 10.9 &  & 3.7 & 14.0 \\
			$\cRef$                     & 0.1 & negl.  &  & 0.2  & 0.2  &  & 0.2 & negl. \\
			Prior                       & 5.9 & 15.9 &  & negl.  & 1.7  &  & 0.1 & 0.9 \\
			Regularization parameter    & 2.4 & 4.8  &  & 0.2  & 5.4  &  & 0.8 & 0.6 \\
			Binning                     & 1.7 & 1.8  &  & 0.9  & 0.9  &  & 1.4 & 0.5 \\
                Unfolding method         & 6.0 & 6.0  &  & X & X &  & X & X \\
			Total uncertainty           & 9.5 & 20.4 &  & 11.9 & 12.3 &  & 4.1 & 14.1 \\
			\midrule
			\multicolumn{1}{c}{$R=0.5$}  & [7, 10]~\gev \phantom{xx}  & [110, 140] \phantom{xx}  & & [10, 20] \phantom{xx} & [50, 100] \phantom{xx} & & [10, 20] \phantom{xx} & [50, 100] \phantom{xx} \\
			\midrule
			Tracking efficiency         & 3.3\% & 10.0 &  & 8.2 & 18.1 &  & 8.6 & 7.5 \\
			$\cRef$                     & 0.4 & negl  &  & 1.4 & negl.  &  & 1.8 & 0.2 \\
			Prior                       & 6.9 & 13.6 &  & 0.1 & 0.9  &  & negl. & 1.0 \\
			Regularization parameter    & 12.5 & 6.4  &  & 1.0 & 3.0  &  & 0.3 & 0.9 \\
			Binning                     & 0.7 & 0.8  &  & 0.6 & 0.8  &  & 3.7 & 0.2 \\
                Unfolding method         & 4.4 & 5.9  &  & X & X &  & X & X \\
            Total uncertainty           & 15.3 & 19.0 &  & 8.4 & 18.4 &  & 9.5 & 7.6 \\
			\midrule
		\end{tabular}
	\end{center}
\label{tab:Syspp}
\end{table}

\begin{table}[htbp]
\small
\caption{Same as Table~\ref{tab:Syspp}, for \PbPb\ collisions. Where the uncertainty is less than 0.1\%, it is specified as `negl.' in the table. Where the uncertainty is not applicable, it is specified as `X'.}
	\begin{center}
		\def\arraystretch{1.2}\tabcolsep=2pt    
		\begin{tabular}{lcccccccc} 
			\toprule
			& \multicolumn{2}{c}{\DrecoilpT} & & \multicolumn{2}{c}{\Drecoilphi;$\dphi\approx\pi/2$} & & \multicolumn{2}{c}{\Drecoilphi;$\dphi\approx\pi$} \\
			\cmidrule{2-3} \cmidrule{5-6} \cmidrule{8-9} 
			\multicolumn{1}{c}{$R=0.2$}  & [7, 10]~\gev \phantom{xx}  & [110, 140] \phantom{xx}  & & [10, 20] \phantom{xx}  & [50, 100] \phantom{xx}  & & [10, 20] \phantom{xx}  & [50, 100] \phantom{xx} \\
			\midrule
			Tracking efficiency         & 1.3\% & 	6.7 &  & 4.1   & 7.8 &  & 3.3 & 5.5 \\
			\cRef\                      & 1.7 & 	negl. &  & 3.5 & 1.3 &  & 2.2 & 0.4 \\
            Prior                       & 36.2 & 	0.3 &  & 28.1  & 6.4 &  & 29.6 & 1.7 \\ 
			Regularization parameter    & 0.8 & 	4.2 &  & 3.2.  & 3.6 &  & 7.3 & 5.0 \\
			Binning                     & 10.3 & 	negl. &  & 5.6 & 5.1 &  & 4.7 & 1.4 \\
            Unfolding method         & 6.5 & 	16.0  &  & X & X &  & X & X \\
			Jet matching                & 10.1 & 	0.6 &  & 11.9  & 0.7 &  & 4.2 & 0.6 \\
      	\drho\                      & 1.5 & 	negl. &  & 2.8 & 2.5 &  & 0.4 & 0.3 \\
			Closure                     & X &   X &  & X & X &  & X & X \\  
   		Total uncertainty           & 39.6 & 	17.9 &  & 31.8 & 12.2 &  & 31.4 & 7.8 \\  
			\midrule
			\multicolumn{1}{c}{$R=0.4$}  & [7, 10]~\gev \phantom{xx}  & [110, 140] \phantom{xx}  & & [10, 20] \phantom{xx}  & [50, 100] \phantom{xx}  & & [10, 20] \phantom{xx}  & [50, 100] \phantom{xx} \\
			\midrule
			Tracking efficiency         & 1.0\% & 	7.2 &  & 2.1 & 10.5 &  & 0.8 & 5.1 \\
			\cRef\                     & 0.1 & 	negl. &  & 2.8 & 1.5 &  & 5.4 & 1.4 \\
			Prior                       & 5.4 & 	2.7 &  & 14.1 & 54.4 &  & 5.8 & 9.7 \\
			Regularization parameter    & 0.6 & 	1.3 &  & 2.5 & 12.9 &  & 1.0 & 5.0 \\
			Binning                     & 9.7 & 	0.8 &  & 8.4 & 6.9 &  & 6.2 & 1.8 \\
            Unfolding method         & 4.4 & 	13.8  &  & X & X &  & X & X \\
			Jet matching                & 1.0 & 	0.7 &  & 6.6 & 0.2 &  & 4.2 & 3.9 \\ 
     	\drho\                       & 5.8 & 	0.4 &  & 8.1 & 3.8 &  & 3.4 & 0.9 \\ 
			Closure                     & 22.6 & 	X &  & X & X &  & X & X \\ 
            Total uncertainty           & 26.3 & 	16.0 &  & 19.9 & 57.4 &  & 11.5 & 12.9 \\ 
			\midrule
			\multicolumn{1}{c}{$R=0.5$}  & [7, 10]~\gev \phantom{xx}  & [110, 140] \phantom{xx}  & & [10, 20] \phantom{xx}  & [50, 100] \phantom{xx}  & & [10, 20] \phantom{xx}  & [50, 100] \phantom{xx} \\
			\midrule
			Tracking efficiency         & 9.1\% & 	9.2 &  & 14.9 & 8.8 &  & 1.0 & 3.5 \\
			\cRef\                     & 5.8 & 	0.4 &  & 11.1 & 19.7 &  & 11.2 & 5.8 \\
			Prior                       & 11.3 & 	12.6 &  & 19.6 & 19.1 &  & 7.2 & 4.1 \\
			Regularization parameter    & 1.2 & 	8.6 &  & 2.8 & 21.8 &  & 1.6 & 2.1 \\
			Binning                     & 5.9 & 	0.4 &  & 5.4 & 7.8 &  & 4.0 & 1.6 \\
            Unfolding method         & 5.6 &  11.7  &  & X & X &  & X & X \\
			Jet matching                & 2.0 & 	0.9 &  & 2.9 & 2.3 &  & 0.3 & 5.3 \\ 
   		\drho\                       & 11.9 & 	0.5 &  & 10.5 & 3.1 &  & 6.2 & 2.2 \\ 
			Closure                     & 20.5 & 	X &  & X & X &  & X & X \\ 
            Total uncertainty           & 29.6 & 	21.3 &  & 29.7 & 37.1 &  & 15.4 & 10.1 \\

			\midrule
		\end{tabular}
	\end{center}
\label{tab:SysPbPb}
\end{table}

The tracking-efficiency uncertainty is estimated by modifying the response matrix used in the unfolding procedure via random rejection of a given fraction of tracks prior to jet finding, with the fraction corresponding to the uncertainty of the single-track efficiency. The single-track efficiency and its corresponding uncertainty is the combination of two contributions. The first contribution originates from the track selection criteria in the TPC. The second contribution originates from the matching of TPC tracks to the ITS hits. In \pp collisions, the single-track efficiency uncertainty is approximately 3\%, while in central \PbPb\ collisions, its value is approximately 6\% for tracks with $\pT = 1$ \gev, decreasing to approximately 3\% for tracks with $\pT = 15$ \gev\ and above. The systematic uncertainty is the relative change in the unfolded result obtained with the modified response matrix with respect to the principal analysis.

The uncertainty in \cRef\ in \PbPb\ collisions is estimated by varying the minimum and maximum values of \pTreco\ used in the fit of the ratios in Fig.~\ref{fig:cref-PbPb-dphi}, as well as the \pTreco\ value used to determine \cRef. The fit range is varied by $\pm2$~\gev\ for $\rr=0.2$,  $\pm 3$~\gev\ for $\rr=0.4$, and $\pm4$~\gev\ for $\rr=0.5$, with the larger range for larger \rr\ accounting for the wider \cRef\ fit range. 
For \pp\ collisions, the uncorrelated background is smaller and the fit range to extract \cRef\ is narrower. The \cRef\ uncertainty is evaluated by varying the range of \dphi\ in the vicinity of $\pT=0$. In both \pp\ and \PbPb\ collisions, the uncertainty reduces as \pTreco\ increases due to the fact that the subtraction of the \TTRef-selected distribution is a smaller relative correction at large \pTreco.

The uncertainties due to the unfolding are assessed by varying its configuration. For \pp\ collisions, the systematic uncertainty due to the prior utilized in the unfolding is determined by varying the value of the power in the functional form used to fit the prior. For \PbPb\ collisions, the prior knowledge of the yield at low \pTjetch\ is not well constrained, and thus the largest uncertainty in the prior arises from the shape of the low-\pTjetch\ distribution within the function used fit it (see Sec.~\ref{sect:unfolding}). The uncertainty on the prior is therefore assessed by varying the value of the parameters $a$ and $K$, while also requiring convergent unfolding without a substantial increase in the required number of iterations with respect to the principle analysis. 
For both collision systems, the unfolding algorithm uncertainty is assessed  by utilizing SVD~\cite{Hocker:1995kb} as an alternative algorithm (where possible when performing 1-dimensional unfolding), 
and by varying the regularization parameter in the iterative Bayesian unfolding by $\pm2$. The uncertainty related to the binning choice was assessed by varying the detector-level \pTjet\ binning, and by varying the minimum and maximum particle-level \pTjet\ bin limits.
For the \PbPb analysis, there is an additional source of systematic uncertainty related to the jet matching criteria when matching the PYTHIA8 detector-level jets with the PYTHIA8+\PbPb\ detector-level jets. This is estimated by varying the matching distance between the PYTHIA8 and PYTHIA8+\PbPb\ detector-level jets between $0.5\rr$ and $0.6\rr$.

In \PbPb collisions, the $\rho$ correction parameter \drho\ (introduced in Sec.~\ref{sect:rhoaligngment}) is determined based on a fit to the ratio of the $\rho$ distributions for \TTSig\- and \TTRef-selected data. However, there remain residual differences between the two distributions. The systematic uncertainty due to \drho\ is determined by varying the shift by $\pm 0.1$~\gev.

The closure test for the $\DrecoilpT$ analysis in \PbPb\ collisions also demonstrates moderate non-closure at low $\pTjet$ (see Sec.~\ref{sect:Closure} for details). In the regions where the closure discrepancy is not covered by the systematic uncertainties listed above, the relative discrepancy is considered as an additional systematic uncertainty. For the $\DrecoilpT$ and $\Drecoilphi$ analyses in \pp\ collisions the unfolding closure is successful for all \pTjet\ and \dphi, so no uncertainty is assigned.

In \pp collisions, the effect of the underlying event subtraction is checked by performing the analysis with and without underlying event subtraction. For $\pTreco>5$~\gev, the \Drecoil\ distributions with and without this subtraction are fully consistent within the statistical uncertainties, and no uncertainty is therefore assigned. 

The total uncertainty is the quadrature sum of the systematic uncertainties from each distinct source. For the yield ratios \IAApT\ (presented in Sec.~\ref{sect:ppPbPb_oneD}) and \IAAphi\ (discussed in Sec.~\ref{sect:Acoplanarity}), the systematic uncertainties in the \PbPb\ and \pp\ measurements are assumed to be uncorrelated, and the uncertainty of \IAApT\ is computed as their quadrature sum.

\section{Closure test}
\label{sect:Closure}

A closure test is carried out on the \PbPb\ analysis to validate the full correction procedure and systematic uncertainties presented in Secs.~\ref{sect:Corrections} and 
\ref{sect:SysUncert}. The closure test utilizes hybrid events, in which detector-level \pp collisions generated by PYTHIA8 are embedded into real \PbPb\ data events. The raw pseudo-data distributions from the hybrid events are then modified bin-by-bin to introduce the statistical precision of the real data, by adding jitter to the central value using a Gaussian-distributed random function whose $\sigma$ is the statistical uncertainty of data in the bin.

A full analysis is carried out on these events, with the trigger track required to originate from the PYTHIA8 \pp\ event. 
Recoil jets are reconstructed, and the $\rho$ correction and $\Delta\rho$ calibration are applied using tracks from the full hybrid event. The raw distributions are then unfolded, the efficiency corrections are applied, and the systematic uncertainties are determined as in the analysis of real data. However, the systematic uncertainty due to the tracking efficiency is not included, since the tracking efficiency entering the raw-level distribution and the response matrix are both from the PYTHIA8 simulation, and therefore are consistent by definition.

This smearing procedure is carried out 20 times. The mean value of the resulting distribution of unfolded solutions is assigned as the central value of the unfolded result, and the RMS of the unfolded solutions is assigned as the statistical uncertainty.
The quality of the closure is then judged by comparing the fully-corrected \Drecoil\ distributions with the particle-level \Drecoil\ distribution used as input.

\begin{figure}[tb]

    \begin{center}
    \includegraphics[width = 0.95\textwidth] {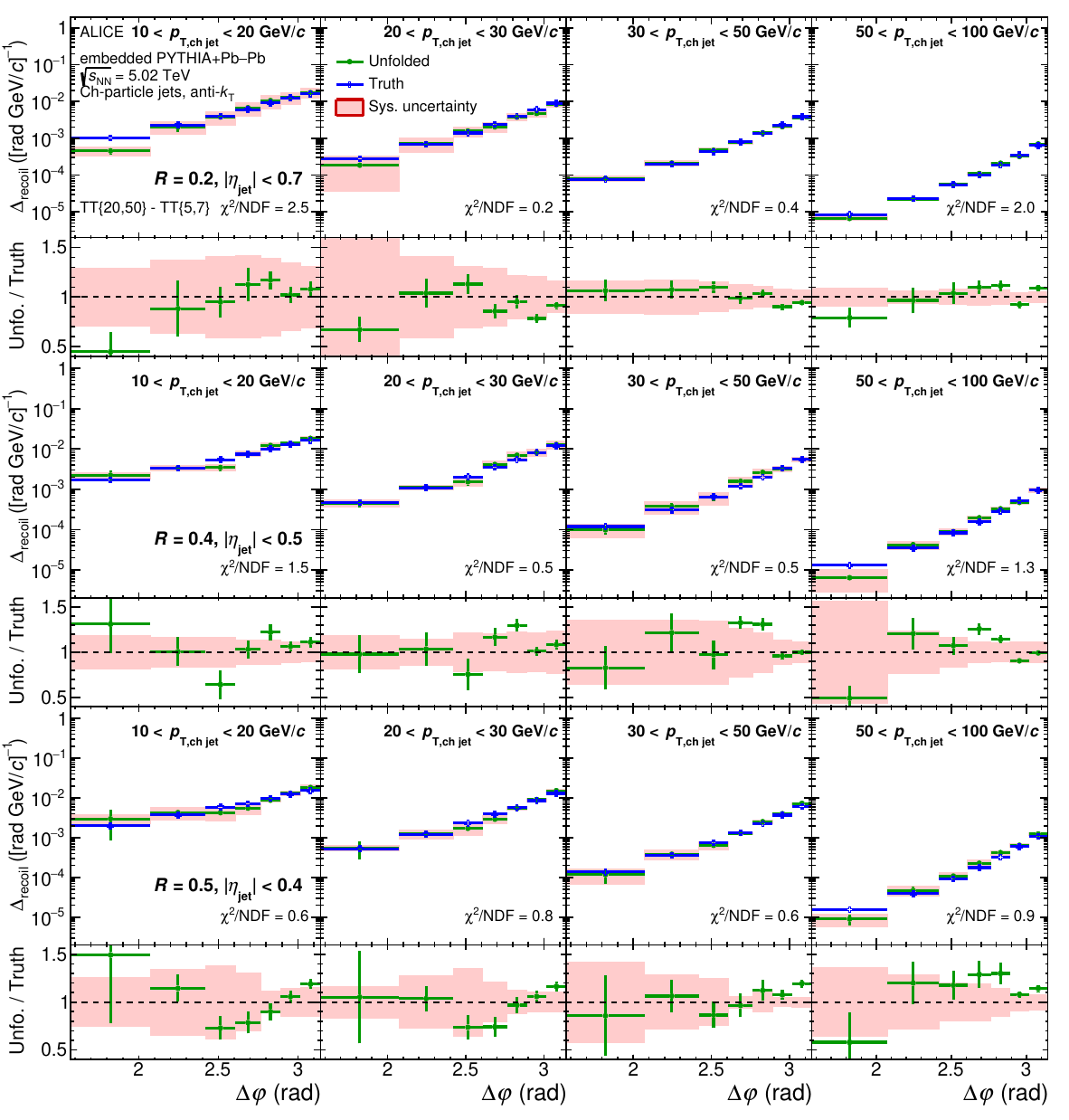}
  \end{center}
    \caption{Closure test of \Drecoilphi\ analysis for central \PbPb\ collisions, for $\rr=0.2$, 0.4, and 0.5 in selected \pTjetch\ bins.
    }
    \label{fig:closure-PbPb-dphi}
\end{figure}

Figure~\ref{fig:closure-PbPb-dphi} shows the result of the closure test for the \Drecoilphi\ analysis of central \PbPb\ collisions, comparing the unfolded result with the PYTHIA8-generated particle-level distribution. 
Each panel shows $\chi^2$/NDF from the comparison of the unfolded and particle-level distributions. While moderate discrepancies occur in the tail of the \dphi\ distribution, farthest from $\dphi\approx\pi$,  the distributions are in agreement within uncertainties, indicating successful closure.

\begin{figure}[tb]

    \begin{center}
    \includegraphics[width = 0.95\textwidth] {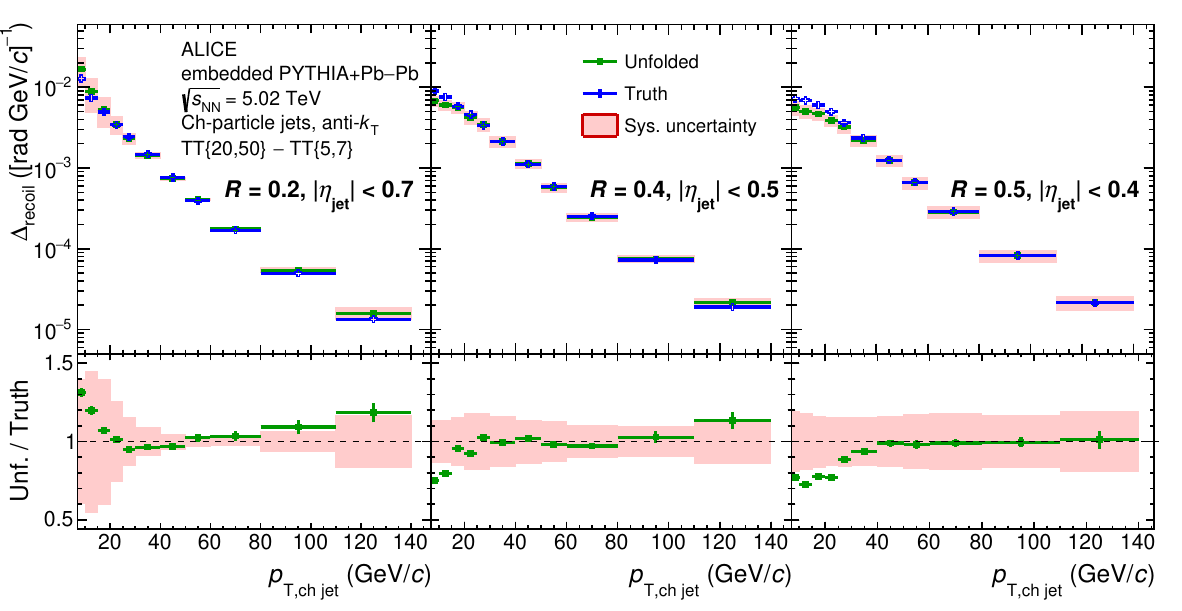}
  \end{center}
    \caption{Closure test of \DrecoilpT\ analysis for central \PbPb\ collisions. }
    \label{fig:closure-PbPb-pt}
\end{figure}

Figure~\ref{fig:closure-PbPb-pt} shows the result of the closure test for the \DrecoilpT\ analysis of central \PbPb\ collisions. Good closure is likewise achieved over most of the \pTjet\ range. However, for $\rr=0.5$ and $\rr=0.4$ at low-\pTjet, the distribution falls outside of the uncertainty of the unfolded result; in this case the relative magnitude of the discrepancy is included as a source of systematic uncertainty, as reported in Tab.~\ref{tab:SysPbPb}.

\section{Results}
\label{sect:Results}

This section presents corrected \DrecoilpTchphi\ distributions for \pp\ and central \PbPb\ collisions and physics results from their comparison. Key results are presented in the companion Letter~\cite{ShorthJetpaper}.

\subsection{\pmb{\Drecoil} in pp collisions}

\label{sect:ResultsDrecoilpp}

\begin{figure}[tbh]
    \begin{center}
    \includegraphics[width = 1.0\textwidth] {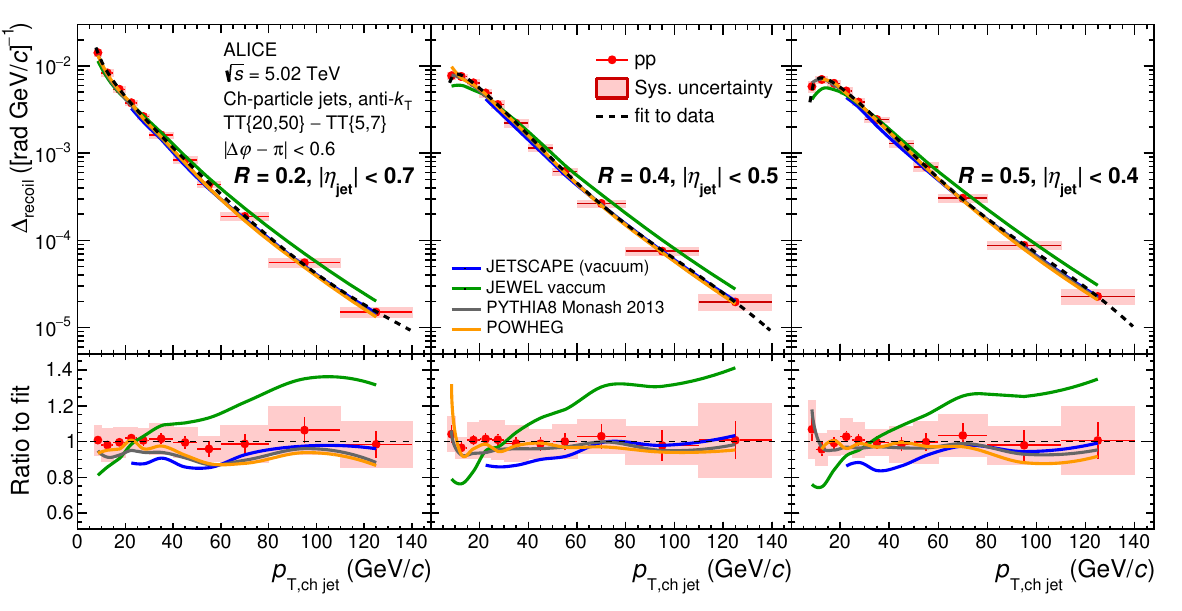}
  \end{center}
\caption{Upper panels: corrected \DrecoilpTch\ distributions measured for $\rr=0.2$ (left), 0.4 (middle), and 0.5 (right) in \pp\ collisions at $\sqrts=5.02$ TeV, compared to calculations from 
JETSCAPE ~\cite{Putschke:2019yrg}, JEWEL ~\cite{Zapp:2008gi,Zapp:2013vla}, PYTHIA8 ~\cite{Sjostrand:2014zea,Skands:2014pea}, and POWHEG ~\cite{Nason:2004rx,Frixione:2007vw,Alioli:2010xd,Buckley:2016bhy}. Lower panels: ratio of the data and calculations to a functional fit of the measured \DrecoilpTch\ distributions.} 
\label{fig:pp_Drecoil_pT}
\end{figure}

\begin{figure}[tbhp]
  \begin{center}
  \includegraphics[width = .95\textwidth] {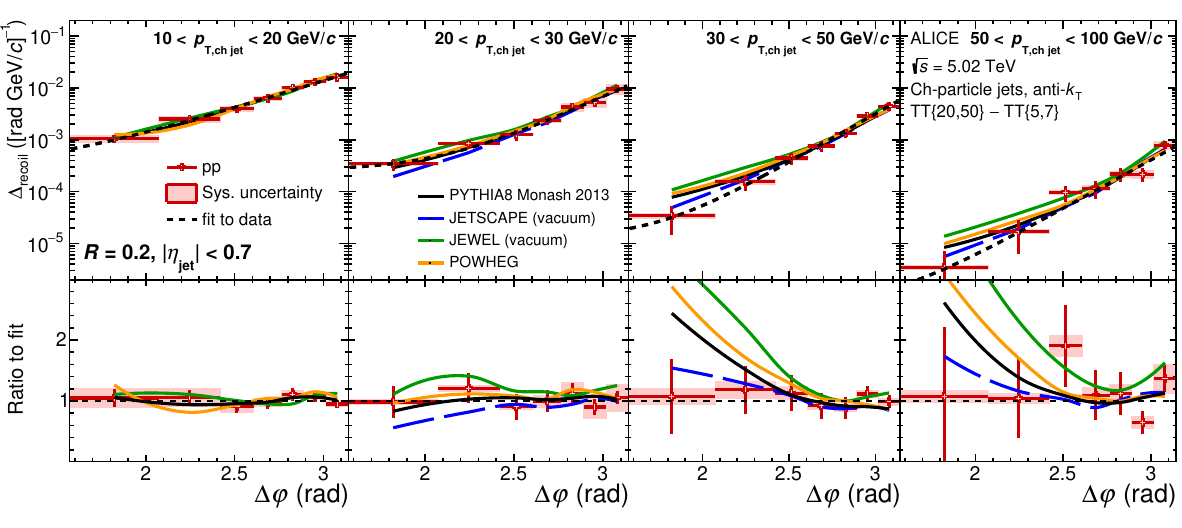}
  \includegraphics[width = .95\textwidth] {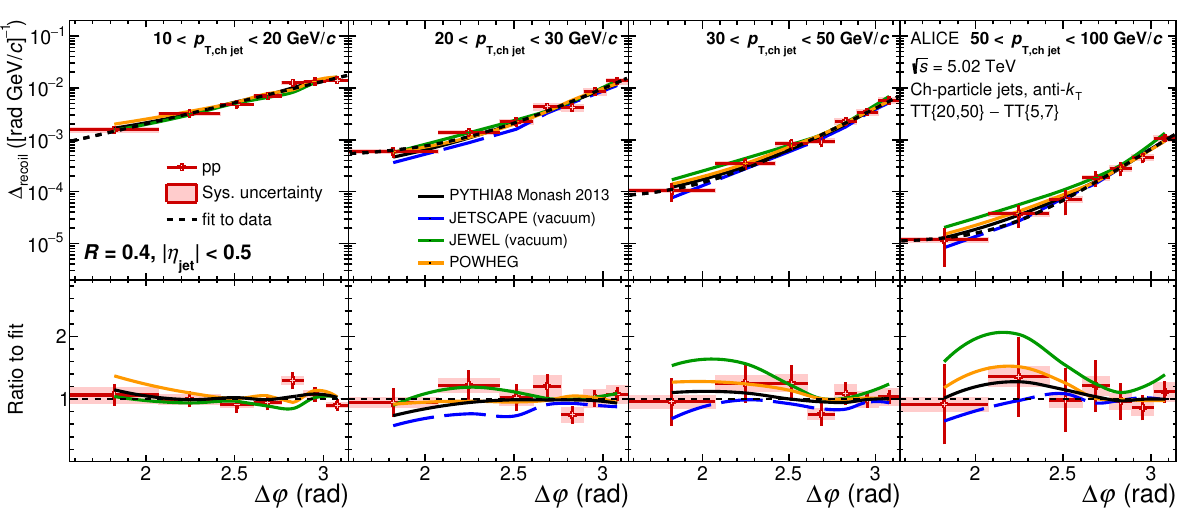}
  \includegraphics[width = .95\textwidth] {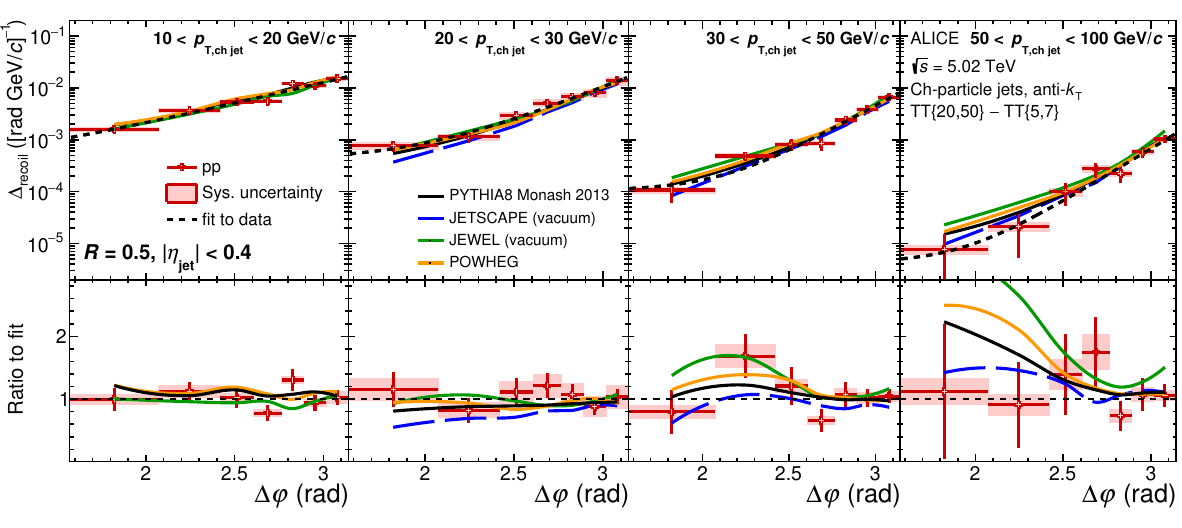}
     \end{center}
\caption{Corrected \Drecoilphi\ distributions for \pp\ collisions at $\sqrts=5.02$ TeV for $\rr=0.2$ (top), 0.4 (middle), and 0.5 (bottom) in \pTjetch\ bins (left to right): [10,20], [20,30], [30,50], and [50,100] \gev. JETSCAPE, JEWEL, PYTHIA8, and POWHEG calculations are also shown.
Upper sub-panels show the individual distributions, while lower sub-panels show their ratio to a functional fit of the measured data.}
\label{fig:pp_Drecoil_Dphi_theory}
\end{figure}

Figure~\ref{fig:pp_Drecoil_pT}, upper panels, show fully-corrected \DrecoilpTch\ distributions for $\rr=0.2,~0.4$, and 0.5 measured in \pp\ collisions at $\sqrts=5.02$ TeV, together with comparison to model calculations based on PYTHIA8 Monash 2013 tune~\cite{Sjostrand:2014zea,Skands:2014pea},
 JEWEL (vacuum)~\cite{Zapp:2012ak,Zapp:2013vla}, 
JETSCAPE (vacuum)~\cite{JETSCAPE:2019udz}, and POWHEG~\cite{Alioli:2010xa}. Figure~\ref{fig:pp_Drecoil_pT}, lower panels, show the ratio of the distributions in the upper panels to the fit of a smooth function to the data, in order to suppress fluctuations in the data for comparison purposes. The same smoothing procedure is used in the lower panels of Figs.~\ref{fig:pp_Drecoil_Dphi_theory},~\ref{fig:PbPb_Drecoil_pT}, and~\ref{fig:PbPb_Drecoil_Dphi_theory}.

The PYTHIA8 and JETSCAPE calculations agree with the data within experimental uncertainties over the full \pTjetch\ range. These calculations are related, since JETSCAPE utilizes PYTHIA8 for hard process generation and string fragmentation, with independent procedures for final-state parton showering and hadronization. These independent processes are expected to have little effect on jet distributions, however. The POWHEG calculations likewise describe the data well over the full \pTjetch\ range. The JEWEL calculation does not describe \pTjetch-dependence of $\DrecoilpTch$ well, overestimating the data for \pTjetch\ $>30$~\gev, with $\approx$40\% disagreement at high \pTjetch. 

Figure~\ref{fig:pp_Drecoil_Dphi_theory} shows corrected \Drecoilphi\ distributions for $\rr=0.2$, 0.4, and 0.5 measured in \pp\ collisions at $\sqrts=5.02$ TeV in various \pTjetch\ bins, together with comparisons to theoretical calculations. The JETSCAPE calculation agrees with the data within uncertainties in all panels. The other calculations also agree with the data within uncertainties except for $\dphi\lesssim2.5$ in the ranges $\pTjetch>30$ \gev\ for $\rr=0.2$ (PYTHIA8, POWHEG, JEWEL), and $\pTjetch>50$ \gev\ for $\rr=0.5$ (POWHEG, JEWEL). These \pp\ data provide the reference for comparison to same distributions measured in \PbPb\ collisions, to explore medium-induced effects.

\subsection{\pmb{\Drecoil} in \PbPb\ collisions}
\label{sect:ResultsDrecoilPbPb}

\begin{figure}[tbhp]
    \begin{center}
    \includegraphics[width = 1.0\textwidth] {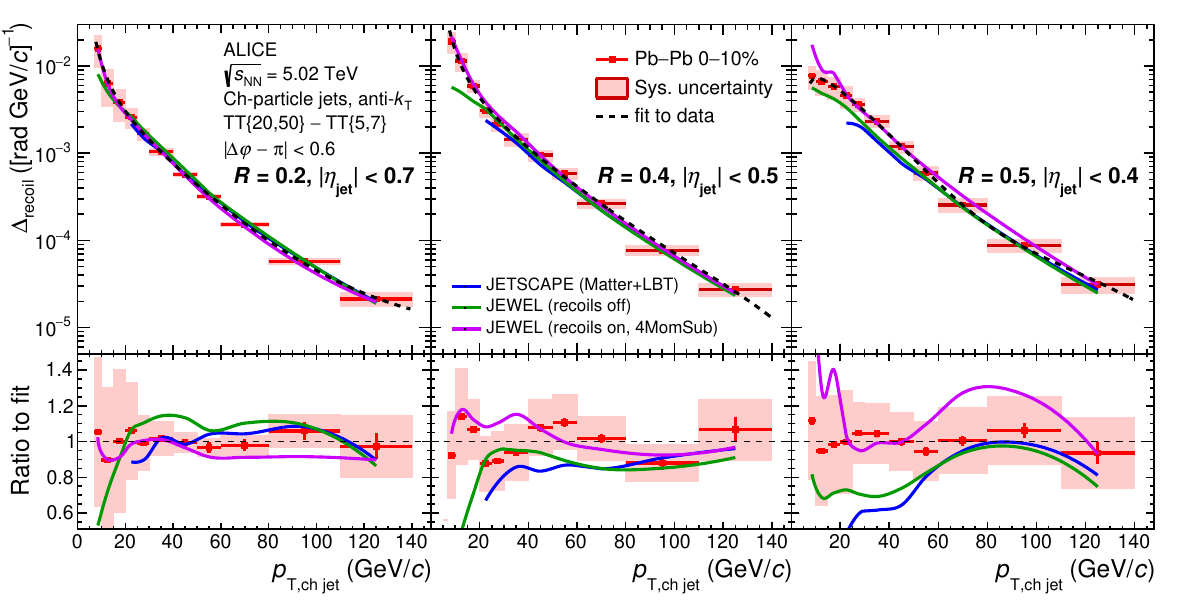}
  \end{center}
\caption{Upper panels: corrected \DrecoilpTch\ distributions measured for $\rr=0.2$ (left), 0.4 (middle), and 0.5 (right) in \PbPb\ collisions at $\sqrtsNN=5.02$ TeV, compared to theoretical calculations from JETSCAPE and JEWEL. Lower panels: ratio of the data and calculations to a functional fit of the measured \DrecoilpTch\ distributions.}
\label{fig:PbPb_Drecoil_pT}
\end{figure}

Figure~\ref{fig:PbPb_Drecoil_pT} shows the corrected \DrecoilpTch\ distributions measured in \PbPb\ collisions at $\sqrtsNN=5.02$ TeV, with comparison to theoretical calculations from JETSCAPE and JEWEL (both recoils on and recoils off). JETSCAPE reproduces the measurements well over the full \pTjetch\ range for $\rr=0.2$ and 0.4, and underpredicts the data for $\rr=0.5$. JEWEL (recoils off) agrees with the data at high \pTjetch\ for all \rr, while it underpredicts the data at low \pTjetch\ for $\rr=0.4$ and 0.5. JEWEL (recoils on) describes the data better in the lowest \pTjetch\ region of the measurement for $\rr=0.2$ and 0.4, while it tends to overshoot the data at low-\pTjetch\ for $\rr=0.5$.
Overall, these models provide a reasonable description of the \DrecoilpTch\ distribution in \PbPb collisions for $\rr=0.2$, while for $\rr=0.4$ and $\rr=0.5$, JEWEL (recoils on) best captures the main features of the data. However, none of the models quantitatively describes the data over the full \pTjetch\ range for all values of \rr.

\begin{figure}[tbhp]
  \begin{center}
  \includegraphics[width = 0.9\textwidth] {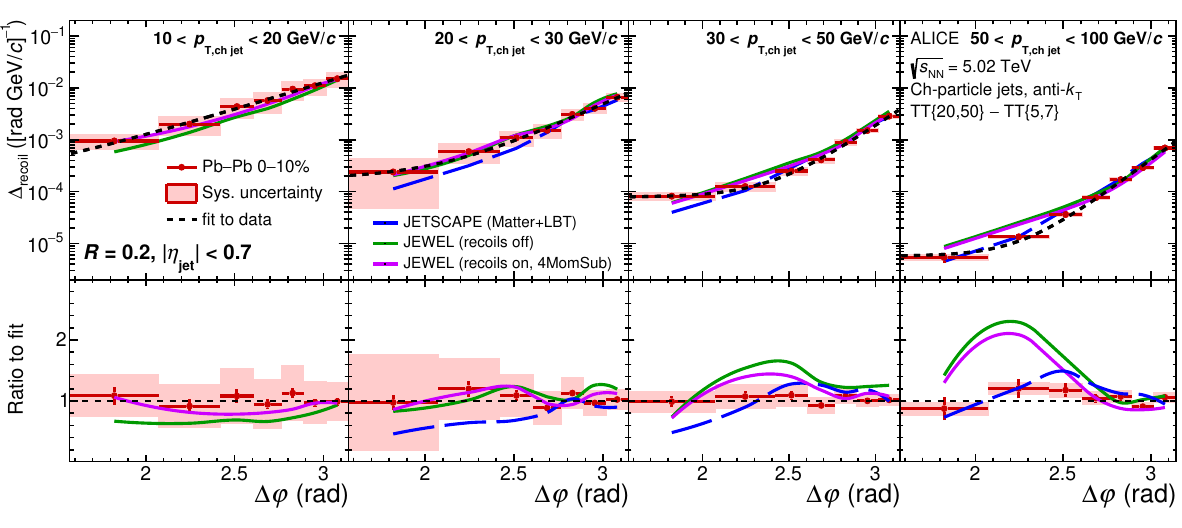}
  \includegraphics[width = 0.9\textwidth] {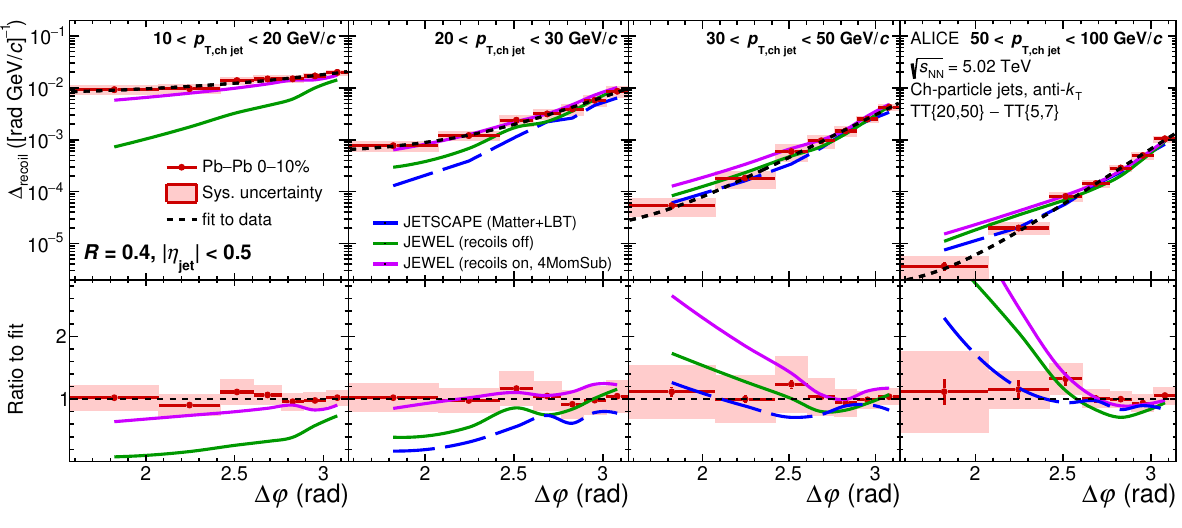}
  \includegraphics[width = 0.9\textwidth] {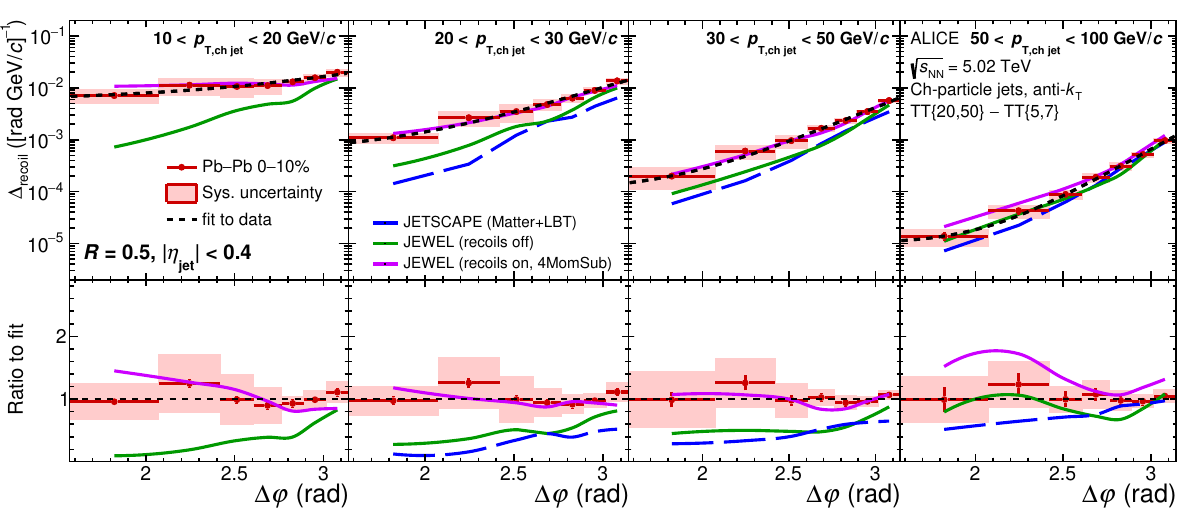}
\end{center}
\caption{Corrected \Drecoilphi\ distributions for \PbPb\ collisions at $\sqrtsNN=5.02$ TeV, for $\rr=0.2$ (top), 0.4 (middle), and 0.5 (bottom) in \pTjetch\ bins (left to right): [10,20], [20,30], [30,50], and [50,100] \gev. JETSCAPE and JEWEL calculations are also shown. Upper sub-panels show the individual distributions, while lower sub-panels show their ratio to a functional fit of the data.}
\label{fig:PbPb_Drecoil_Dphi_theory}
\end{figure}

Figure~\ref{fig:PbPb_Drecoil_Dphi_theory} shows the corrected \Drecoilphi\ distributions for \PbPb\ collisions at $\sqrtsNN=5.02$ TeV for $\rr=0.2$, 0.4, and 0.5 in selected \pTjetch\ bins, with comparison to model calculations. JETSCAPE describes the $\dphi$ distributions fairly well in the ranges $\pTjetch>20$ \gev\ for $\rr=0.2$ and $\pTjetch>30$ \gev\ for $\rr=0.4$, underpredicting the data in $20<\pTjetch<30$ \gev\ for $\rr=0.4$ and $\pTjetch>20$ \gev\ for $\rr=0.5$. JEWEL (recoils off) describes the data well only for $10<\pTjetch<30$ \gev\ for $\rr=0.2$ and $50<\pTjetch<100$ \gev\ for $\rr=0.5$, with poorer agreement in all other panels. JEWEL (recoils on) describes the data well except for  the range $30<\pTjetch<100$ \gev\ for $\rr=0.2$ and 0.4.

Overall, JETSCAPE provides the best description of the data for high \pTjetch\ and small \rr. JEWEL (recoils on) does best for low \pTjetch\ and large \rr, while JEWEL (recoils off) and JETSCAPE significantly underestimate the tails of these distributions. These features indicate that the observable is sensitive to treatment of the medium response, which differs between the models. Further insight into in-medium jet modification can be gained by direct comparison of the \PbPb\ and \pp\ distributions, which is presented next.

\subsection{\pmb{\IAApT}}
\label{sect:ppPbPb_oneD}

\begin{figure}[tbhp]
    \begin{center}
    \includegraphics[width = 0.6\textwidth] {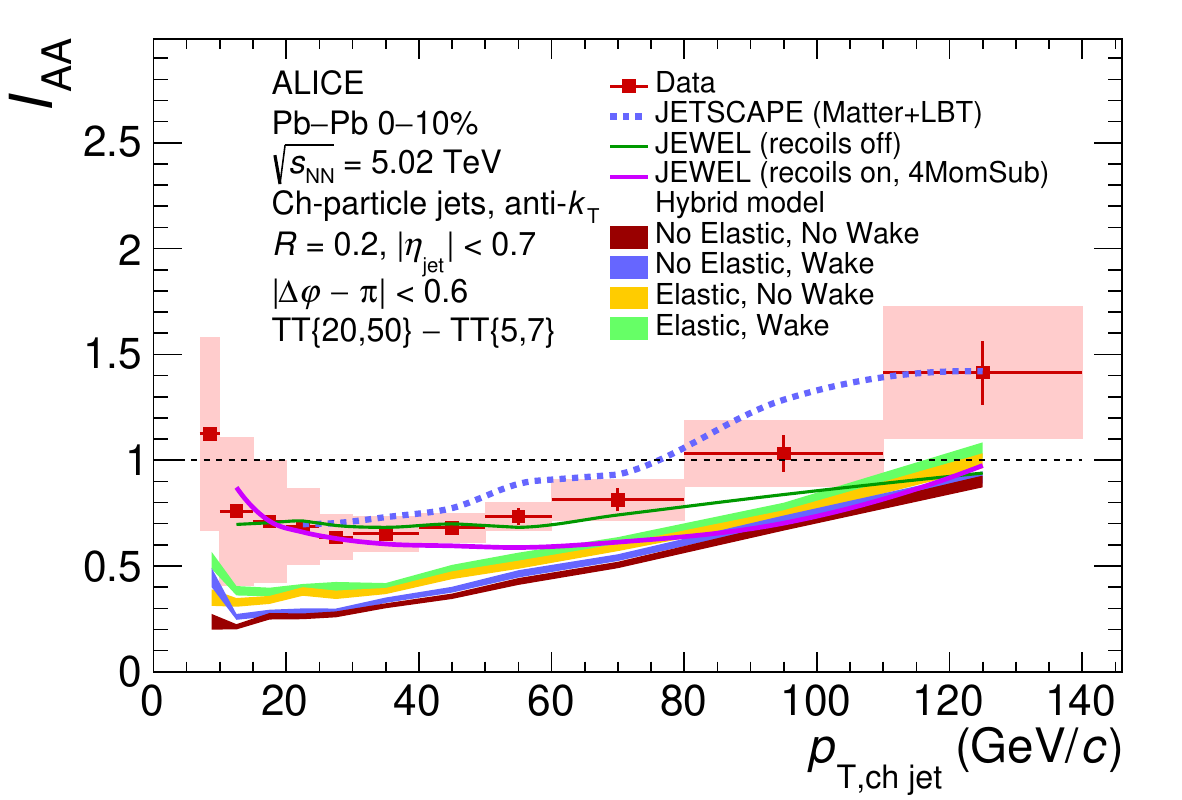}
    \includegraphics[width = 0.6\textwidth] {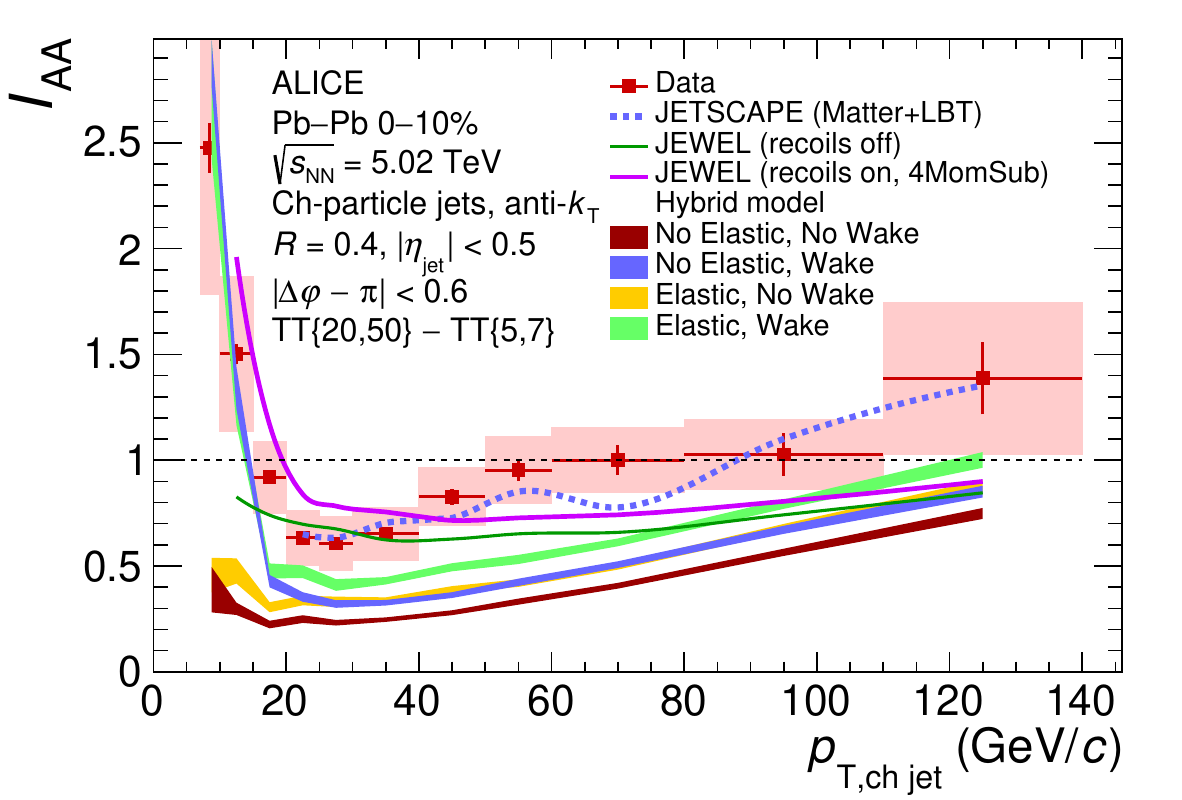}
    \includegraphics[width = 0.6\textwidth] {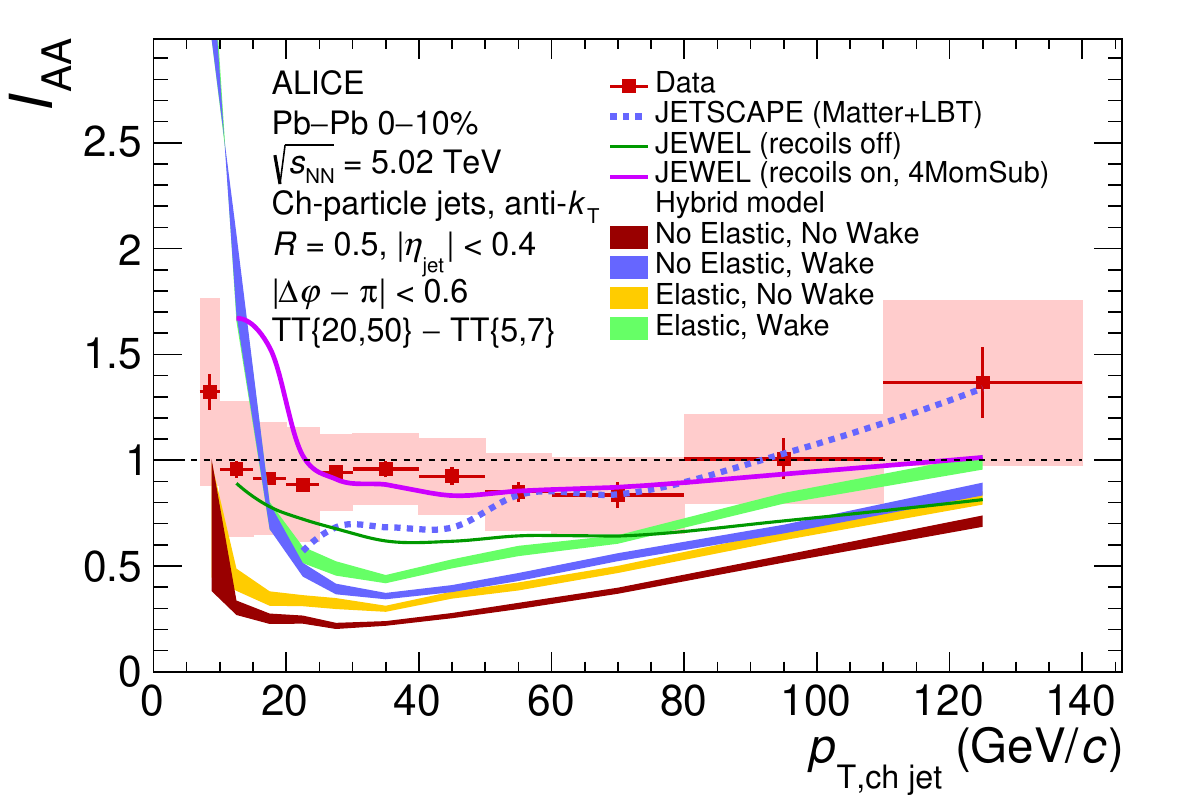}
  \end{center}
\caption{\IAApT\ from the \DrecoilpT\ distributions measured for $\rr=0.2$ (top), 0.4 (middle), and 0.5 (bottom) in central \PbPb\ (Fig.~\ref{fig:PbPb_Drecoil_pT}) and \pp\ collisions (Fig.~\ref{fig:pp_Drecoil_pT}).
JETSCAPE, JEWEL, and the Hybrid Model calculations are also shown.}
\label{fig:IAA}
\end{figure}

Medium-induced yield modification is measured by $\IAApT = \Drecoil(\PbPb)/\Drecoil(\pp)$, the ratio of the \DrecoilpT\ distributions measured in \PbPb\ and \pp\ collisions. Figure~\ref{fig:IAA} shows \IAApT, determined using the \DrecoilpTch\ measurements in
Figs.~\ref{fig:pp_Drecoil_pT} and~\ref{fig:PbPb_Drecoil_pT}.

The \IAApT\ distributions have significant dependence on \pTjetch\ and \rr. For $\pTjetch<20$ \gev, \IAApT\ either increases above or is consistent with unity for all \rr. For $\rr=0.2$ and 0.4, \IAApT\ is lower than unity in the region $20<\pTjetch<60$ \gev, corresponding to medium-induced yield suppression due to energy loss, rising to be consistent with or larger than unity at larger \pTjetch. Note that $\IAA > 1$ does not necessarily imply that jets in that \pTjetch\ range do not experience energy loss; indeed, calculations in Ref.~\cite{He:2024rcv} indicate that energy loss of trigger-side jets can also enhance \IAA\ significantly.
In contrast, \IAApT\ for $\rr=0.5$ is consistent with unity over the range $7<\pTjetch<110$ \gev, indicating that the angular scale of medium-induced energy loss is less than 0.5 rad. Measurements of \IAApT\ for central \AuAu\ collisions at $\sqrtsNN=200$ GeV with direct photon and \pizero\ triggers have recently been reported~\cite{STAR:2023pal,STAR:2023ksv}, with a similar observation of less suppression for $\rr=0.5$ than for $\rr=0.2$, likewise indicating a similar angular scale of jet energy redistribution due to quenching at RHIC collision energies.

The JETSCAPE calculation describes well the measured \IAApT\ distributions for $\rr=0.2$ and 0.4 in $\pTjetch>20$ \gev, including the rising trend for $\pTjetch>60$ \gev. JETSCAPE predicts a similar \pTjetch-dependence of \IAApT\ for $\rr=0.5$, which however is not consistent with the measurement. 

The JEWEL calculations, both recoils-off and recoils-on, describe the \IAApT\ distribution for $\rr=0.2$ at low \pTjetch, but do not capture the \pTjetch\ dependence of the data and underpredict them at higher \pTjetch. For $\rr=0.4$, both versions underestimate the data at high \pTjetch. For $\rr=0.4$ and 0.5, JEWEL (recoils on) shows a significant increase in \IAApT\ towards low \pTjetch\ for $\pTjetch<20~\gev$, similar to the trend in the data for $\rr=0.4$. This increase is not seen for recoils-off. The larger value of \IAApT\ in $20<\pTjetch<60$~\gev\ for $\rr=0.5$ seen in the data is reproduced  by JEWEL with recoils-on but not recoils-off. This \rr-dependence is due to the implementation of medium response in JEWEL, in which energy is carried by recoiling partons at large angles to the jet centroid~\cite{KunnawalkamElayavalli:2017hxo}.

Hybrid Model calculations of \IAApT\ underestimate the magnitude of the data for all settings, although they do reproduce the rising trend with increasing \pTjetch\ seen in the data for $\pTjetch > 20$~\gev for $\rr=0.2$ and $\rr=0.4$. The Hybrid Model with wake turned on likewise captures the sharply rising trend in the data with decreasing \pTjetch\ at low \pTjetch\ for $\rr=0.4$, while no rising trend is seen when the wake is turned off, independent of the elastic scattering component. The model also exhibits a rising trend for $\rr=0.5$, which in this case is not seen in the data within the experimental uncertainties.

Overall, JETSCAPE most accurately describes both the magnitude and \pTjetch\ dependence of \IAApT\ in the range $\pTjetch>20$ \gev\ for $\rr=0.2$ and $\rr=0.4$, while JEWEL most accurately describes it in the same \pTjetch\ region for $\rr=0.5$. The rising trend in data towards low \pTjetch\ for $\rr=0.4$ in $\pTjetch<20$ \gev\ is described by both the Hybrid Model and JEWEL, but only with the inclusion of medium-response effects. These models do not, however, describe the flatter trend seen for $\rr=0.5$.

\subsection{\pmb{\IAApT} and trigger bias}
\label{sect:IAAtrigbias}

A common picture of inclusive high-\pT\ hadron production in the presence of jet quenching in central \aaa\ collisions is that the geometric distribution of vertices which generate such hadrons is biased, due to the interplay of jet energy loss and the shapes of the jet \pT\ spectrum and fragmentation function~\cite{Baier:2002tc,Zhang:2007ja,Renk:2012ve,Bass:2008rv,Adam:2015doa}. In this picture, observed high-\pT\ hadrons arise predominantly from jets generated in hard partonic scattering processes occurring at the surface of the hot QGP, headed outward (``surface bias''). For the semi-inclusive observable in this measurement, this surface bias implies a longer average path length for the jet population recoiling from a high-\pT\ hadron trigger than for the unbiased inclusive jet population.

This analysis indeed selects events based on the presence of a trigger hadron, which is a leading fragment of a ``trigger'' jet. To characterize this measurement the observable \ztilde\ is defined as

\begin{equation}
\ztilde=\frac{\pTtrig}{\pTjetch}, 
\label{eq:ztilde}
\end{equation}

\noindent
which is the ratio between the trigger hadron \pT\ and the recoiling charged-particle jet \pT. Dijet production at LO is a $2\rightarrow2$ process, generating a jet pair that is azimuthally back-to-back at the same value of \pTjetch. If LO processes dominate hadron production, and the measurement includes the full (rather than charged-particle) energy of the recoil jet, then $\ztilde<1$ for all jets contributing to the \IAA\ distribution. However, higher-order processes are in general not negligible, and the recoil jets in this measurement are based on charged particles only. Initial-state \kT\ effects can also modify \ztilde. Nevertheless, \ztilde\ provides a useful qualitative categorization of different kinematic regions in Fig.~\ref{fig:IAA}. 

The surface-bias argument outlined above is based on the shape of the high-$z$ (high momentum fraction) tail of the fragmentation function. However, for $\pTtrig=20$ \gev, which is the lower bound of the \TTSig\ trigger interval \TT{20}{50}, $\pTjetch=100$ \gev\ corresponds to $\ztilde=0.2$, and its value would be yet smaller if the fully-reconstructed jet energy were used. It is evident that, for events containing jets that contribute to the highest \pTjetch\ bins in Fig.~\ref{fig:IAA}, the assumptions underlying the surface-bias argument may not pertain; in particular, the trigger hadron that satisfies the  \TT{20}{50} selection may not be a leading fragment of its parent (``trigger'') jet. 

\begin{figure}[tbhp]
    \begin{center}
    \includegraphics[width = 0.6\textwidth] {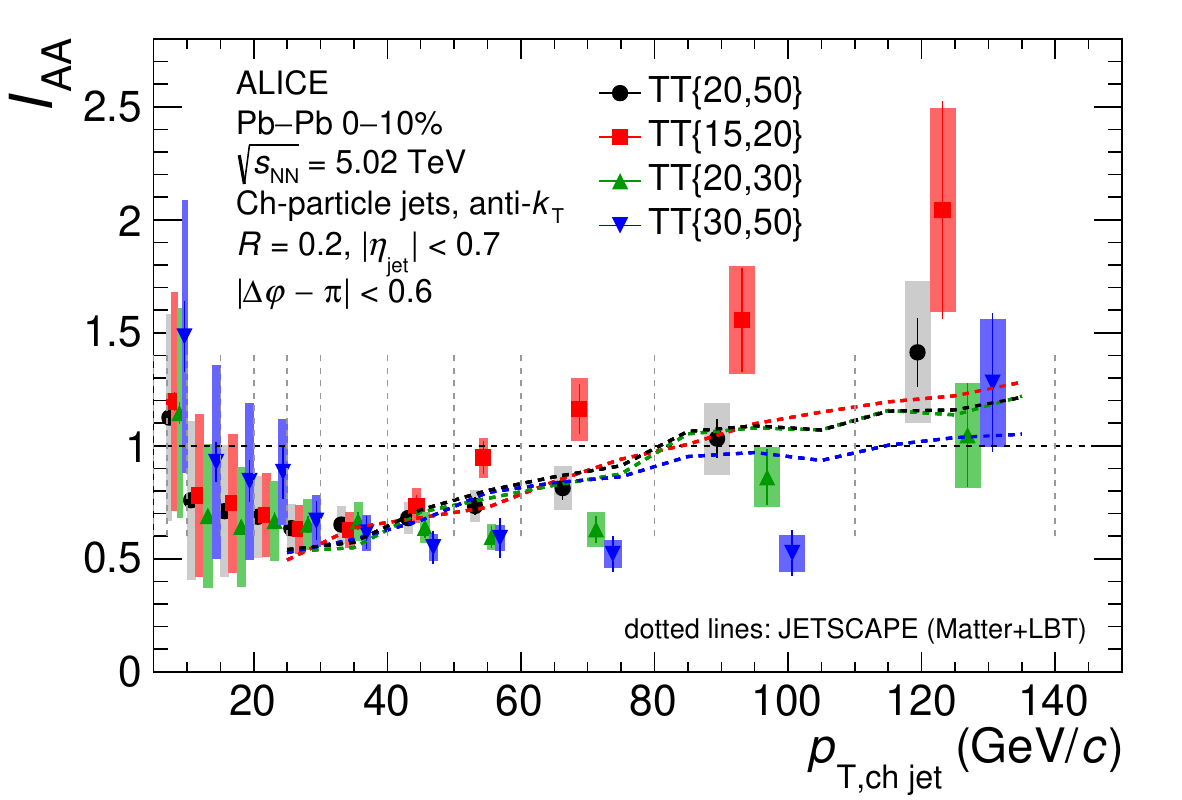}
    \includegraphics[width = 0.6\textwidth] {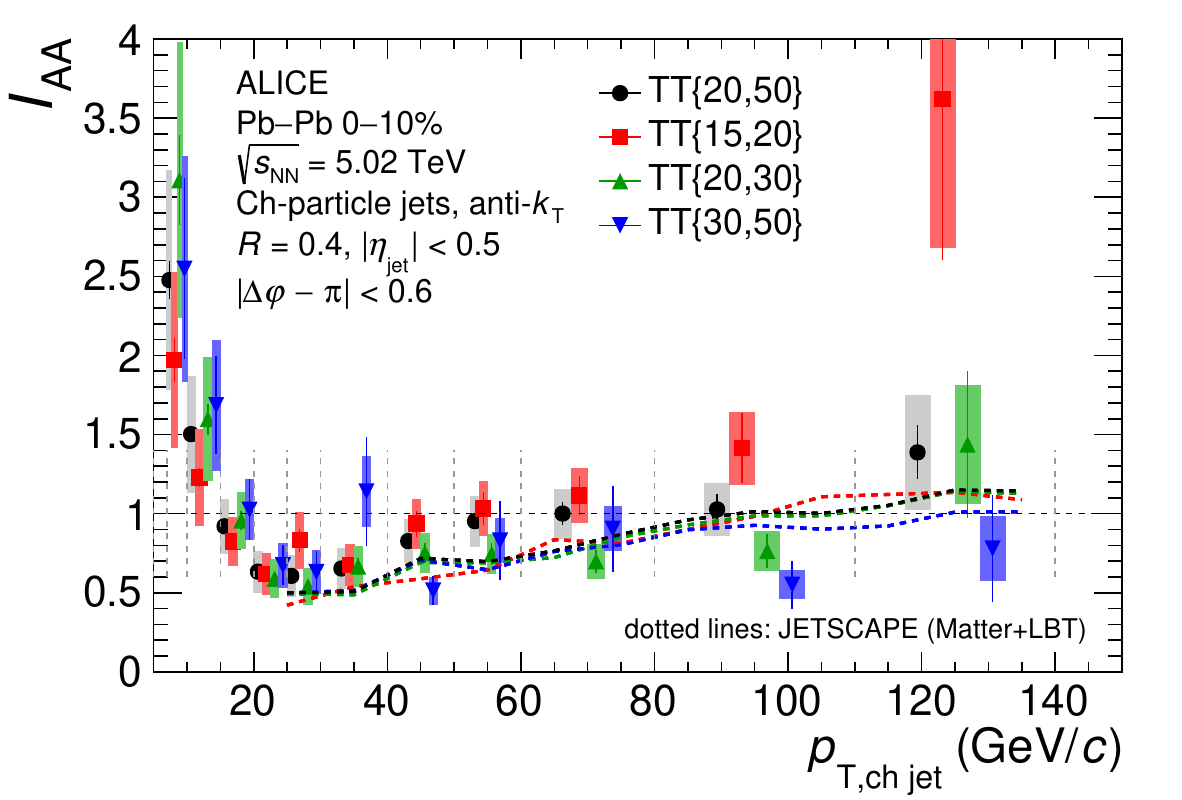}
  \end{center}
\caption{\IAApT\ for $\rr=0.2$ and 0.4, for various \TTSig\ selections. The data points have been displaced for clarity, and the vertical dashed grey lines indicate the bin edges. JETSCAPE predictions for the corresponding \TTSig\ selections are also shown.}
\label{fig:IAA-TT}
\end{figure}

Further insight into the rising trend of \IAApT\ at large \pTjetch\ seen in Fig.~\ref{fig:IAA} can be obtained by varying the hadron trigger \pT\ range for the \TTSig-selected event population, i.e.~by varying the distribution of \ztilde\ for fixed recoil \pTjetch. Figure~\ref{fig:IAA-TT} shows the \IAApT\ distribution for $\rr=0.2$ and 0.4 measured for several choices of \pTtrig\ interval for the \TTSig\ event selection. A higher \pTtrig\ threshold corresponds to larger \ztilde, where the assumptions underlying the surface-bias picture may better apply. The results show that, as the lower \pTtrig\ bound is raised, the rate of increase in \IAApT\ at large \pTjetch\ diminishes. 

As noted above, large \ztilde\ may correspond to larger average in-medium path length of the recoiling jet, with corresponding larger recoil yield suppression due to quenching. Figure~\ref{fig:IAA-TT} shows that increasing \ztilde\ indeed results in larger  recoil yield suppression, consistent with this geometrical picture. The figure also shows JETSCAPE calculations, which exhibit a slightly slower rise in \IAApT\ at large \pTjetch\ for higher \pTtrig\ intervals (larger \ztilde), although the variation is quantitatively smaller than that seen in the data. The results in Fig.~\ref{fig:IAA-TT} provide new insight into the conjecture of surface bias for inclusive high-\pT\ hadron production, and the interplay of jet quenching effects as observed via hadronic and reconstructed jet observables. 

\subsection{Jet shape modification: \rr-dependence of \pmb{\DrecoilpTch}}
\label{sect:rdepyields}

The ratio of inclusive jet cross sections or semi-inclusive jet yields at different values of \rr\ provides a precise probe of jet shape, since there is significant cancellation of correlated uncertainties in the ratio for both experimental measurements and theoretical calculations ~\cite{Soyez:2011np,Abelev:2013fn,Dasgupta:2014yra,Acharya:2019jyg}. In \pp\ collisions, \rr-dependent ratios are sensitive to high-order pQCD effects~\cite{Abelev:2013fn, CMS:2014nvq,Acharya:2019jyg, Dasgupta:2016bnd}. In \aaa\ collisions, such ratios provide experimentally robust probes of  medium-induced modification of jet shapes over a broad kinematic range, including low \pTjet~\cite{Adam:2015doa,Adamczyk:2017yhe,STAR:2020xiv}.

\subsubsection{pp collisions}
\label{sect:rdepyieldspp}

\begin{figure}[tbhp]
\begin{center}
\includegraphics[width = 0.49\textwidth] {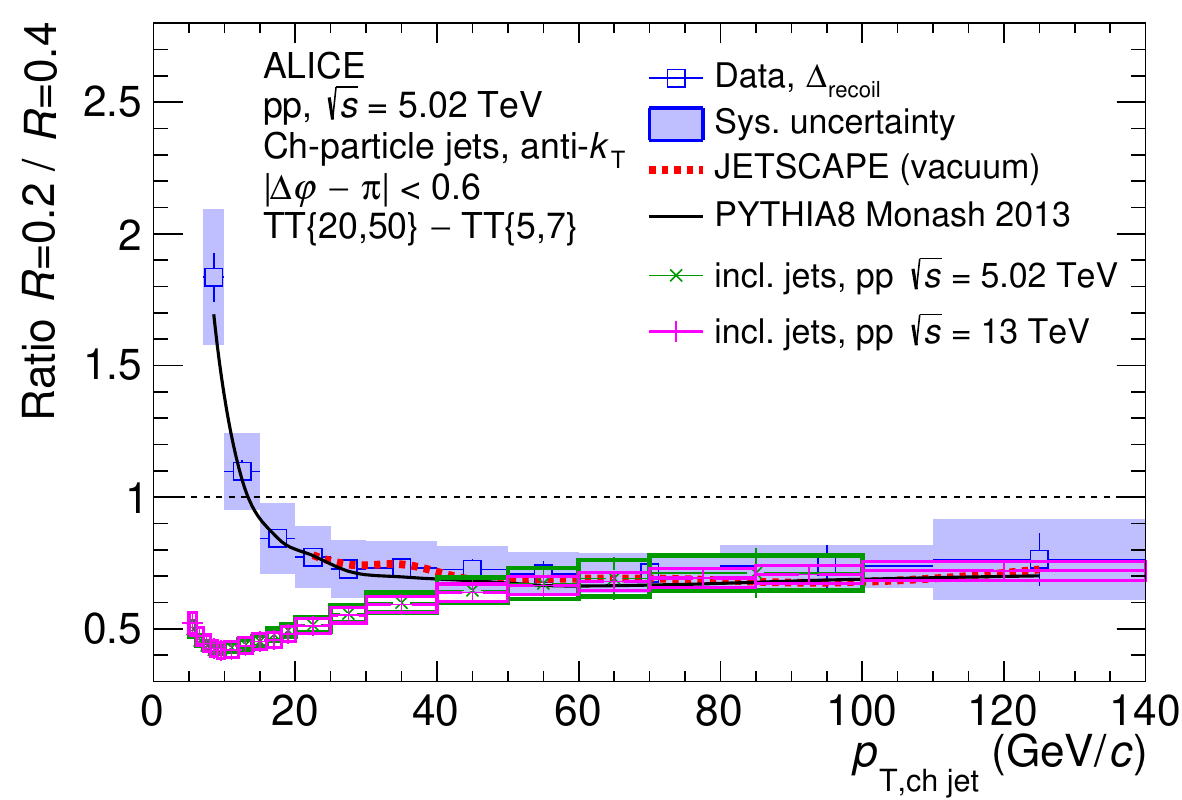}
\includegraphics[width = 0.49\textwidth] {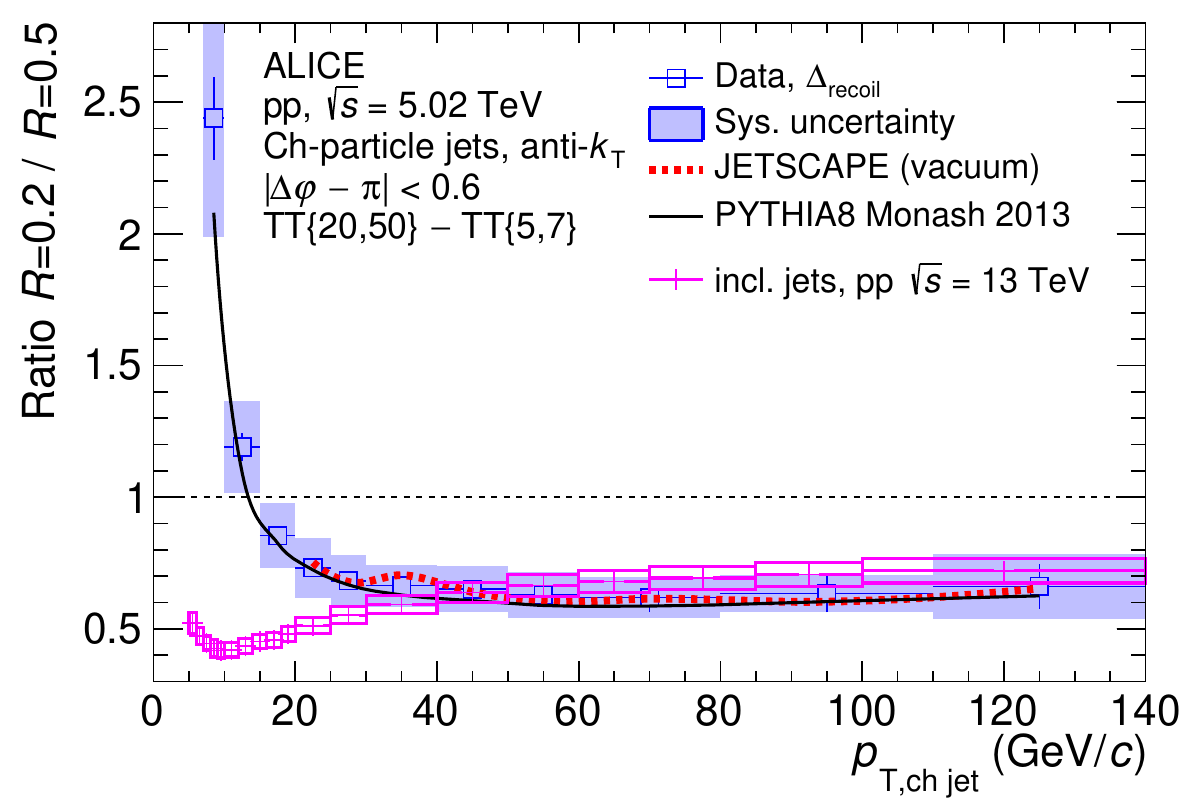}
 \end{center}
\caption{Ratio of \DrecoilpTch\ distributions in \pp\ collisions at $\sqrts=5.02$ TeV using the data from Fig.~\ref{fig:pp_Drecoil_pT}, for $\rr=0.2/\rr=0.4$ (left) and $\rr=0.2/\rr=0.5$ (right), compared to calculations from PYTHIA8 and JETSCAPE. The corresponding ratios of cross sections for inclusive jets are also shown for \pp\ collisions at $\sqrts = 5.02$~\TeV~\cite{Acharya:2019tku} and $\sqrts = 13$~\TeV~\cite{ALICE:2022jbp}. The uncertainties in the ratio take into account the correlation of uncertainties between numerator and denominator.} 
\label{fig:RRatios-pp}
\end{figure}

Figure~\ref{fig:RRatios-pp} shows the ratio of \DrecoilpTch\ distributions for $\rr=0.2$ over that for $\rr=0.4$ or $\rr=0.5$ in \pp\ collisions at $\sqrts=5.02$ TeV, using the data in Fig.~\ref{fig:pp_Drecoil_pT}. The ratio is below unity for $\pTjetch>15$ \gev, consistent with the expected intra-jet energy distribution in which significant energy is carried at distances larger than 0.2 radians relative to the jet axis. The ratio rises towards low \pTjetch\ and crossing unity at $\pTjetch\approx10$ \gev. The measured distributions are well-reproduced by PYTHIA8 and JETSCAPE (vacuum) calculations. This \pTjetch-dependence and the good agreement of this observable with PYTHIA8 were also observed in Ref.~\cite{Adam:2015doa}, where an NLO pQCD calculation was likewise shown to reproduce the measured \pTjetch-dependence of the ratio, although not its absolute magnitude.

Figure~\ref{fig:RRatios-pp} also shows the ratios of the inclusive charged-jet cross sections for different values of \rr\ in \pp\ collisions at $\sqrts = 5.02$~\TeV~\cite{Acharya:2019tku} and $\sqrts = 13$~\TeV~\cite{ALICE:2022jbp}. At high \pTjetch\, these ratios are consistent within uncertainties with the corresponding \DrecoilpTch\ ratios. However, the observed increase in the \DrecoilpTch\ ratio with decreasing \pTjetch\ is opposite to the behavior of the ratios for inclusive jet cross sections, which decrease with decreasing \pTjetch. Note that the inclusive charged jet cross section ratios are also well-described by pQCD and Monte Carlo model calculations~\cite{Acharya:2019tku,ALICE:2022jbp}. Since PYTHIA8 accurately reproduces the \rr-dependent ratios for both populations, this difference evidently originates in QCD processes that are incorporated in PYTHIA8. Similar phenomenology of \rr-dependent yield ratios has also been observed in semi-inclusive measurements with direct photon and \pizero\ triggers in \pp\ collisions at $\sqrts=200$ GeV ~\cite{STAR:2023ksv}.

The difference in low-\pTjetch\ behavior of the inclusive and coincidence channel \rr-ratios may be understood in terms of the \ztilde\ observable defined in Eq.~\ref{eq:ztilde}. Since the lower bound of the \TTSig\ selection is $\pTtrig=20$ \gev, the region $\pTjetch<20$ \gev\ in Fig.~\ref{fig:RRatios-pp} corresponds to $\ztilde>1$, where LO processes are suppressed and jet production is dominated by higher-order processes that incorporate hard gluon radiation. For an NLO process in which hard radiation is emitted at angles $\approx0.2-0.4$ rad relative to the jet axis, jet reconstruction may result in two rather than one jet (jet splitting) preferentially for $\rr=0.2$ relative to $\rr=0.4$. In this semi-inclusive analysis, in which the number of jets in a triggered event is simply counted, such splittings will enhance the \rr-dependent \DrecoilpT\ ratio at low \pTjetch, consistent with Fig.~\ref{fig:RRatios-pp}.
The inclusive population is not subject to a bias like that induced by \pTtrig\ for the coincidence population, and its \rr-dependent inclusive cross section ratio at low \pTjet\ therefore reflects the ensemble-averaged shape of the unbiased jet distribution. 

This picture can be further explored using model calculations. If LO production processes indeed are suppressed in the phase space $\ztilde>1$, this may provide a new tool for controlled generation of a population of ``wide'' jets, which are predicted to interact differently on average with the QGP as compared to a more inclusive population~\cite{Casalderrey-Solana:2019ubu}. 

\subsubsection{\PbPb\ collisions}
\label{sect:rdepyieldsPbPb}

\begin{figure}[tbhp]
\begin{center}
\includegraphics[width = 0.55\textwidth] {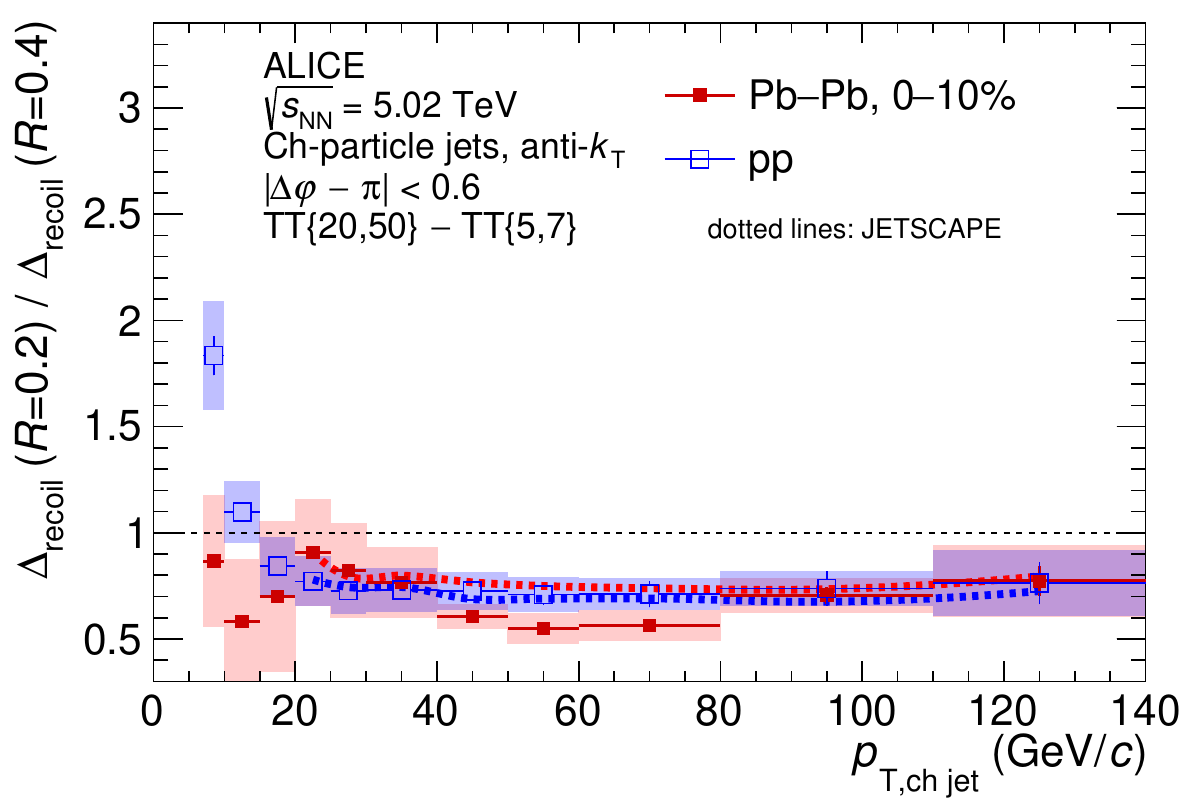}
\includegraphics[width = 0.55\textwidth] {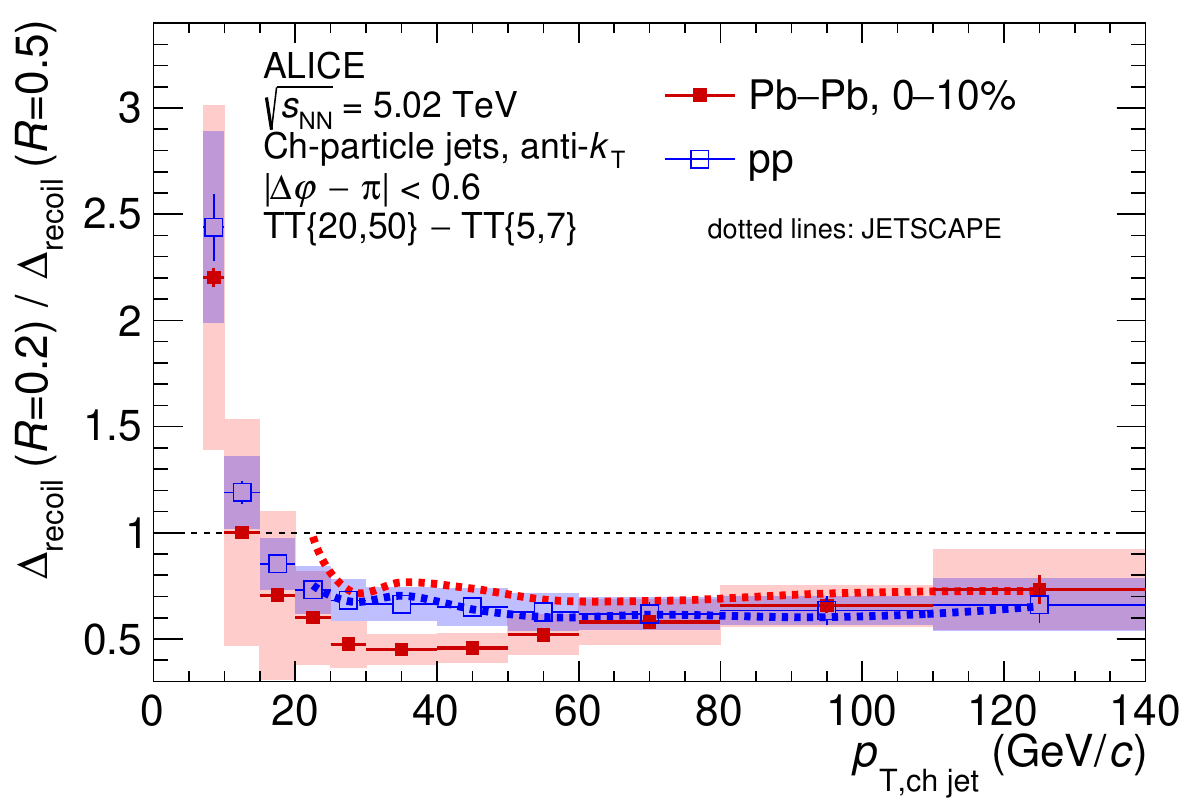}
\includegraphics[width = 0.55\textwidth] {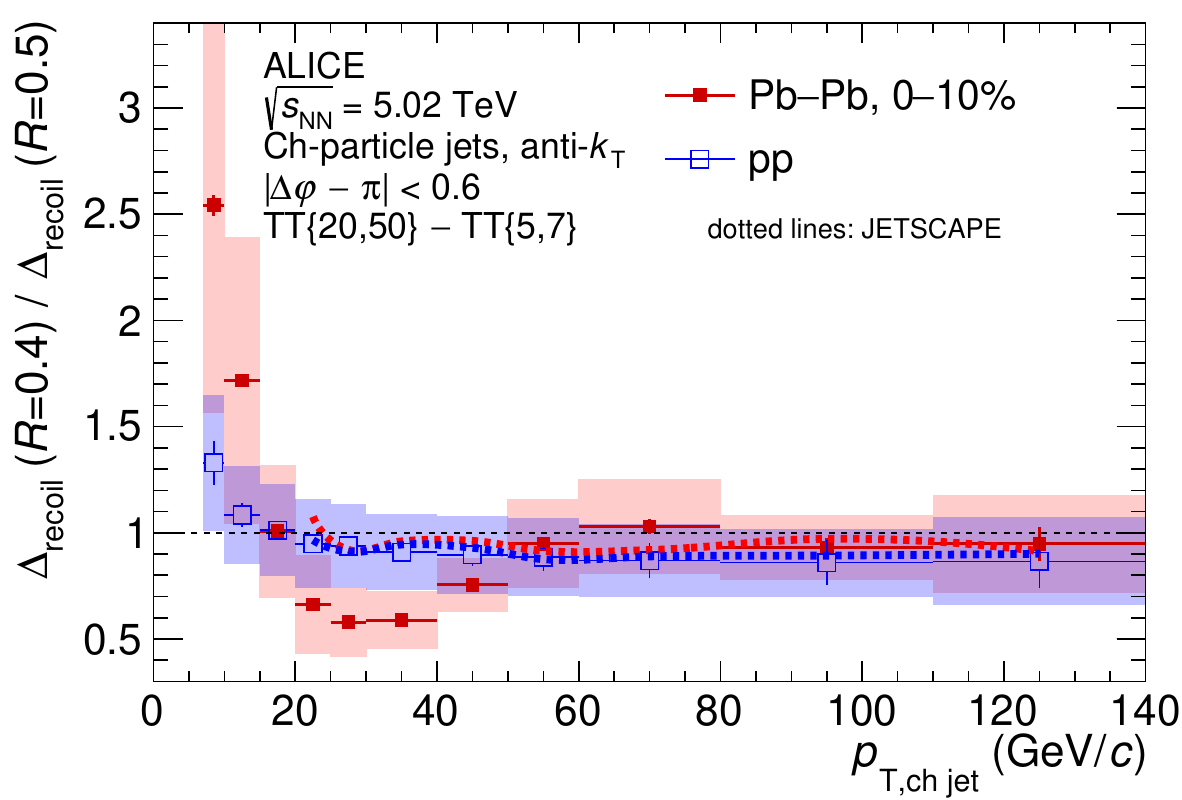}
\end{center}
\caption{
Ratio of \DrecoilpTch\ distributions with different \rr\ for \PbPb\ collisions at $\sqrtsNN=5.02$ TeV  from the data in Fig.~\ref{fig:PbPb_Drecoil_pT}, together with the ratios for \pp\ collisions from Fig.~\ref{fig:RRatios-pp}, for $\rr=0.2/\rr=0.4$ (upper), $\rr=0.2/\rr=0.5$ (middle), and $\rr=0.4/\rr=0.5$ (lower). The uncertainty in the ratio takes into account the correlation of uncertainties between numerator and denominator. JETSCAPE calculations for \pp\ and \PbPb\ collisions are also shown.}
 \label{fig:RRatios-PbPb}
\end{figure}

Figure~\ref{fig:RRatios-PbPb} shows the \rr-dependent ratio of \DrecoilpTch\ distributions measured for \PbPb\ and for \pp\ collisions. These ratios quantify the difference in shape of the individual \Drecoil\ distributions seen qualitatively in Figs.~\ref{fig:PbPb_Drecoil_pT} and ~\ref{fig:pp_Drecoil_pT}. For $\rr=0.2$ in the numerator and $\rr=0.4$ (upper) or $\rr=0.5$ (middle) in the denominator,
at intermediate values of \pTjetch\ the ratios for \PbPb\ collisions are lower than those for \pp\ collisions, indicating significant medium-induced intra-jet broadening in that region. 

The bottom panel shows the ratio of $\rr=0.4$ and $\rr=0.5$. 
For \pp\ collisions this ratio is consistent with unity within uncertainties at all values of \pTjetch. For \PbPb\ collisions it is consistent with unity for $\pTjetch>50$ \gev, but is significantly less than unity in the range $20<\pTjetch<50$ \gev\ and less than the \pp\ ratio. For $\pTjetch<20$ \gev, the ratio then rises, with the central points exceeding unity in the lowest \pTjetch\ region, although still consistent with the \pp\ ratio within 1$\sigma$. 
A difference could arise between the $\rr=0.4$ and $\rr=0.5$ jet populations in \PbPb\ collisions if there is significant jet quenching which broadens and softens transverse jet structure, so that jet area $\approx{\rr^2}$ is the most relevant factor in determining the \pTjet\ distribution.

For these ratios, the results of JETSCAPE calculations incorporating jet quenching in \PbPb\ collisions are larger than those in \pp\ collisions. This is in contrast to the data, where the ratios in \PbPb\ collisions are instead smaller than those in \pp\ collisions at intermediate \pTjetch. JETSCAPE also overpredicts the $\rr=0.2/\rr=0.4$ and $\rr=0.4/\rr=0.5$ ratios for $\pTjetch<50$ \gev. This indicates that medium-induced intra-jet modification is not accurately modeled in JETSCAPE.

The medium-induced suppression of the \rr-dependent ratio in Fig.~\ref{fig:RRatios-PbPb}, corresponding to medium-induced intra-jet broadening, is in contrast to a similar measurement of the inclusive jet population~\cite{ALICE:2023waz} which finds medium-induced jet narrowing in a similar kinematic range. The jet populations of these two measurements differ, however, and they cannot be compared directly. Exploration of this difference requires the calculation of both observables within the same model framework.

\subsection{Acoplanarity}
\label{sect:Acoplanarity}

\begin{figure}[tbhp]
\begin{center}
\includegraphics[width = 0.95\textwidth] {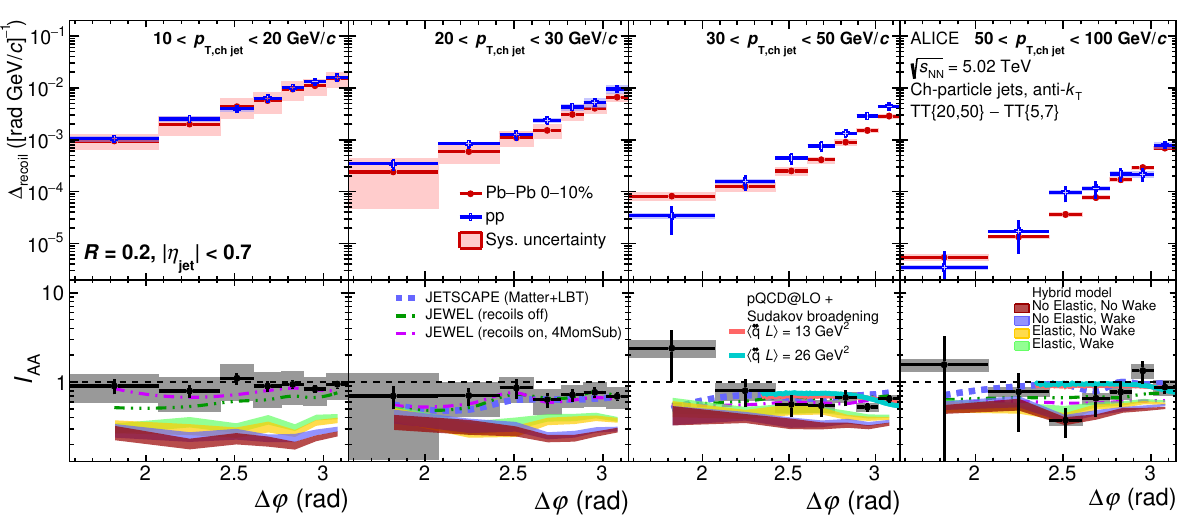}
\end{center}
\caption{Distributions as a function of \dphi\ for $\rr=0.2$. Upper panels: \Drecoilphi\ in intervals of \pTjetch\ measured in \pp\ and \PbPb\ collisions. Lower panels: \IAAphi, the ratio of the \pp\ and \PbPb\ distributions in the corresponding upper panel. Predictions from JETSCAPE, JEWEL, Hybrid model, and a pQCD calculation are also shown.}
\label{fig:DeltaPhi-pp_PbPb_02}
\end{figure}

\begin{figure}[tbhp]
\begin{center}
\includegraphics[width = 0.95\textwidth] {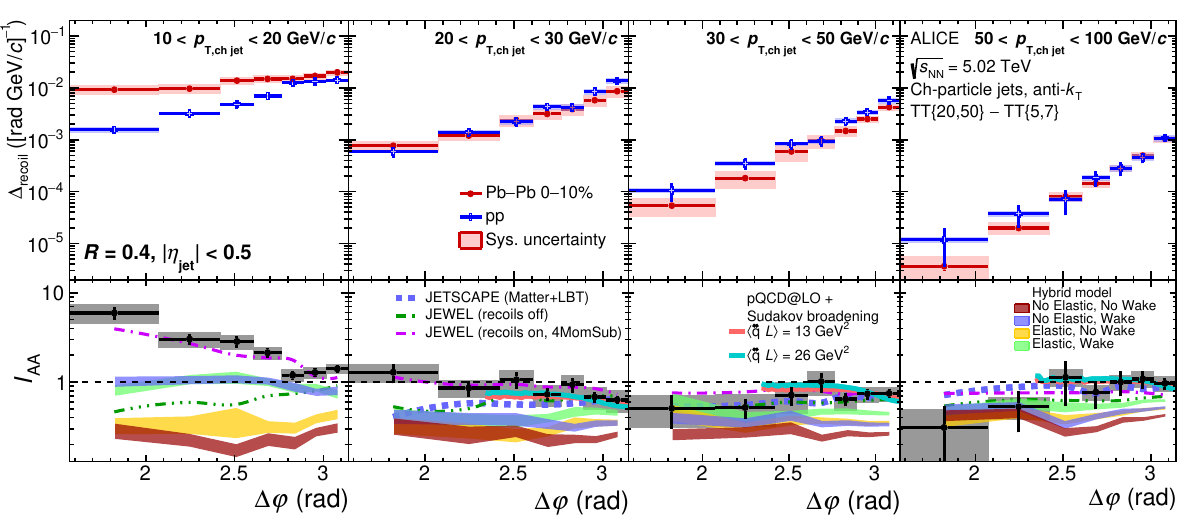}
\end{center}
\caption{Same as Fig.~\ref{fig:DeltaPhi-pp_PbPb_02}, for $\rr=0.4$.}
\label{fig:DeltaPhi-pp_PbPb_04}
\end{figure}

\begin{figure}[tbhp]
\begin{center}
\includegraphics[width = 0.95\textwidth] {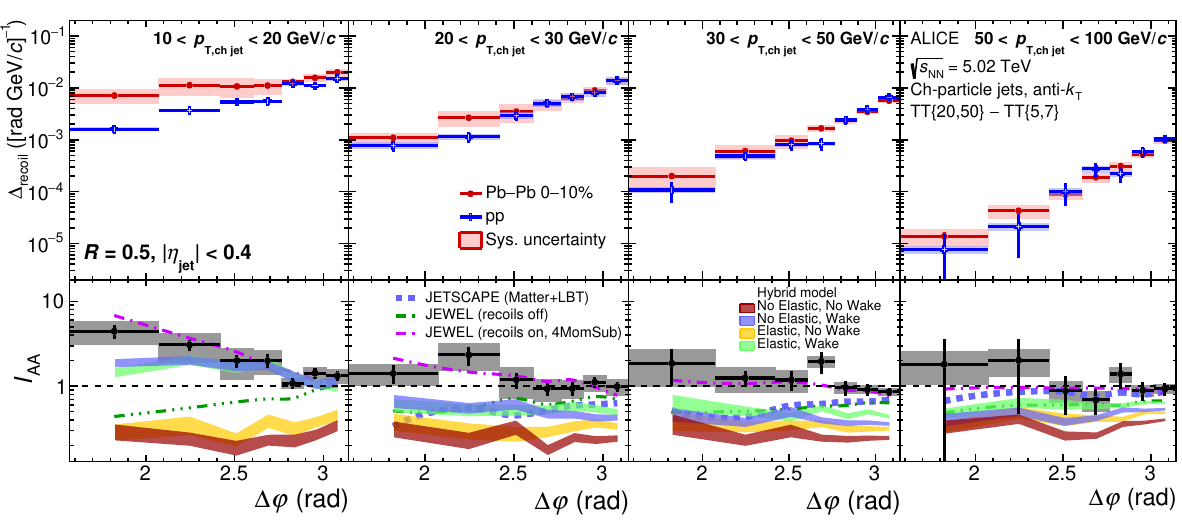}
\end{center}
\caption{Same as Fig.~\ref{fig:DeltaPhi-pp_PbPb_02}, for $\rr=0.5$.}
\label{fig:DeltaPhi-pp_PbPb_05}
\end{figure}

Figures~\ref{fig:DeltaPhi-pp_PbPb_02},~\ref{fig:DeltaPhi-pp_PbPb_04}, and~\ref{fig:DeltaPhi-pp_PbPb_05} show \Drecoilphi\ (acoplanarity) distributions measured in different \pTjetch\ intervals for \pp\ and \PbPb\ collisions, together with their ratio \IAAphi, for $\rr=0.2, 0.4$, and $0.5$, respectively. The key physics conclusions from these results, including phenomenological discussion and model comparisons for $\rr=0.4$, are presented in the companion Letter~\cite{ShorthJetpaper}. This paper additionally reports acoplanarity distributions in the range $50<\pTjetch<100$ \gev, and comparison to model calculations for $\rr=0.2$ and $\rr=0.5$.

Suppression of \IAAphi\ below unity can be seen in some \pTjetch\ regions, consistent with the yield suppression in Fig.~\ref{fig:IAA}. In the range $\pTjetch>10$~\gev\ for $\rr=0.2$, and the range $\pTjetch>20$~\gev\ for $\rr=0.4$ and $\rr=0.5$, the shape of the \PbPb\ \Drecoilphi\ distributions are consistent with that of the \pp\ distributions, corresponding to no significant in-medium acoplanarity broadening within the experimental uncertainties. In contrast, significant enhancement in \IAAphi\ at \dphi\ values far from $\pi$ is observed in the region $10 < \pTjetch < 20$~\gev\  for $\rr=0.4$ and 0.5, corresponding to medium-induced broadening of the acoplanarity distribution.

While these figures present fully-corrected distributions in regions of small signal to background at low \pTjetch, it is important to determine whether significant correlated signal in these regions is already evident in the raw data distributions, and is not introduced primarily by the correction procedures. This point is discussed in Sec.~\ref{sect:EstimatedKine}. As an additional cross-check, the \pp unfolded distributions were folded with the \PbPb response and compared to the corresponding raw \PbPb distributions. Significant broadening of the \dphi distribution with respect to this reference was also observed at low \pTjetch.

Figures~\ref{fig:DeltaPhi-pp_PbPb_02},~\ref{fig:DeltaPhi-pp_PbPb_04}, and~\ref{fig:DeltaPhi-pp_PbPb_05} also compare the measured \IAAphi\ to theoretical calculations.
The JETSCAPE calculation describes the $\rr=0.2$ data for $\pTjetch>20$ \gev, where the results of the calculation are available, while it underestimates the $\rr=0.4$ data for $20<\pTjetch< 30$~\gev\ and $\rr=0.5$ data for $20 < \pTjetch < 50$~\gev, with larger discrepancy farther from $\dphi=\pi$. 
The JEWEL calculation (both options) also describes the $\rr=0.2$ data for all \pTjetch\ intervals, with minimal difference between recoils-on and recoils-off. For larger \rr, JEWEL (recoils on) describes the data for all \pTjetch\ intervals and jet \rr, while JEWEL (recoils off) significantly underpredicts the data in the region $\pTjet<20$~\gev\ for $\rr=0.4$ and $\pTjet<50$~\gev\ for $\rr=0.5$, most significantly in the tails of the distributions. 
The Hybrid model underpredicts the magnitude of the \IAAphi\ for $\rr=0.2$ and $\rr=0.4$, for all model settings. The inclusion of wake effects increases the \IAAphi\ at low \pTjetch\ for $\rr=0.2$ and $\rr=0.4$, while the inclusion of elastic scattering moderately increases the prediction close to $\dphi=\pi$ in all \pTjetch\ intervals for $\rr=0.2$, and for $\pTjetch>20$~\gev\ for $\rr=0.4$ and $\rr=0.5$, bringing the predictions closer to data. Similar to the JEWEL calculation, the significant azimuthal broadening seen at low \pTjetch\ for $\rr=0.4$ and $\rr=0.5$ is qualitatively reproduced when including wake effects in the Hybrid model, although the magnitude of the broadening is underpredicted for $\rr=0.4$.
The pQCD calculations at LO reproduce the measured \IAApT\ distributions in the range $\pTjetch>20$ \gev\ for $\rr=0.2$ and 0.4, though over a restricted range in acoplanarity, $2.4<\dphi<\pi$. The data do not discriminate between the two values of quenching parameter in the calculation, $\langle\qhat L\rangle=$ 13 and 26 GeV$^2$. A higher-order calculation is required to extend the range of \dphi, with correspondingly greater discrimination of quenching parameters.

Overall, JEWEL (recoils on) describes the data the best over the full \rr\ and \pTjetch\ range, including the significant azimuthal broadening for low \pTjetch\ and large \rr. However, none of the models considered successfully describes the full set of measured data. The physics consequences of the systematic dependencies of medium-induced effects on \pTjetch\ and \rr, and the comparison of models to these data, are discussed in Ref.~\cite{ShorthJetpaper}

A measurement of energetic di-jets in \PbPb\ collisions at $\sqrtsNN=2.76$ TeV has also revealed significant broadening and softening of recoil-jet structure~\cite{CMS:2011iwn}. Such measurements, the results of this analysis, and inclusive jet production and jet substructure measurements, each probe a different aspect of the jet--medium interaction. A successful model of jet quenching must describe this full set of data correctly. A global analysis is required to ascertain whether a fully consistent description of all such data can be achieved by a suitable choice of model parameters, or whether the jet quenching mechanisms encoded in the model can be excluded by such a comprehensive comparison to multi-messenger jet quenching data.

\subsection{{\pmb{\IAApT}}: cross-check of 1-d and 2-d unfolding}
\label{sect:IAAUnfoldingCheck}

\begin{figure}[tbhp]
    \begin{center}
    \includegraphics[width = 0.6\textwidth] {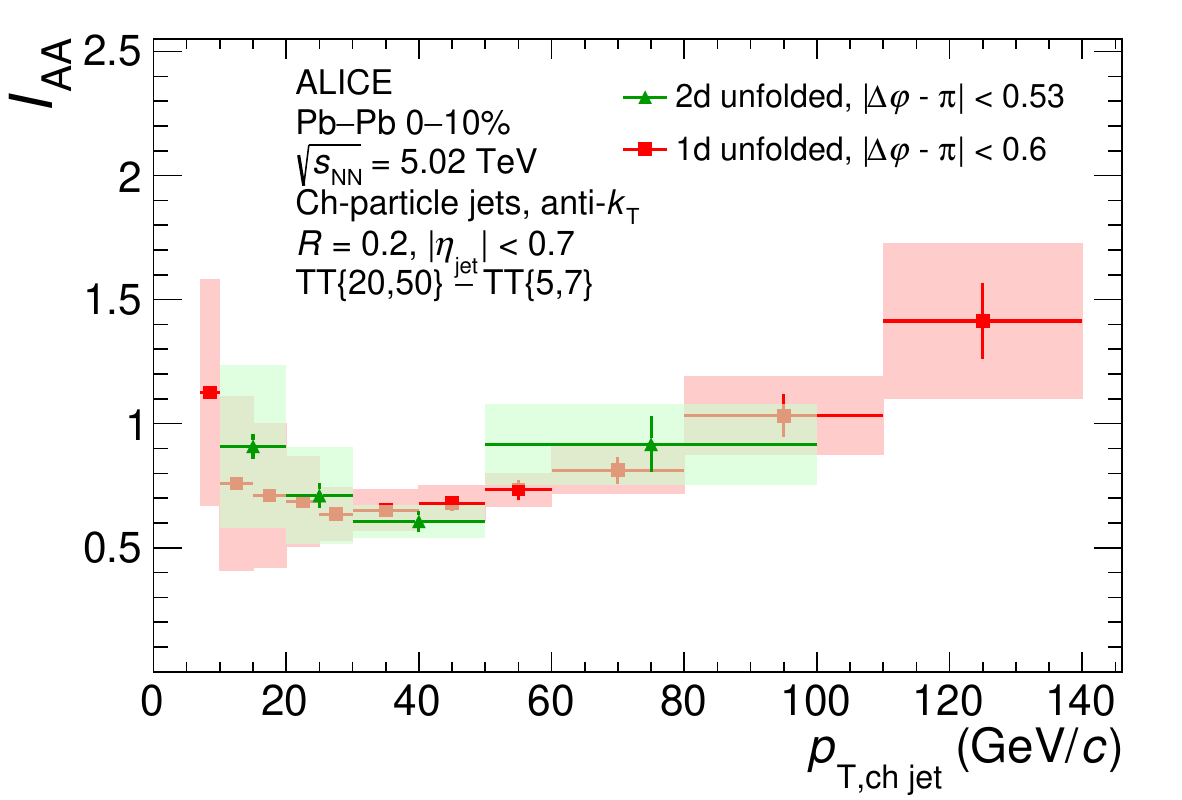}
    \includegraphics[width = 0.6\textwidth] {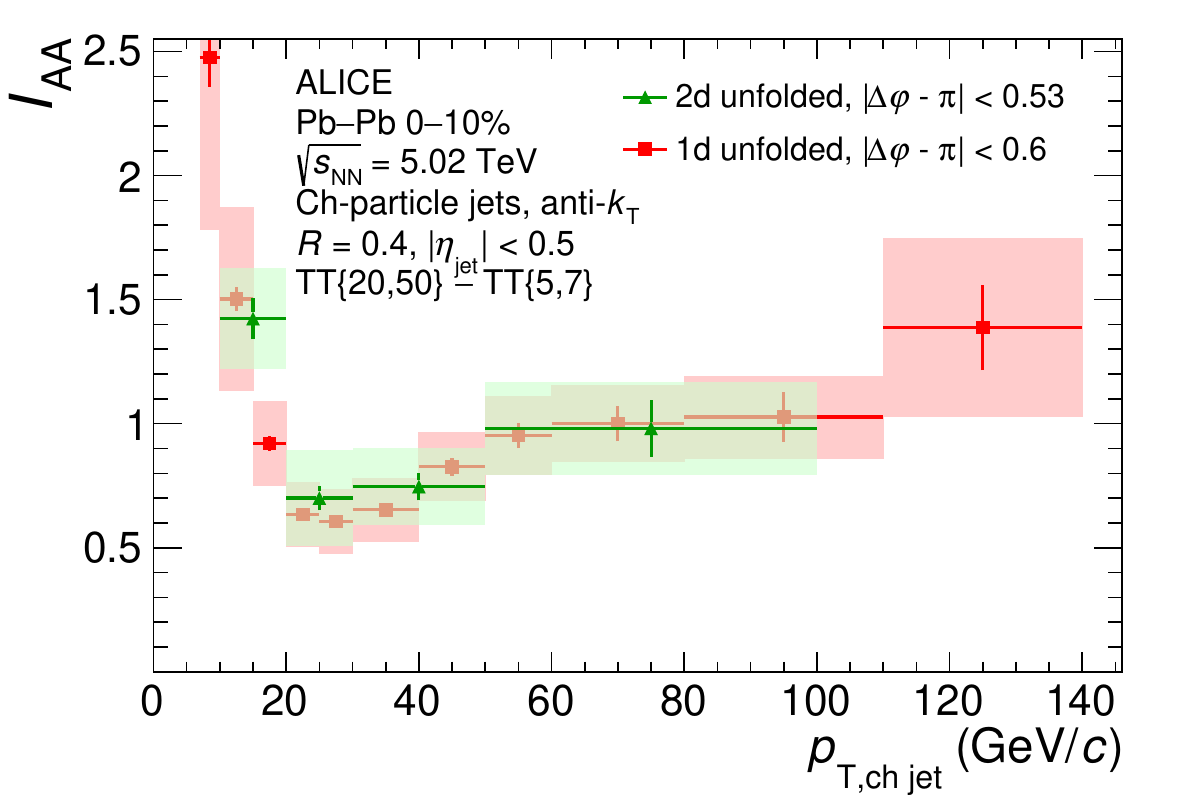}
    \includegraphics[width = 0.6\textwidth] {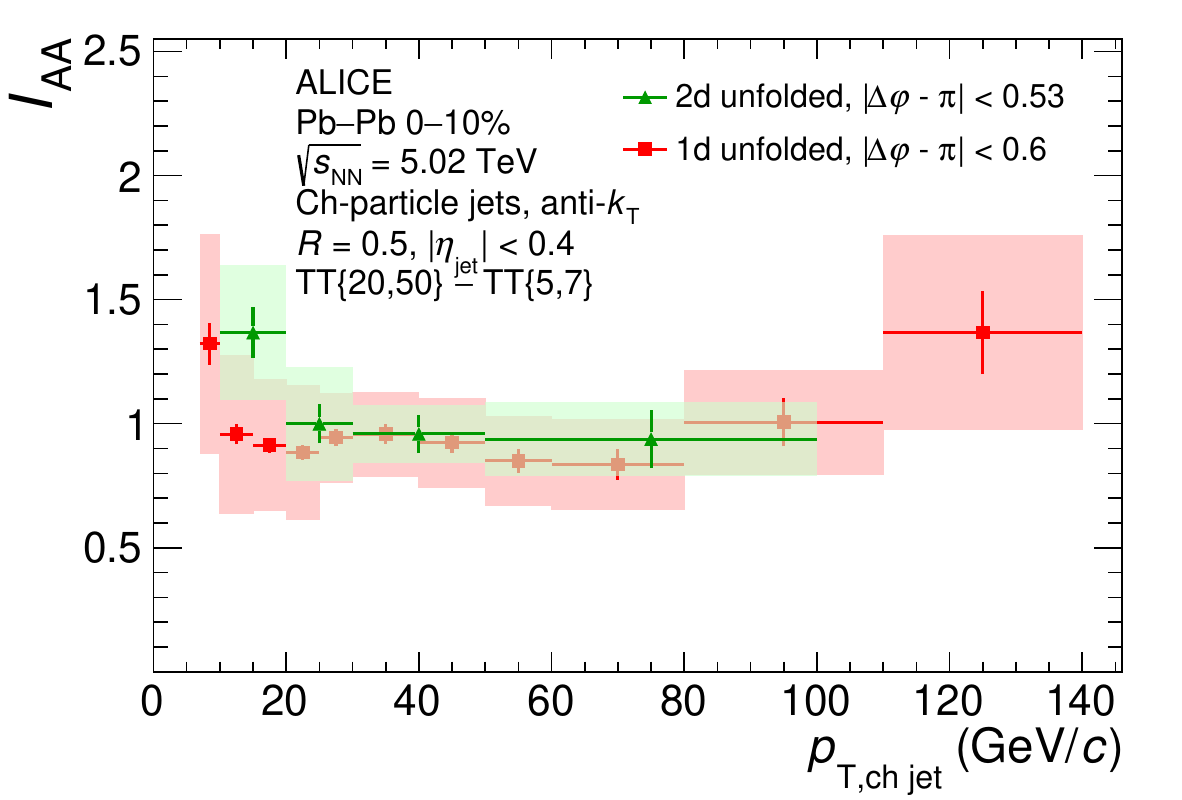}
  \end{center}
\caption{\IAApT\ from the \DrecoilpT\ distributions and \Drecoilphi\ distributions projected within similar \dphi\ intervals, measured for $\rr=0.2$ (top), 0.4 (middle), and 0.5 (bottom) in central \PbPb\ and \pp\ collisions.}
\label{fig:IAA1d2d}
\end{figure}

The distributions of \IAApT\ in Fig.~\ref{fig:IAA} appear to differ for $\rr=0.4$ and 0.5 in the range $\pTjetch<40$ \gev, whereas the distributions of \IAAphi\ in Figs.~\ref{fig:DeltaPhi-pp_PbPb_04} and~\ref{fig:DeltaPhi-pp_PbPb_05} appear to be similar for $\rr=0.4$ and 0.5 at low \pTjet\ (see also~\cite{ShorthJetpaper}). As discussed in Sect.~\ref{sect:CorrJets}, the analyses for these two sets of distributions differ in that, in order to achieve fine binning, corrected \DrecoilpT\ distributions are obtained by unfolding only in one dimension as a function of \pTjetch, taking advantage of small relative smearing in \dphi, whereas corrected \Drecoilphi\ distributions are obtained by unfolding in two dimensions, as a function of both \pTjetch\ and \dphi. 

However, \IAApT\ and \IAAphi\ are one-dimensional slices and projections of the same parent two-dimensional distribution, and physically must be consistent. In order to explore whether the \IAApT\ and \IAAphi\ distributions at low \pTjetch\ for $\rr=0.4$ and 0.5 are operationally consistent in the analysis, in light of the different choice of unfolding algorithm, Fig.~\ref{fig:IAA1d2d} compares \IAApT\ for $\rr = 0.2$, $0.4$, and $0.5$ determined from one-dimensional unfolding and from two-dimensional unfolding, where the latter \IAApT\ distribution is determined by taking a slice within a similar region $|\dphi-\pi|<0.53$ and projecting onto \pTjetch.

The distributions from the two different unfolding approaches are seen to be consistent within experimental uncertainties, with the results for $\rr=0.5$ in the lower \pTjetch\ region of the ``2d unfolded'' result being at the upper edge of the uncertainty band of the ``1d unfolded'' result. Note that the dominant uncertainties, which are due to unfolding and to the $\cRef$ correction, are largely uncorrelated between the two analyses.

\section{Summary and Outlook}
\label{sect:Summary}

This paper reports measurements of semi-inclusive distributions of charged-particle jets recoiling from a high-\pT\ hadron trigger in \pp\ and central \PbPb\ collisions at $\sqrtsNN=5.02$ TeV, using the large data samples recorded with the ALICE detector during LHC Run 2. The results are reported as a function of both \pTjet, the recoil jet transverse momentum, and \dphi, the azimuthal separation between the trigger and recoil jet. A statistical, data-driven method previously developed by the ALICE Collaboration is used to mitigate the large uncorrelated background jet yield in central \PbPb\ collisions, enabling measurements of jet quenching in a kinematic region previously unexplored by reconstructed jets at the LHC, including low $\pTjet\approx10$ \gev\ with jet resolution parameter $\rr = 0.5$. The observed phenomena explore several different aspects of jet production in \pp\ and \PbPb\ collisions.

The \pTjet\ and azimuthal distributions measured in \pp\ collisions provide a precise reference to explore medium-induced modifications to jet production in central \PbPb\ collisions, and are well described by pQCD-based calculations over the entire measured ranges. The ratio of recoil jet yields in \pp\ collisions for $\rr=0.2$ to that for $\rr=0.4$ or 0.5 is below unity at high \pT, reflecting the well-established transverse profile of energy within a jet in vacuum. However, this ratio is observed to increase as \pTjet\ is reduced below the value of \pTtrig, the trigger hadron \pT, in marked contrast to the behavior of a similar ratio measured for inclusive jet cross sections. Both sets of measurements are well described by pQCD calculations, suggesting that these opposing effects may arise from different jet production mechanisms, in particular suppression of leading order processes for the semi-inclusive population of jets recoiling from a high-\pT\ hadron trigger, in which \pTtrig\ provides an additional scale.

The measured values of \IAApT, the ratio of recoil yield for \PbPb\ and \pp\ collisions for the same jet \rr\ as a function of \pTjetch, exhibit a dependence on \pTjet\ and \rr. For $\rr=0.5$, the ratio is consistent with unity within uncertainties over the entire measured range, indicating that medium-induced jet modifications as probed by this observable are largely constrained to angular scales less than 0.5 radians. The ratio of recoil jet yield in \PbPb\ collisions for $\rr=0.2$ to that for $\rr=0.4$ or 0.5 is below that for \pp\ collisions at intermediate \pTjet, indicating medium-induced intra-jet broadening within this angular scale. 

For $\rr=0.2$ and 0.4, the value of \IAApT\ is below unity at intermediate \pTjet, increasing to unity at both lower and higher \pTjet. For $\rr=0.4$, \IAApT\ exceeds unity at the lowest value of \pTjet\ reported here. Comparison to models indicates that the low-\pTjet\ behavior may be due to the recovery of energy lost to the medium by higher-\pTjet\ jets that are likewise correlated with the trigger. The high-\pTjet\ behavior may arise from the interplay between the energy loss due to jet quenching and the geometric bias induced by using a hadron trigger.

The measured values of \IAAphi, the ratio of recoil yield for \PbPb\ and \pp\ collisions for the same jet \rr\ as a function of \dphi, provide the first measurement of significant in-medium jet acoplanarity broadening in \PbPb\ collisions, for $\rr=0.4$ and 0.5 at low \pTjetch. 

These measurements and model calculations are discussed further in the companion Letter. Global analyses incorporating these data will enable quantitative, multi-messenger studies to further elucidate the nature of the QGP as probed by jets.


\newenvironment{acknowledgement}{\relax}{\relax}
\begin{acknowledgement}
\section*{Acknowledgements}
We thank Daniel Pablos, Krishna Rajagopal, and Zachary Hulcher for providing the Hybrid Model calculations, Shuyi Wei and Guangyou Qin for providing the pQCD@LO calculations, the JETSCAPE collaboration for guidance in using the JETSCAPE framework, and Raghav Kunnawalkam Elayavalli for providing the medium parameters for JEWEL simulations.

The ALICE Collaboration would like to thank all its engineers and technicians for their invaluable contributions to the construction of the experiment and the CERN accelerator teams for the outstanding performance of the LHC complex.
The ALICE Collaboration gratefully acknowledges the resources and support provided by all Grid centres and the Worldwide LHC Computing Grid (WLCG) collaboration.
The ALICE Collaboration acknowledges the following funding agencies for their support in building and running the ALICE detector:
A. I. Alikhanyan National Science Laboratory (Yerevan Physics Institute) Foundation (ANSL), State Committee of Science and World Federation of Scientists (WFS), Armenia;
Austrian Academy of Sciences, Austrian Science Fund (FWF): [M 2467-N36] and Nationalstiftung f\"{u}r Forschung, Technologie und Entwicklung, Austria;
Ministry of Communications and High Technologies, National Nuclear Research Center, Azerbaijan;
Conselho Nacional de Desenvolvimento Cient\'{\i}fico e Tecnol\'{o}gico (CNPq), Financiadora de Estudos e Projetos (Finep), Funda\c{c}\~{a}o de Amparo \`{a} Pesquisa do Estado de S\~{a}o Paulo (FAPESP) and Universidade Federal do Rio Grande do Sul (UFRGS), Brazil;
Bulgarian Ministry of Education and Science, within the National Roadmap for Research Infrastructures 2020--2027 (object CERN), Bulgaria;
Ministry of Education of China (MOEC) , Ministry of Science \& Technology of China (MSTC) and National Natural Science Foundation of China (NSFC), China;
Ministry of Science and Education and Croatian Science Foundation, Croatia;
Centro de Aplicaciones Tecnol\'{o}gicas y Desarrollo Nuclear (CEADEN), Cubaenerg\'{\i}a, Cuba;
Ministry of Education, Youth and Sports of the Czech Republic, Czech Republic;
The Danish Council for Independent Research | Natural Sciences, the VILLUM FONDEN and Danish National Research Foundation (DNRF), Denmark;
Helsinki Institute of Physics (HIP), Finland;
Commissariat \`{a} l'Energie Atomique (CEA) and Institut National de Physique Nucl\'{e}aire et de Physique des Particules (IN2P3) and Centre National de la Recherche Scientifique (CNRS), France;
Bundesministerium f\"{u}r Bildung und Forschung (BMBF) and GSI Helmholtzzentrum f\"{u}r Schwerionenforschung GmbH, Germany;
General Secretariat for Research and Technology, Ministry of Education, Research and Religions, Greece;
National Research, Development and Innovation Office, Hungary;
Department of Atomic Energy Government of India (DAE), Department of Science and Technology, Government of India (DST), University Grants Commission, Government of India (UGC) and Council of Scientific and Industrial Research (CSIR), India;
National Research and Innovation Agency - BRIN, Indonesia;
Istituto Nazionale di Fisica Nucleare (INFN), Italy;
Japanese Ministry of Education, Culture, Sports, Science and Technology (MEXT) and Japan Society for the Promotion of Science (JSPS) KAKENHI, Japan;
Consejo Nacional de Ciencia (CONACYT) y Tecnolog\'{i}a, through Fondo de Cooperaci\'{o}n Internacional en Ciencia y Tecnolog\'{i}a (FONCICYT) and Direcci\'{o}n General de Asuntos del Personal Academico (DGAPA), Mexico;
Nederlandse Organisatie voor Wetenschappelijk Onderzoek (NWO), Netherlands;
The Research Council of Norway, Norway;
Commission on Science and Technology for Sustainable Development in the South (COMSATS), Pakistan;
Pontificia Universidad Cat\'{o}lica del Per\'{u}, Peru;
Ministry of Education and Science, National Science Centre and WUT ID-UB, Poland;
Korea Institute of Science and Technology Information and National Research Foundation of Korea (NRF), Republic of Korea;
Ministry of Education and Scientific Research, Institute of Atomic Physics, Ministry of Research and Innovation and Institute of Atomic Physics and University Politehnica of Bucharest, Romania;
Ministry of Education, Science, Research and Sport of the Slovak Republic, Slovakia;
National Research Foundation of South Africa, South Africa;
Swedish Research Council (VR) and Knut \& Alice Wallenberg Foundation (KAW), Sweden;
European Organization for Nuclear Research, Switzerland;
Suranaree University of Technology (SUT), National Science and Technology Development Agency (NSTDA), Thailand Science Research and Innovation (TSRI) and National Science, Research and Innovation Fund (NSRF), Thailand;
Turkish Energy, Nuclear and Mineral Research Agency (TENMAK), Turkey;
National Academy of  Sciences of Ukraine, Ukraine;
Science and Technology Facilities Council (STFC), United Kingdom;
National Science Foundation of the United States of America (NSF) and United States Department of Energy, Office of Nuclear Physics (DOE NP), United States of America.
In addition, individual groups or members have received support from:
European Research Council, Strong 2020 - Horizon 2020 (grant nos. 950692, 824093), European Union;
Academy of Finland (Center of Excellence in Quark Matter) (grant nos. 346327, 346328), Finland;
\end{acknowledgement}

\bibliographystyle{utphys}   
\bibliography{Common/bibliography}

\newpage
\appendix

\section{The ALICE Collaboration}
\label{app:collab}
\begin{flushleft} 
\small

S.~Acharya\,\orcidlink{0000-0002-9213-5329}\,$^{\rm 128}$, 
D.~Adamov\'{a}\,\orcidlink{0000-0002-0504-7428}\,$^{\rm 87}$, 
G.~Aglieri Rinella\,\orcidlink{0000-0002-9611-3696}\,$^{\rm 33}$, 
M.~Agnello\,\orcidlink{0000-0002-0760-5075}\,$^{\rm 30}$, 
N.~Agrawal\,\orcidlink{0000-0003-0348-9836}\,$^{\rm 52}$, 
Z.~Ahammed\,\orcidlink{0000-0001-5241-7412}\,$^{\rm 136}$, 
S.~Ahmad\,\orcidlink{0000-0003-0497-5705}\,$^{\rm 16}$, 
S.U.~Ahn\,\orcidlink{0000-0001-8847-489X}\,$^{\rm 72}$, 
I.~Ahuja\,\orcidlink{0000-0002-4417-1392}\,$^{\rm 38}$, 
A.~Akindinov\,\orcidlink{0000-0002-7388-3022}\,$^{\rm 142}$
% ...

\section*{Affiliation Notes}

$^{\rm I}$ Deceased\\
$^{\rm II}$ Also at: Max-Planck-Institut fur Physik, Munich, Germany\\
$^{\rm III}$ Also at: Italian National Agency for New Technologies, Energy and Sustainable Economic Development (ENEA), Bologna, Italy\\
$^{\rm IV}$ Also at: Dipartimento DET del Politecnico di Torino, Turin, Italy\\
$^{\rm V}$ Also at: Yildiz Technical University, Istanbul, T\"{u}rkiye\\
$^{\rm VI}$ Also at: Department of Applied Physics, Aligarh Muslim University, Aligarh, India\\
$^{\rm VII}$ Also at: Institute of Theoretical Physics, University of Wroclaw, Poland\\
$^{\rm VIII}$ Also at: An institution covered by a cooperation agreement with CERN\\

\section*{Collaboration Institutes}

$^{1}$ A.I. Alikhanyan National Science Laboratory (Yerevan Physics Institute) Foundation, Yerevan, Armenia\\
$^{2}$ AGH University of Krakow, Cracow, Poland\\
$^{3}$ Bogolyubov Institute for Theoretical Physics, National Academy of Sciences of Ukraine, Kiev, Ukraine\\
$^{4}$ Bose Institute, Department of Physics  and Centre for Astroparticle Physics and Space Science (CAPSS), Kolkata, India\\
$^{5}$ California Polytechnic State University, San Luis Obispo, California, United States\\
$^{6}$ Central China Normal University, Wuhan, China\\
$^{7}$ Centro de Aplicaciones Tecnol\'{o}gicas y Desarrollo Nuclear (CEADEN), Havana, Cuba\\
$^{8}$ Centro de Investigaci\'{o}n y de Estudios Avanzados (CINVESTAV), Mexico City and M\'{e}rida, Mexico\\
$^{9}$ Chicago State University, Chicago, Illinois, United States\\
$^{10}$ China Institute of Atomic Energy, Beijing, China\\
$^{11}$ China University of Geosciences, Wuhan, China\\
$^{12}$ Chungbuk National University, Cheongju, Republic of Korea\\
$^{13}$ Comenius University Bratislava, Faculty of Mathematics, Physics and Informatics, Bratislava, Slovak Republic\\
$^{14}$ COMSATS University Islamabad, Islamabad, Pakistan\\
$^{15}$ Creighton University, Omaha, Nebraska, United States\\
$^{16}$ Department of Physics, Aligarh Muslim University, Aligarh, India\\
$^{17}$ Department of Physics, Pusan National University, Pusan, Republic of Korea\\
$^{18}$ Department of Physics, Sejong University, Seoul, Republic of Korea\\
$^{19}$ Department of Physics, University of California, Berkeley, California, United States\\
$^{20}$ Department of Physics, University of Oslo, Oslo, Norway\\
$^{21}$ Department of Physics and Technology, University of Bergen, Bergen, Norway\\
$^{22}$ Dipartimento di Fisica, Universit\`{a} di Pavia, Pavia, Italy\\
$^{23}$ Dipartimento di Fisica dell'Universit\`{a} and Sezione INFN, Cagliari, Italy\\
$^{24}$ Dipartimento di Fisica dell'Universit\`{a} and Sezione INFN, Trieste, Italy\\
$^{25}$ Dipartimento di Fisica dell'Universit\`{a} and Sezione INFN, Turin, Italy\\
$^{26}$ Dipartimento di Fisica e Astronomia dell'Universit\`{a} and Sezione INFN, Bologna, Italy\\
$^{27}$ Dipartimento di Fisica e Astronomia dell'Universit\`{a} and Sezione INFN, Catania, Italy\\
$^{28}$ Dipartimento di Fisica e Astronomia dell'Universit\`{a} and Sezione INFN, Padova, Italy\\
$^{29}$ Dipartimento di Fisica `E.R.~Caianiello' dell'Universit\`{a} and Gruppo Collegato INFN, Salerno, Italy\\
$^{30}$ Dipartimento DISAT del Politecnico and Sezione INFN, Turin, Italy\\
$^{31}$ Dipartimento di Scienze MIFT, Universit\`{a} di Messina, Messina, Italy\\
$^{32}$ Dipartimento Interateneo di Fisica `M.~Merlin' and Sezione INFN, Bari, Italy\\
$^{33}$ European Organization for Nuclear Research (CERN), Geneva, Switzerland\\
$^{34}$ Faculty of Electrical Engineering, Mechanical Engineering and Naval Architecture, University of Split, Split, Croatia\\
$^{35}$ Faculty of Engineering and Science, Western Norway University of Applied Sciences, Bergen, Norway\\
$^{36}$ Faculty of Nuclear Sciences and Physical Engineering, Czech Technical University in Prague, Prague, Czech Republic\\
$^{37}$ Faculty of Physics, Sofia University, Sofia, Bulgaria\\
$^{38}$ Faculty of Science, P.J.~\v{S}af\'{a}rik University, Ko\v{s}ice, Slovak Republic\\
$^{39}$ Frankfurt Institute for Advanced Studies, Johann Wolfgang Goethe-Universit\"{a}t Frankfurt, Frankfurt, Germany\\
$^{40}$ Fudan University, Shanghai, China\\
$^{41}$ Gangneung-Wonju National University, Gangneung, Republic of Korea\\
$^{42}$ Gauhati University, Department of Physics, Guwahati, India\\
$^{43}$ Helmholtz-Institut f\"{u}r Strahlen- und Kernphysik, Rheinische Friedrich-Wilhelms-Universit\"{a}t Bonn, Bonn, Germany\\
$^{44}$ Helsinki Institute of Physics (HIP), Helsinki, Finland\\
$^{45}$ High Energy Physics Group,  Universidad Aut\'{o}noma de Puebla, Puebla, Mexico\\
$^{46}$ Horia Hulubei National Institute of Physics and Nuclear Engineering, Bucharest, Romania\\
$^{47}$ HUN-REN Wigner Research Centre for Physics, Budapest, Hungary\\
$^{48}$ Indian Institute of Technology Bombay (IIT), Mumbai, India\\
$^{49}$ Indian Institute of Technology Indore, Indore, India\\
$^{50}$ INFN, Laboratori Nazionali di Frascati, Frascati, Italy\\
$^{51}$ INFN, Sezione di Bari, Bari, Italy\\
$^{52}$ INFN, Sezione di Bologna, Bologna, Italy\\
$^{53}$ INFN, Sezione di Cagliari, Cagliari, Italy\\
$^{54}$ INFN, Sezione di Catania, Catania, Italy\\
$^{55}$ INFN, Sezione di Padova, Padova, Italy\\
$^{56}$ INFN, Sezione di Pavia, Pavia, Italy\\
$^{57}$ INFN, Sezione di Torino, Turin, Italy\\
$^{58}$ INFN, Sezione di Trieste, Trieste, Italy\\
$^{59}$ Inha University, Incheon, Republic of Korea\\
$^{60}$ Institute for Gravitational and Subatomic Physics (GRASP), Utrecht University/Nikhef, Utrecht, Netherlands\\
$^{61}$ Institute of Experimental Physics, Slovak Academy of Sciences, Ko\v{s}ice, Slovak Republic\\
$^{62}$ Institute of Physics, Homi Bhabha National Institute, Bhubaneswar, India\\
$^{63}$ Institute of Physics of the Czech Academy of Sciences, Prague, Czech Republic\\
$^{64}$ Institute of Space Science (ISS), Bucharest, Romania\\
$^{65}$ Institut f\"{u}r Kernphysik, Johann Wolfgang Goethe-Universit\"{a}t Frankfurt, Frankfurt, Germany\\
$^{66}$ Instituto de Ciencias Nucleares, Universidad Nacional Aut\'{o}noma de M\'{e}xico, Mexico City, Mexico\\
$^{67}$ Instituto de F\'{i}sica, Universidade Federal do Rio Grande do Sul (UFRGS), Porto Alegre, Brazil\\
$^{68}$ Instituto de F\'{\i}sica, Universidad Nacional Aut\'{o}noma de M\'{e}xico, Mexico City, Mexico\\
$^{69}$ iThemba LABS, National Research Foundation, Somerset West, South Africa\\
$^{70}$ Jeonbuk National University, Jeonju, Republic of Korea\\
$^{71}$ Johann-Wolfgang-Goethe Universit\"{a}t Frankfurt Institut f\"{u}r Informatik, Fachbereich Informatik und Mathematik, Frankfurt, Germany\\
$^{72}$ Korea Institute of Science and Technology Information, Daejeon, Republic of Korea\\
$^{73}$ KTO Karatay University, Konya, Turkey\\
$^{74}$ Laboratoire de Physique Subatomique et de Cosmologie, Universit\'{e} Grenoble-Alpes, CNRS-IN2P3, Grenoble, France\\
$^{75}$ Lawrence Berkeley National Laboratory, Berkeley, California, United States\\
$^{76}$ Lund University Department of Physics, Division of Particle Physics, Lund, Sweden\\
$^{77}$ Nagasaki Institute of Applied Science, Nagasaki, Japan\\
$^{78}$ Nara Women{'}s University (NWU), Nara, Japan\\
$^{79}$ National and Kapodistrian University of Athens, School of Science, Department of Physics , Athens, Greece\\
$^{80}$ National Centre for Nuclear Research, Warsaw, Poland\\
$^{81}$ National Institute of Science Education and Research, Homi Bhabha National Institute, Jatni, India\\
$^{82}$ National Nuclear Research Center, Baku, Azerbaijan\\
$^{83}$ National Research and Innovation Agency - BRIN, Jakarta, Indonesia\\
$^{84}$ Niels Bohr Institute, University of Copenhagen, Copenhagen, Denmark\\
$^{85}$ Nikhef, National institute for subatomic physics, Amsterdam, Netherlands\\
$^{86}$ Nuclear Physics Group, STFC Daresbury Laboratory, Daresbury, United Kingdom\\
$^{87}$ Nuclear Physics Institute of the Czech Academy of Sciences, Husinec-\v{R}e\v{z}, Czech Republic\\
$^{88}$ Oak Ridge National Laboratory, Oak Ridge, Tennessee, United States\\
$^{89}$ Ohio State University, Columbus, Ohio, United States\\
$^{90}$ Physics department, Faculty of science, University of Zagreb, Zagreb, Croatia\\
$^{91}$ Physics Department, Panjab University, Chandigarh, India\\
$^{92}$ Physics Department, University of Jammu, Jammu, India\\
$^{93}$ Physics Program and International Institute for Sustainability with Knotted Chiral Meta Matter (SKCM2), Hiroshima University, Hiroshima, Japan\\
$^{94}$ Physikalisches Institut, Eberhard-Karls-Universit\"{a}t T\"{u}bingen, T\"{u}bingen, Germany\\
$^{95}$ Physikalisches Institut, Ruprecht-Karls-Universit\"{a}t Heidelberg, Heidelberg, Germany\\
$^{96}$ Physik Department, Technische Universit\"{a}t M\"{u}nchen, Munich, Germany\\
$^{97}$ Politecnico di Bari and Sezione INFN, Bari, Italy\\
$^{98}$ Research Division and ExtreMe Matter Institute EMMI, GSI Helmholtzzentrum f\"ur Schwerionenforschung GmbH, Darmstadt, Germany\\
$^{99}$ Saga University, Saga, Japan\\
$^{100}$ Saha Institute of Nuclear Physics, Homi Bhabha National Institute, Kolkata, India\\
$^{101}$ School of Physics and Astronomy, University of Birmingham, Birmingham, United Kingdom\\
$^{102}$ Secci\'{o}n F\'{\i}sica, Departamento de Ciencias, Pontificia Universidad Cat\'{o}lica del Per\'{u}, Lima, Peru\\
$^{103}$ Stefan Meyer Institut f\"{u}r Subatomare Physik (SMI), Vienna, Austria\\
$^{104}$ SUBATECH, IMT Atlantique, Nantes Universit\'{e}, CNRS-IN2P3, Nantes, France\\
$^{105}$ Sungkyunkwan University, Suwon City, Republic of Korea\\
$^{106}$ Suranaree University of Technology, Nakhon Ratchasima, Thailand\\
$^{107}$ Technical University of Ko\v{s}ice, Ko\v{s}ice, Slovak Republic\\
$^{108}$ The Henryk Niewodniczanski Institute of Nuclear Physics, Polish Academy of Sciences, Cracow, Poland\\
$^{109}$ The University of Texas at Austin, Austin, Texas, United States\\
$^{110}$ Universidad Aut\'{o}noma de Sinaloa, Culiac\'{a}n, Mexico\\
$^{111}$ Universidade de S\~{a}o Paulo (USP), S\~{a}o Paulo, Brazil\\
$^{112}$ Universidade Estadual de Campinas (UNICAMP), Campinas, Brazil\\
$^{113}$ Universidade Federal do ABC, Santo Andre, Brazil\\
$^{114}$ Universitatea Nationala de Stiinta si Tehnologie Politehnica Bucuresti, Bucharest, Romania\\
$^{115}$ University of Cape Town, Cape Town, South Africa\\
$^{116}$ University of Derby, Derby, United Kingdom\\
$^{117}$ University of Houston, Houston, Texas, United States\\
$^{118}$ University of Jyv\"{a}skyl\"{a}, Jyv\"{a}skyl\"{a}, Finland\\
$^{119}$ University of Kansas, Lawrence, Kansas, United States\\
$^{120}$ University of Liverpool, Liverpool, United Kingdom\\
$^{121}$ University of Science and Technology of China, Hefei, China\\
$^{122}$ University of South-Eastern Norway, Kongsberg, Norway\\
$^{123}$ University of Tennessee, Knoxville, Tennessee, United States\\
$^{124}$ University of the Witwatersrand, Johannesburg, South Africa\\
$^{125}$ University of Tokyo, Tokyo, Japan\\
$^{126}$ University of Tsukuba, Tsukuba, Japan\\
$^{127}$ Universit\"{a}t M\"{u}nster, Institut f\"{u}r Kernphysik, M\"{u}nster, Germany\\
$^{128}$ Universit\'{e} Clermont Auvergne, CNRS/IN2P3, LPC, Clermont-Ferrand, France\\
$^{129}$ Universit\'{e} de Lyon, CNRS/IN2P3, Institut de Physique des 2 Infinis de Lyon, Lyon, France\\
$^{130}$ Universit\'{e} de Strasbourg, CNRS, IPHC UMR 7178, F-67000 Strasbourg, France, Strasbourg, France\\
$^{131}$ Universit\'{e} Paris-Saclay, Centre d'Etudes de Saclay (CEA), IRFU, D\'{e}partment de Physique Nucl\'{e}aire (DPhN), Saclay, France\\
$^{132}$ Universit\'{e}  Paris-Saclay, CNRS/IN2P3, IJCLab, Orsay, France\\
$^{133}$ Universit\`{a} degli Studi di Foggia, Foggia, Italy\\
$^{134}$ Universit\`{a} del Piemonte Orientale, Vercelli, Italy\\
$^{135}$ Universit\`{a} di Brescia, Brescia, Italy\\
$^{136}$ Variable Energy Cyclotron Centre, Homi Bhabha National Institute, Kolkata, India\\
$^{137}$ Warsaw University of Technology, Warsaw, Poland\\
$^{138}$ Wayne State University, Detroit, Michigan, United States\\
$^{139}$ Yale University, New Haven, Connecticut, United States\\
$^{140}$ Yonsei University, Seoul, Republic of Korea\\
$^{141}$  Zentrum  f\"{u}r Technologie und Transfer (ZTT), Worms, Germany\\
$^{142}$ Affiliated with an institute covered by a cooperation agreement with CERN\\
$^{143}$ Affiliated with an international laboratory covered by a cooperation agreement with CERN.\\

\end{flushleft} 
  
\end{document}